\documentclass{article}
\usepackage{eurosym}
\usepackage[utf8]{inputenc}
\usepackage[T1]{fontenc}
\usepackage[english]{babel}
\usepackage{amsmath}
\usepackage{amssymb}
\usepackage{graphicx}
\usepackage[margin=2cm]{geometry}
\usepackage{url}
\usepackage{placeins}

\newtheorem{theorem}{Theorem}

\newenvironment{proof}[1][Proof]{\noindent\textbf{#1.} }{\ \rule{0.5em}{0.5em}}
\input{tcilatex}
\begin{document}
	
	\begin{center}
		\bigskip
		
		\bigskip
		
		\textbf{ Instant cost and delayed reward. Demographic eco-evolutionary game dynamics under the impact of the delay resulting from the offspring maturation time.}\bigskip
		
		\bigskip
		
		Krzysztof Argasinski*,
		
		Faculty of Mathematics Informatics and Mechanics
		
		University of Warsaw
		
		ul. Stefana Banacha 2
		
		02-097 Warszawa
		
		Poland
		
		\bigskip
		
		Ryszard Rudnicki
		
		Institute of Mathematics of Polish Academy of Sciences
		
		ul. \'{S}niadeckich 8
		
		00-656 Warszawa
		
		\bigskip
		
		Robert Szczelina
		
		Faculty Of Mathematics and Computer Science
		
		Jagiellonian University, ul.\ Łojasiewicza 6, 30-348 Kraków, Poland
		
		Poland
	\end{center}
	
	*corresponding author: \textit{argas1@wp.pl}\bigskip
	
	\textbf{Acknowledgments}: We would like to thank Mark Broom for support of the project. This paper was supported by the
	Polish National Science Centre Grant No.OPUS 2020/39/B/NZ8/03485 (KA) and in part by Polish National Science Centre Grant 2023/49/B/ST6/02801, and in part by NAWA Bekker Scholarship Program under grant no. BPN/BEK/2023/1/00170 (RS). \textbf{THIS IS A WORK IN PROGRESS VERSION, IN THE CASE OF DISCOVERED ERRORS PLEASE CONTACT THE CORRESPONDING AUTHOR}
	
	\bigskip \bigskip
	
	\begin{abstract}
		In this paper, we extend the demographic eco-evolutionary game approach, based on explicit birth and death dynamics instead of abstract "fitness" interpreted as an abstract "Malthusian parameter", by the introduction of the delay resulting from the juvenile maturation time. This leads to the application of the Delay Differential Equations (DDE). We show that delay seriously affects the resulting dynamics and may lead to the loss of stability of equilibria when critical delay is exceeded. We provide theoretical tools for the assessment of the critical delays and the parameter values when this may happen. Our results emphasize the importance of the mechanisms of density dependence. We analyze the impact of three different suppression modes based on: adult mortality, juvenile recruitment survival after the maturation period (without delay), and juvenile recruitment at birth (with the delay). The last mode leads to extreme patterns such as bifurcations, complex cycles, and chaotic dynamics. However, surprisingly, this mode leads to extension of the duration of the temporary transient metastable states known as "ghost attractors". In addition, we also focus on the problem of resilience of the analyzed systems against external periodic perturbations and feedback-driven factors such as additional predator pressure.   
	\end{abstract}

	.\newpage

\section{Introduction}

George Evelyn Hutchinson's seminal essay, "The Ecological Theater and the
Evolutionary Play" \cite{hutch}, underscored the vital connection between ecology
and evolutionary theory. Presently, evolutionary biology research
prioritizes eco-evolutionary synthesis, with a particular focus on
understanding eco-evolutionary feedbacks \cite{Govaert,Post,Pelletier,Hanski,Hendry1,Hendry2}. An example
of such feedback within evolutionary game theory, was revealed by studies on
demographic games \cite{argbr1,argbr2,argbr3}. This approach
explored the interplay between ecological factors shaping the population
dynamics and selection dynamics.

In contrast to standard evolutionary game theory \cite{maynard1,maynard2,weibull,hofsig1,hofsig2,BroomRychtar,Sinervo}, which links "payoffs" to
"Darwinian fitness", the demographic approach considers two payoff functions
related to reproductive success and the risk of death to the player. The definition of Darwinian fitness through demography and explicit balance between birth and death rates was advocated by \cite{doebeli,Metcalf1, Metcalf2, Bertram} and it solves the methodological issue known as "tautology problem" in evolutionary theory (the "fittest" survive and succesfully reproduce, while the fittest are those with best survival and reproductive output, \cite{peters1976tautology,peters1978predictable,van1983methodological,von2016survival}).
The new approach emphasizes the importance of the density dependent growth
limiting factors. In the standard replicator dynamics, the growth
suppression can act through adult mortality, and cancels out in the
replicator equation. In contrast, the aproach of Argasinski and Broom
assumes that the density dependence is driven by juvenile recruitment
survival, described by logistic term acting on the fertility rate.

Recognizing the importance of density-dependent juvenile recruitment,
further research explored more realistic suppression models than the
classical logistic approach \cite{argRudNSL1,argRudNSL2,argRudState}, The
demographic approach allowed for the development of replicator
dynamics models, with explicit age structure \cite{argbr4age} and
state based replicator dynamics \cite{argRudState} related to the
approach of Houston and McNamara \cite{McNamaraState}. This paved the way towards
integrating the evolutionary game theory (focused on the evolution of
behaviour) and life history theory (focused on the evolution of individual
life cycles, \cite{Roff,Stearns}). Due to clear mechanistic interpretation of parameters the demographic framework can be easily completed by individual based \cite{VolkerGrimm} or Monte Carlo simulations \cite{Araujo} acting as \emph{in silico} experiments.

In this paper we will investigate the impact of the important life history
parameter, which is maturation time, on the population dynamics and the
trajectories of selection. Because models with explicit age structure \cite{argbr4age}
 are complicated, in our initial research we will
use much simpler but still interesting approach based on delay differential
equations. Delay differential equations are already used in evolutionary
game theory (\cite{Alboszta,Ben-Khalifa,Zhong1,Zhong2,MiekiszBodnar,JavadBodnar} and the references therein).
However, combination of delays and demographic approach can be very
profitable. When we add delay to ferility payoff functions it can be
interpreted as the maturation or egg hatching time, very important parameter
from the biological point of view. This will introduce aspect related to
life history theory \cite{Roff,Stearns} into game theoretic framework,
which was advocated by \cite{McNamaraGame2013}. However, the resulting impact will
be much broader and important from the point of view of the questions on the
definition of Darwinian fitness and it's role in modelling and theoretical
framework.

\newpage

List of important symbols:

\begin{tabular}{ll}
$n_{i}$ & number of $i$-strategy indiviuals \\ 
$e_{i}$ &  $i$-th cannonical unit vector \\ 
$n=n_{1}+n_{2}$ & populations size \\ 
$q_{i}=n_{i}/(n_{1}+n_{2})$ & frequency of the $i$-th strategy \\ 
$\tau =1$ & virtual interaction rate allowing for demographic interpretation
\\ 
& of payoffs in differential equation \\ 
$V$ & fertility matrix, entries $V_{i,j}$ describe number of offspring \\ 
& produced during game round between $i$ and $j$ strategy carriers \\ 
$D$ & mortality payoff matrix, entries $D_{i,j}$ describe probability of
death \\ 
& of $i$th strategy player during the game round against $j$th strategy \\ 
& opponent \\ 
$\Phi $ & background fertility rate (intensity of the background birth) \\ 
$\Psi $ & background mortality rate (intensity of the background death) \\ 
$\left( 1-\dfrac{n(t)}{K}\right)$ & logistic juvenile recruitment survival,
proportional to the fraction of free nest sites \\ 
$K$ & carrying capacity interpreted as the number of nest sites \\ 
\ $\Omega n(t)$ & phenomenological linear density dependent mortality \\ 
$u^{n(t-\tau )/z}$ & phenomenological juvenile survival with delay in
argument \\ 
& $u\in \left\langle 0,1\right\rangle $ and $z$ is the scale parameter
describing \\ 
& the population size where juvenile survival equals $u$ \\ 
& 
\end{tabular}

\section{\protect Methods and technical details from previous papers:
}

\bigskip

The general model from previous papers \cite{argbr1,argbr2,argbr3} is as follows:

Assume two competing strategies and we have $n_{i}$ individuals of each
strategy.

\textbf{To be used in differential equations, demographic payoff functions $V$  (describiing the number of offspring resulting from a game round) and $D $ (probability of death during a game round) should be multiplied by virtual interaction rate $\tau $ describing the
intensity of the occurrence of the focal game rounds.} Background mortality
and fertility rates are adjusted to the timescale where $\tau =1$. In effect,
this parameter can be not explicitly present in the equations, while producing the correct unit of the.

Starting point is the system of the Malthusian growth equations:
\begin{equation}
\frac{dn_{i}}{dt}=n_{i}(t)\left[ \left( \tau e_{i}Vq^{T}(t)+\Phi \right) 
\mathbf{D}(n(t))-\tau e_{i}Dq^{T}(t)-\Psi \right] \text{ for }i=1,2
\label{malth}
\end{equation}
\bigskip where:

$e_{i}$  $i$-th cannonical unit vector

$n=n_{1}+n_{2}$ populations size

$q_{i}=n_{i}/n$ $i$-th strategy frequency

$\tau =1$ virtual interaction rate allowing for demographic interpretation
of payoffs in differential equation, which is described below

$V$ fertility matrix, entries $V_{i,j}$ describe number of offspring
produced during game round

$D$ mortality payoff matrix, entries $D_{i,j}$ describe probability of death
of $i$th strategy player during the game round against $j$th strategy
opponent

$\Phi $ background fertility rate (intensity of the background birth)

$\Psi $ background mortality rate (intensity of the background death)

$\mathbf{D}(n(t))$-juvenile recruitment survival. Previous works 
used the logistic term $\left( 1-\dfrac{n(t)}{K}\right) $, proportional to
the fraction of free nest sites. Thus every newborn checks single random
nest site. If it is free then it survives if it is occupied, it dies.
Generalizations are possible.

$K$ carrying capacity interpreted as the number of nest sites \cite{hui}\bigskip

Then we can rescale equations (\ref{malth}) to replicator dynamics by change
of coordinates where $q=[q_{1},1-q_{1}]$ is the\ vector of strategy
frequencies where $q_{1}=n_{1}/(n_{1}+n_{2})$ and $n=n_{1}+n_{2}$ is
population size\bigskip

\begin{equation}
\frac{dq_{1}}{dt}=q_{1}(t)\left[ \left(
e_{1}Vq^{T}(t)-q(t)Vq^{T}(t)\right) \mathbf{D}(n(t))-\left(
e_{1}Dq^{T}(t)-q(t)Dq(t)\right) \right]  \label{rep}
\end{equation}
\begin{equation}
\frac{dn}{dt}=n(t)\left[ q(t)Vq^{T}(t)\mathbf{D}(n(t))-q(t)Dq(t)+\Phi 
\mathbf{D}(n(t))-\Psi \right]  \label{popsize}
\end{equation}

Zeros of the bracketed terms from equations (\ref{rep}) and (\ref{popsize})
will constitute frequency and density nullclines: $\tilde{q}(n)$ and $\tilde{
n}(q)$. Intersections of those surfaces will constitute stable and unstable
rest points.

\subsection{Resilience against seasonal perturbations and external feedback driven by predator pressure}

We can analyze our model from the point of view of the resilience of the system (sensu Holling), defined as the propensity to absorb the perturbation altering the state while maintaining the functions of the system \cite{ResHolling1, ResHolling2, ResGund, ResMeyer, ResKrak, ResReed}. Thus the question is how the system will react to the external perturbation caused by external enviromnental factors or additional feedbacks. In the so-called  "flow-kick" approach  \cite{ResMeyer}, perturbation is represented by a discrete shift of the population state followed by a continuous return phase, modeled by a standard autonomous differential equation. This approach is termed impulse differential equation \cite{Lakshmi}. In contrast to that, we will incorporate the perturbation as the explicit part of the model, expressed as external mortality pressures (such as periodic mortality or explicit predator pressure), similarly to the Lotka-Volterra models with forcing \cite{Sahoo,Sabin,Inoue,Rinaldi,Tang,Liu}. This approach to the resilience analysis is conceptually related to the structural stability \cite{Peixoto,Arnold}, but within the biological context of the model, since perturbation should have biological interpretation. It is also linked to the concept of permanence (\cite{Schuster,Sigmund,Schreiber,Schreiber2,Hofbauer,Jansen}). It is defined as the existence of the attracting region in the interior of the phase space bounded from extinction. Then the system can resist frequent small perturbations and rare big events. Therefore, this system can be completed by addition of terms describing the impact of strategically neutral (i.e. acting on all strategies in the same way)
external environmental factors, such as seasonality or predator pressure.
Seasonality may act as a periodic mortality factor $\alpha \left( 1+\sin (\theta t)\right)$ added to the population size equation (\ref
{popsize}), where $\alpha $ describes mortality amplitude and $\theta $ is the
duration of the season. Impact of the predator population feedback (denoted by
variable $p(t)$) can act on the equation (\ref{popsize}) via another per capita mortality factor $
px(t)$ and the additional equation for
the dynamics of the predator population size
\begin{equation}
\frac{dx}{dt}=x(t)b_{p}n(t)-x(t)d_{p}.  \label{predator}
\end{equation}

\subsection{Hawk-Dove example without delays}

Classical Hawk-Dove game can be used as the illustrative example. In the
classical case when population grows exponentially or the growth is
suppressed by some selectively neutral adult mortality, matrix $V-d$ is the
classic Hawk-Dove matrix%
\begin{equation*}
	U=\left( 
	\begin{array}{c|cc}
		& H & D \\ \hline
		H & 0.5\left( F-d\right)  & F \\ 
		D & 0 & 0.5F
	\end{array}
	\right) .
\end{equation*}

According to the demographic interpretation in every Hawk vs. Hawk contest
winner consumes fertility reward $F$ while loser can die with probability $d$
. Thus payoff matrix $U=V-D$ can be decomposed into separate fertility and
mortality payoff matrices: 
\begin{equation*}
	V=\left( 
	\begin{array}{c|cc}
		& H & D \\ \hline
		H & 0.5F & F \\ 
		D & 0 & 0.5F
	\end{array}
	\right) \text{ \ \ \ \ and \ \ \ \ }D=\left( 
	\begin{array}{c|cc}
		& H & D \\ \hline
		H & 0.5d & 0 \\ 
		D & 0 & 0
	\end{array}
	\right) \text{\ \ }
\end{equation*}

Then the demographic payoffs are:

\begin{eqnarray}
	V_{h} &=&\left[ (1-q_{d})0.5+q_{d}\right] F=0.5\left( 1+q_{d}\right) F\text{
		\ \ \ \ \ \ \ Hawk fertility payoff}  \label{hawkpayoff} \\
	V_{d} &=&q_{d}0.5F+(1-q_{d})0=q_{d}0.5F\text{ \ \ \ \ \ Dove fertility payoff
	}  \label{dovepayoff} \\
	\bar{V} &=&(1-q_{d})V_{h}+q_{d}V_{d}=0.5F\,\ \ \ \ \text{average fertility
		payoff}  \label{averpayoff}
\end{eqnarray}

and 
\begin{eqnarray}
	D_{h} &=&(1-q_{d})0.5d\text{ \ \ \ \ \ \ Hawk mortality payoff}
	\label{hawkmort} \\
	D_{d} &=&0\text{ \ \ \ \ \ \ \ \ Dove mortality payoff}  \label{dovemort} \\
	\bar{D} &=&(1-q_{d})^{2}0.5d\text{ \ \ \ \ \ \ average mortality payoff}
	\label{avermort}
\end{eqnarray}
\bigskip 

After substitution of the above payoff functions to equations (\ref{rep},\ref{popsize})

\begin{eqnarray}
	\frac{dq_{d}}{dt} &=&0.5q_{d}(t)\left( 1-q_{d}(t)\right) \left( \left(
	1-q_{d}(t)\right) d-\mathbf{D}(n(t))F\right)  \label{repq} \\
	\frac{dn}{dt} &=&n(t)\left( \left( 0.5F+\Phi \right) \mathbf{D}(n(t))-\left[
	\left( 1-q_{d}(t)\right) ^{2}0.5d+\Psi \right] \right) .  \label{repn}
\end{eqnarray}

Flat trivial nullclines for the values $q_{d}=0$, $q_{d}=1$ and $n=0$
describe the boundaries of the phase space and extinction of respectively
Doves, Hawks and the whole population. Nontrivial frequency and density
nullclines for this system are following\bigskip 
\begin{gather}
	\tilde{q}_{d}(n)=1-\mathbf{D}(n(t))\frac{F}{d} \\
	\mathbf{D}(n(t))=\frac{\left( 1-q_{d}(t)\right) ^{2}0.5d+\Psi }{0.5F+\Phi }
\end{gather}

and for any monototnusly decreasing, density dependent juvenile recruitment
survival $\mathbf{D}(n(t))$ the density nullcline is 
\begin{equation*}
	\tilde{n}(q)=\mathbf{D}^{-1}\left( \frac{\left( 1-q_{d}(t)\right)
		^{2}0.5d+\Psi }{0.5F+\Phi }\right) ,
\end{equation*}

and both nullclines intersect when
\begin{equation*}
	\Delta =\left( \frac{\Phi }{F}+0.5\right) ^{2}-2\frac{\Psi }{d}>0,
\end{equation*}
and we have two intersections
\begin{eqnarray*}
	\check{q}_{d} &=& 0.5-\frac{\Phi }{F}-\sqrt{\Delta}=0.5-\frac{\Phi }{F}-\sqrt{\left( \dfrac{\Phi }{F}\right)^{2}+\dfrac{\Phi }{F}-2\dfrac{\Psi }{d}+0.25} \\
	\hat{q}_{d} &=&0.5-\frac{\Phi }{F}+\sqrt{\Delta}=0.5-\frac{\Phi }{F}+\sqrt{\left( \dfrac{\Phi }{F}\right) ^{2}+
		\dfrac{\Phi }{F}-2\dfrac{\Psi }{d}+0.25},
\end{eqnarray*}

and above frequencies of intersections do not depend on the form of the juvenile survival $\mathbf{D}(n(t))$. Then the respective population sizes are:

\begin{eqnarray*}
	\check{n} &=& \mathbf{D}^{-1}\left( \frac{d}{f}\left( 0.5+\frac{\Phi }{f}\right)- \sqrt{\Delta } \right)
	= \mathbf{D}^{-1}\left( \frac{d}{f}\left( 0.5+\frac{\Phi }{f}\right)- \sqrt{\left( \dfrac{\Phi }{F}\right)^{2}+\dfrac{\Phi }{F}-2\dfrac{\Psi }{d}+0.25 } \right) \\
	\hat{n} &=&\mathbf{D}^{-1}\left( \frac{d}{f}\left( 0.5+\frac{\Phi }{f}\right)+ \sqrt{\Delta } \right) 
	=\mathbf{D}^{-1}\left( \frac{d}{f}\left( 0.5+\frac{\Phi }{f}\right)+ \sqrt{\left( \dfrac{\Phi }{F}\right)^{2}+\dfrac{\Phi }{F}-2\dfrac{\Psi }{d}+0.25 } \right)\\
\end{eqnarray*}

For the logistic suppression $\mathbf{D}(n(t))=\left( 1-\frac{n(t)}{K}
\right) $, the density nullcline (also termed stationary density surface \cite{CressmanGaray}) is:
\begin{equation*}
	\tilde{n}(q)=\left[ 1-\frac{\left( 1-q_{d}\right) ^{2}0.5d+\Psi }{0.5F+\Phi }
	\right] K,
\end{equation*}
and the population sizes at the rest points will be
\begin{eqnarray*}
	\check{n} &=&\left[ 1-\frac{d}{f}\left( 0.5+\frac{\Phi }{f}\right) +\sqrt{
		\Delta }\right] K \\
	\hat{n} &=&\left[ 1-\frac{d}{f}\left( 0.5+\frac{\Phi }{f}\right) -\sqrt{
		\Delta }\right] K
\end{eqnarray*}
\bigskip 

Detailed derivation is in the Appendix 1.\bigskip

Those intersections constitute the rest points of the system if they are
contained in the phase space $\left[ 0,1\right] \times \left[ 0.K\right] $. When $\Delta$ has a small negative value then both nullclines disconnect and
form a narrow channel attracting trajectories and trapping them for a longer
time. Then the system pretends to be in the stable state and afterwards it
rapidly switches to the real stable attractor. Narrow channel trapping the
trajectories is called a ghost attractor and the resulting pattern in time
is termed a long transient behaviour \cite{Morozov,Hastings1,Hastings2}.
As we mentioned before. Equation (\ref{repn}) can be completed by additional
mortality factors describing seasonality, or predator impact (then the system (\ref{repq},\ref{repn}) can be completed by equation (\ref{predator}) describing the predator population). Similarly, the seasonality can be described by periodic background
mortality, which will indirectly affect the frequency dynamics. Both factors
will be presented in detail in the Results section.\bigskip 
 
\section{Results}

\subsection{Abstract "fitness" or explicit births and deaths? Interaction rates, pair formation and the timing of interaction events}

Traditionally payoff functions are interpreted as differences in the resulting growth rates $U_i=V_i-D_i$, without explicit distinction between fertility and mortality rates. This leads to the
conceptual limitations of the classical framework. Darwinian fitness (thus a game payoff) is defined as
the long term growth rate, while in replicator equation, growth
rate is defined as proportional to the payoff. To summarize, since fitness
is defined as the growth rate, then, in this model, growth rate is defined as
"growth rate". This is related to the widely discussed the so-called tautology problem in
evolutionary theory 
\cite{peters1976tautology,peters1978predictable,van1983methodological,von2016survival}
 (the survival and reproduction of the fittest are those with the greatest
survival and reproduction). Rates of increase $n_{i}U_{i}$, are expressed as
numbers of individuals per unit time. Therefore, a unit of payoff $U_{i}$
(i.e., the per capita fitness) should be 1/(time unit). Since for small $
\Delta t$ the population increases as $n_{i}(t+\Delta
t)-n_{i}(t)=n_{i}(t)U_{i}\Delta t$, the phenomenological dimensionless
scaling factor $U_{i}\Delta t$ describes how many more individuals will be "$
\Delta t$ later". We cannot measure this value at time $t$, we can try to
estimate it post factum at $t+\Delta t$. This value should be rather a prediction generated by a model with
parameters measurable, at time $t$, i.e. "now". In such a framework we
cannot build a fully predictive model based on some properties of the
interacting agents. We must attribute to them some abstract growth rates,
which we can try to estimate post factum since we cannot measure them at
time $t$. \bigskip

This leads to a fundamental question about the biological/physical
meaning of the cost $d$ and benefit $f$ parameters in the payoff matrix.
They turns out to be unspecified increments and decrements of some
unspecified growth rate (thus cost describes unborn descendents). Such vague
connection to biological realism, in particular, experimental measurements
of parameters limit the quantitative description and the explanatory power
of the models \cite{McNamaraGame2013}.\bigskip

Therefore, decomposition of the constant abstract fitness into separate
birth and death rates $U_{i}=\tau V_{i}-\tau D_{i}$ allows for mechanistic
interpretation and increases the framework's predictive power. Note that the
model still can be expressed in terms of differences in constant "fitnesses" 
$U_{i}$. However, when we add the maturation delay $\gamma $, birth rate
becomes $n(t-\gamma )V_{i}$ while the death rate is still $n(t)D_{i}$. then
the per capita growth rate has the form:

\begin{equation*}
n_{i}(t)U_{i}=n(t)\left[ \frac{n(t-\gamma )}{n(t)}V_{i}-D_{i}\right] .
\end{equation*}

Therefore, this parameter has the different values when the population is
growing or declining. Thus, in this case we cannot derive the constant and
independent of time "fitness effect" resulting from the game interaction.
This shows that the additive quantity called "fitness" simply do not exists in
that case.

\subsection{Meaning and interpretation of the time unit}

For the interaction rate $\tau $ the per capita average time between game
interaction equals $1/\tau $. By definition, interaction intensity is the
derivative of per capita probability of playing the game round. However,
this suggests that each individual can be described as the independent
Poisson process. The problem is that for each game round we need two
players. Therefore, the modelling framework should explicitly consider the
dynamics of pair formation. From the perspective of the whole population,
number of individuals that played at least single game round is described by
equation $\dot{n}(t)=n(t)\tau $. For small time intervals $\Delta t$ this
equation can be approximated by first order Taylor expansion

\begin{eqnarray}
n(t+\Delta t) &=&n(t)+\dot{n}(t)\Delta t \\
n(t+\Delta t)-n(t) &=&n(t)\tau \Delta t
\end{eqnarray}

Therefore $\tau $ is equal to the fraction of the population that
participated in at least single game round during small $\Delta t$.\ Since for
each game round we need two players, during $\Delta t$ interval $\tau
n(t)\Delta t/2$ game rounds will be played. For the timescale where interaction rate equals 1
average per capita time between interaction events also equals 1.\ This
interpretation of the time unit will be important for the evaluation of the
value of the delay resulting from maturation time. Hence the fraction of
individuals that played at least a single game round during a short interval 
$\Delta t\ll 1$ equals to $\Delta t$ . Then, the dynamics of the game rounds
occurrence (with birth and deaths ignored) follows the exponential decay $
n_{i}(t_{0})e^{-t}$ (a solution of the equation $dn_{i}/dt=-\tau n_{i}$ with
intensity $\tau =1$). This implies that during single time unit, fraction $
1-e^{-1}$ of individuals present at the $t_{0}$, have not played the focal game
(however, they can die and reproduce due to the background events).

\subsection{Methods used in analysis of the impact of the delays}

In this section we will introduce the methods used for the analysis of the
impact of delay on stability of the rest points.

\subsubsection{Analytical toolbox}

Let us introduce the method to calculate of the critical
bifurcation delay as the function of the model parameters in general two dimensional systems described by variables $x,y$. First Let $x_{\gamma }(t)=x(t-\gamma )$, $y_{\gamma
}(t)=y(t-\gamma )$. Consider the general system of the form 

\begin{eqnarray}
&&x^{\prime }=f_{1}(x,x_{\gamma },y,y_{\gamma }),  \label{eq-no} \\
&&y^{\prime }=f_{2}(x,x_{\gamma },y,y_{\gamma }),  \label{eq-n1o}
\end{eqnarray}

where $f_{1}$ and $f_{2}$ are continuously differentiable functions. Let $
(x^{\ast },y^{\ast })$ be a rest point of this system, i.e. $x^{\ast }$ and $y^{\ast }$ satisfy the conditions
$f_{1}(x^{\ast},x^{\ast },y^{\ast },y^{\ast })=0$ and $f_{2}(x^{\ast },x^{\ast},y^{\ast },y^{\ast })=0$. Let 

\begin{equation*}
b_{11}=\frac{\partial f_{1}}{\partial x},\quad b_{12}=\frac{\partial f_{1}}{
\partial x_{\gamma }},\quad b_{13}=\frac{\partial f_{1}}{\partial y}
,\quad b_{14}=\frac{\partial f_{1}}{\partial y_{\gamma }},
\end{equation*}

\begin{equation*}
b_{21}=\frac{\partial f_{2}}{\partial x},\quad b_{22}=\frac{\partial f_{1}}{
\partial x_{\gamma }},\quad b_{23}=\frac{\partial f_{2}}{\partial y}
,\quad b_{24}=\frac{\partial f_{2}}{\partial y_{\gamma }},
\end{equation*}

\begin{equation*}
a_{11}=b_{11}+b_{12},\quad a_{12}=b_{13}+b_{14},
\end{equation*}
\begin{equation*}
a_{21}=b_{21}+b_{22},\quad a_{22}=b_{23}+b_{24}.
\end{equation*}
In the formulas for the coefficients $b_{ij}$, we calculate the partial
derivatives of the function $f$ at the point $(x^{\ast },x^{\ast
},y^{\ast },y^{\ast })$.

\begin{theorem}
\label{th:stab+bif} Let $\mathbf{A}$ be the matrix of the form $\mathbf{A}
=[a_{ij}]$, $1\leq i,j\leq 2$. Assume that the rest point $(x^{\ast
},y^{\ast })$ of system (\ref{eq-no})--(\ref{eq-n1o}) is locally stable
for $\gamma =0$. Let 
\begin{eqnarray*}
&&A=b_{12}b_{24}-b_{22}b_{14}, \\
&&B=b_{11}b_{24}+b_{12}b_{23}-b_{13}b_{22}-b_{14}b_{21}, \\
&&C=b_{11}b_{23}-b_{13}b_{21},
\end{eqnarray*}
and
\begin{eqnarray*}
&&P=2\left( (b_{11}+b_{23})^{2}-2C\right) -(b_{12}+b_{24})^{2}, \\
&&Q=\left( (b_{11}+b_{23})^{2}-2C\right) ^{2}-2(b_{12}+b_{24})\left[
(b_{11}+b_{23})B-\left( A+C\right) (b_{12}+b_{24})\right] \\
&&\hskip15pt{}+2\left( C^{2}-A^{2}\right) -\left(
(b_{11}+b_{23})(b_{12}+b_{24})-B\right) ^{2}, \\
&&R=2\left( C^{2}-A^{2}\right) \left( (b_{11}+b_{23})^{2}-2C\right) -2B(A-C) 
\left[ B-(b_{11}+b_{23})(b_{12}+b_{24})\right] \\
&&\hskip15pt{}-\left[ (A+C)(b_{12}+b_{24})-(b_{11}+b_{23})B\right] ^{2}, \\
&&S=(C^{2}-A^{2})^{2}-B^{2}(A-C)^{2}.
\end{eqnarray*}
Let $u>0$ be a constant which satisfies the equation 
\begin{equation}
u^{4}+Pu^{3}+Qu^{2}+Ru+S=0  \label{eq-xgamma4}
\end{equation}
The stationary solution loses its stability ($\gamma $ is the bifurcation
point) for the smallest positive $\gamma $ given by 
\begin{equation}
\gamma =\frac{1}{\sqrt{u}}\arccos \left( \frac{-B(A-\left( C-u\right)
)+(b_{11}+b_{23})(b_{12}+b_{24})u}{A^{2}-\left( C-u\right)
^{2}-(b_{11}+b_{23})^{2}u}\right) .  \label{eq-xgamma5}
\end{equation}
\end{theorem}

Proof is in the Appendix 3.

\bigskip

In the next sections, for simplicity of calculations, our replicator system will be transformed into coordinates $x=n_{1}=q_{d}n$ and $y=n$.

\subsubsection{Numerical toolbox}

Loss of the stability of rest points and the first bifurcation is the beginning of a very
complex behavior. To analyze those events we use the numerical method from 
\cite{Szczelina1} to construct approximate solutions of
Eqs.~(33)-(35), using parameter values as in Section~2.3, while varying the
delay $\gamma $ in the range $\gamma \in \lbrack 1,50]$, with step size $
\Delta \gamma \leq 10^{-2}$ (refined further in more chaotic regions). We
start with a constant initial function for each value of $\gamma $ and
generate the solution $(n^{\gamma }(t),q_{d}^{\gamma }(t))$ over a long time
interval. Since we are interested in the long-term behaviour of solutions
(near the attractor), we discard the initial transients in the time range $
t\in \lbrack -\gamma ,S]$ for some fixed $S$. Then, for some $T\gg S$, we
record all extrema of the solution $n(t)$, i.e., all points $t_{i}\in
\lbrack S,T]$ such that $n^{\prime }(t_{i})=0$. The value of $T$ is chosen
such that at least $N=50$ extrema are identified. In cases where the
solution tends toward a fixed point (for small values of $\gamma $), we
record only one point. This type of diagram is commonly used in the study of
bifurcations in DDEs; (see \cite{Leandro,Duruissea}, and references therein). The
bifurcation diagram based on extrema, typically differs from bifurcation
diagrams obtained from discrete maps like the logistic map or chaotic ODEs
such as the R\"{o}ssler system (see \cite{Gierzkiewicz}
and references therein). The concept of the extrema diagram relies on two
observations: first, during a period-doubling bifurcation, a new extremum
should appear in the diagram; and second, it is particularly useful when no
natural candidate for a Poincar\'{e} section exists. While informative, some
newly appearing periodic points in the extrema bifurcation diagram may be
spurious i.e. not resulting from true
bifurcations, but rather from small perturbations in the underlying
solution, such as the so-called \emph{kinks} \cite
{Leandro}. Nevertheless, the emergence of
these kinks typically precedes chaotic dynamics in DDEs (this method is used in section devoted to bifurcation analysis). All numerical trajectories are generated by JiTCDDE delay differential equations
solver for Python \cite{Ansmann}.

\subsection{Derivation of the general framework for replicator dynamics with
fertility delay resulting from the egg hatching or
maturation time.\protect\bigskip}

For simplicity, at this stage, consider uniform delays. Generalization to non-uniform delays for different strategies is the topic for the future research. To introduce delay to the fertility bracket we should
renormalize the number of newborns per capita. If $\gamma $ is the delay then we have
\begin{equation}
n_{1}(t-\gamma )e_{1}Vq^{T}(t-\gamma )
\end{equation}%
newborns. Then current per capita reproductive success is 
\begin{equation}
\frac{n_{1}(t-\gamma )}{n_{1}(t)}e_{1}Vq^{T}(t-\gamma )
\end{equation}

When we substitute this into the standard replicator equation we obtain the
following fertility bracket in the replicator equation
\begin{equation}
\frac{n(t-\gamma )}{n(t)}\left[ q_{1}(t-\gamma )V_{1}(t-\gamma )-q_{1}(t)
\bar{V}(t-\gamma )\right] .
\end{equation}

The above term is negative when:
\begin{equation}
q_{1}(t)>\frac{q_{1}(t-\gamma )V_{1}(t-\gamma )}{\bar{V}(t-\gamma )}=\frac{
n_{1}(t-\gamma )V_{1}(t-\gamma )}{\sum_{j}n_{j}(t-\gamma )V_{j}(t-\gamma )},
\end{equation}

and the r.h.s of the above condition describes the frequency of the first strategy among newborns. The
dynamics driven by the fertility bracket cannot escape the unit interval and
for two strategies it reduces to:

\begin{equation}
\left[ 1-q_{1}(t)\right] q_{1}(t-\gamma )V_{1}(t-\gamma
)-q_{1}(t)(1-q_{1}(t-\gamma ))V_{2}(t-\gamma ).
\end{equation}

See Appendix 2 for detailed derivation.

\bigskip Note that in this case background fertility rate $\Phi $ (which
averaged over all strategies also equals $\Phi $)\ will not vanish from
replicator equations and will appear as the additional bracketed term:
\begin{eqnarray}
&&\frac{n(t-\gamma )}{n(t)}\left[ q_{1}(t-\gamma )\Phi -q_{1}(t)\Phi \right]
\\
&=&\frac{n(t-\gamma )}{n(t)}\left[ q_{1}(t-\gamma )-q_{1}(t)\right] \Phi .
\end{eqnarray}

It can be combined with fertility bracketed term. Therefore our system of
replicator dynamics will have the following forms (newborns produced $\gamma $ ago hatch
from eggs and look for free nest sites). In a simplified form where $V_{1}$
and $D_{1}$ are demographic payoffs of the first strategy while $\bar{V}$
and $\bar{D}$ are average values: 
\begin{eqnarray}
\frac{dq_{1}}{dt} &=&\frac{n(t-\gamma )}{n(t)}\left( \left[ q_{1}(t-\gamma
)V_{1}(t-\gamma )-q_{1}(t)\bar{V}(t-\gamma )\right] +\left[ q_{1}(t-\gamma
)-q_{1}(t)\right] \Phi \right)  \notag \\
&&-q_{1}(t)\left[ D_{1}(t)-\bar{D}(t)\right]  \label{delrep} \\
\frac{dn}{dt} &=&n(t-\gamma )\left[ \bar{V}(t-\gamma )+\Phi \right] -n(t)
\left[ \bar{D}(t)+\Psi \right]  \label{delpopsize}
\end{eqnarray}

Note that for the delay $\gamma=0$ the above system collapses to the equations (\ref{rep},\ref{popsize}). The expanded form of the above equations will be 
\begin{gather*}
\frac{dq_{1}}{dt}=\frac{n(t-\gamma )}{n(t)} \left( \left[ q_{1}(t-\gamma
)e_{1}Vq^{T}(t-\gamma )-q_{1}(t)\sum_{i}q_{i}(t-\gamma )e_{i}Vq^{T}(t-\gamma
) \right] + \left[ q_{1}(t-\gamma )-q_{1}(t)\right] \Phi \right) \\
-q_{1}(t)\left[ \left( e_{1}Dq^{T}(t)-q(t)Dq(t)\right) \right] \\
\frac{dn}{dt}=n(t-\gamma )\left[ \sum_{i}q_{i}(t-\gamma
)e_{i}Vq^{T}(t-\gamma )+\Phi \right] -n(t)\left[ q(t)Dq(t)+\Psi \right]
\end{gather*}
\bigskip

\subsection{Density dependence}

Now we should consider the problem of density dependent growth supression.
For the models without delay suppression can be realized by mortality acting
on adult individuals. In this setting it does not affect the replicator dynamics (as in the
standard models from textbooks). It can also act as the juvenile recruitment
survival. This can be described by some multiplicative factor of fertility payoffs (for
example logistic term $(1-n/K)$, as in demographic
models from \cite{argbr1,argbr2,argbr3}). Then replicator equation becomes sensitive
for adult mortality factors acting on population size $n(t)$. This impact
may lead to very complex dynamics such as in the case of periodic seasonal
mortality or predator pressure. Introduction of delays complicates the problem even more. Since equation (\ref{delrep}) depends on population size, due to factor $n(t-\gamma )/n(t)$,
adult mortality factors may affect the dynamics of strategy frequencies even
in the case without density dependent juvenile survival. If those factors
are more complicated, such as seasonal mortality or predator pressure, the
dynamics may be also complex.

Therefore, in this case we need some
growth supression factor acting on adult individuals, which will be present
in the population size equation (\ref{delpopsize}). We will use simple
phenomenological linear mortality rate $\Omega n(t)$.

For juvenile survival the situation is also more complicated. We can use the
standard logistic model $(1-n(t)/K)$. Then the recruitment will be realized
after maturation time (for example after egg hatching). The attracting point
of the population size dynamics is 
\begin{equation}
\tilde{n}(t)=\left( 1-\frac{\left[ \bar{D}(t)+\Psi \right] }{\left[ \bar{V}
(t-\gamma )+\Phi \right] }\right) K,
\end{equation}
If we want to use delayed juvenile survival, then we cannot simply use
factor $(1-n(t-\gamma )/K)$, as the obtained dynamics makes no sense in the following manner:
the size $n(t)$ might be close to the limit $K$ at a given $t$, while the dynamics accounts for the survival the value of $n(t-\gamma)$ in the past, which might be much smaller than $K$. In effect the
population size may exceed $K$, changing the sign of the supression bracket $
(1-n(t)/K)$. Which may lead to infinite population growth. This problem is
known as Levins Paradox in the classical population models. Our model
clearly shows that this is not a paradox but ordinary bug since frequency
dynamics escapes the unit interval too. To counter this problem, for delayed recruitment, we
might use some phenomenological decreasing fertility at birth (or egg
laying time) described by multiplicative factor

\begin{equation}
u^{n(t-\tau)/z}=e^{\frac{\ln (u)}{z}n(t-\tau )},
\label{DelSupp}
\end{equation}

where $u\in \left(
0,1\right) $ and $z$ is the scale parameter describing the population
size where juvenile survival equals $u$. This factor should be rather
interpreted as the density dependent fraction of eggs produced from the
maximal potential, for example for some insect population. \bigskip

\subsection{Hawk-Dove game example}

Derivation of the bracketed terms with (\ref{dovepayoff}) and (\ref
{averpayoff}):
\begin{eqnarray}
&&q_{1}(t-\gamma )e_{1}Vq^{T}(t-\gamma )-q_{1}(t)\sum_{i}q_{i}(t-\gamma
)e_{i}Vq^{T}(t-\gamma ) \\
&=&q_{d}^{2}(t-\gamma )0.5F-q_{d}(t)0.5F  \notag \\
&=&\left( q_{d}^{2}(t-\gamma )-q_{d}(t)\right) 0.5F  \notag
\end{eqnarray}

and for mortality bracket with (\ref{dovemort}) and (\ref{avermort}) we
have: 
\begin{equation}
\left( e_{1}dq^{T}(t)-q(t)dq(t)\right) =0-(1-q_{d})^{2}0.5d
\end{equation}

When we substitute the payoff functions of the Hawk-Dove game to equations (\ref{delrep},\ref{delpopsize}) we obtain the system of equations:

\begin{eqnarray}
\frac{dq_{d}}{dt} &=&\frac{n(t-\gamma )}{n(t)}\left[ q_{d}^{2}(t-\gamma
)-q_{d}(t)\right] 0.5F\mathbf{D}(n(t))+  \notag \\
&&\frac{n(t-\gamma )}{n(t)}\left[ q_{d}(t-\gamma )-q_{d}(t)\right] \Phi 
\mathbf{D}(n(t))+q_{d}(t)\left( 1-q_{d}(t)\right) ^{2}0.5d
\end{eqnarray}
\begin{equation}
\frac{dn}{dt}=n(t-\gamma )\left( 0.5F+\Phi \right) \mathbf{D}(n(t))-n(t)
\left[ \left( 1-q_{d}(t)\right) ^{2}0.5d+\Psi +\mathbf{A}+\mathbf{B}+\mathbf{
C}\right]
\end{equation}
\begin{equation}
\frac{dx}{dt}=x(t)b_{p}n(t)-x(t)d_{p}
\end{equation}

where juvenile survival $\mathbf{D}(n(t))$ will have following variants:

a) $\mathbf{D}(n(t))=1$ lack of density dependent juvenile mortality factors;

b) $\mathbf{D}(n(t))=\left( 1-\dfrac{n(t)}{K}\right) $ logistic suppression,
after maturation period every juvenile indiviual can check single random
nest site. If it will be free it will survive and will be succefully
recruited;

c) $\mathbf{D}(n(t))=u^{n(t-\gamma )/z}$ \ recruitment with the delay, where 
$u\in \left\langle 0,1\right\rangle $ and $z$ is the scale parameter
describing the population size where juvenile survival equals $u$;\bigskip

The rest of the optional adult mortality factors will be:

$\mathbf{A=}\alpha \left( 1+\sin (\frac{2\Pi }{\theta }t)\right) $ seasonal
fluctuations of the background mortality, where $\alpha $ describes
amplitude and $\theta $ duration of the season;

$\mathbf{B=}px(t)$ impact of predators;

$\mathbf{C=}\Omega n(t)$ phenomenological linear adult density dependent
mortality;

Note that, for zero delays and logistic juvenile survival the above system
reduces to (\ref{repq},\ref{repn}). \bigskip

\section{Numerical simulations}

\subsection{Case without juvenile recruitment mortality, type a) juvenile
survival}

\begin{figure}[tbp] 
		\includegraphics[height=19cm]{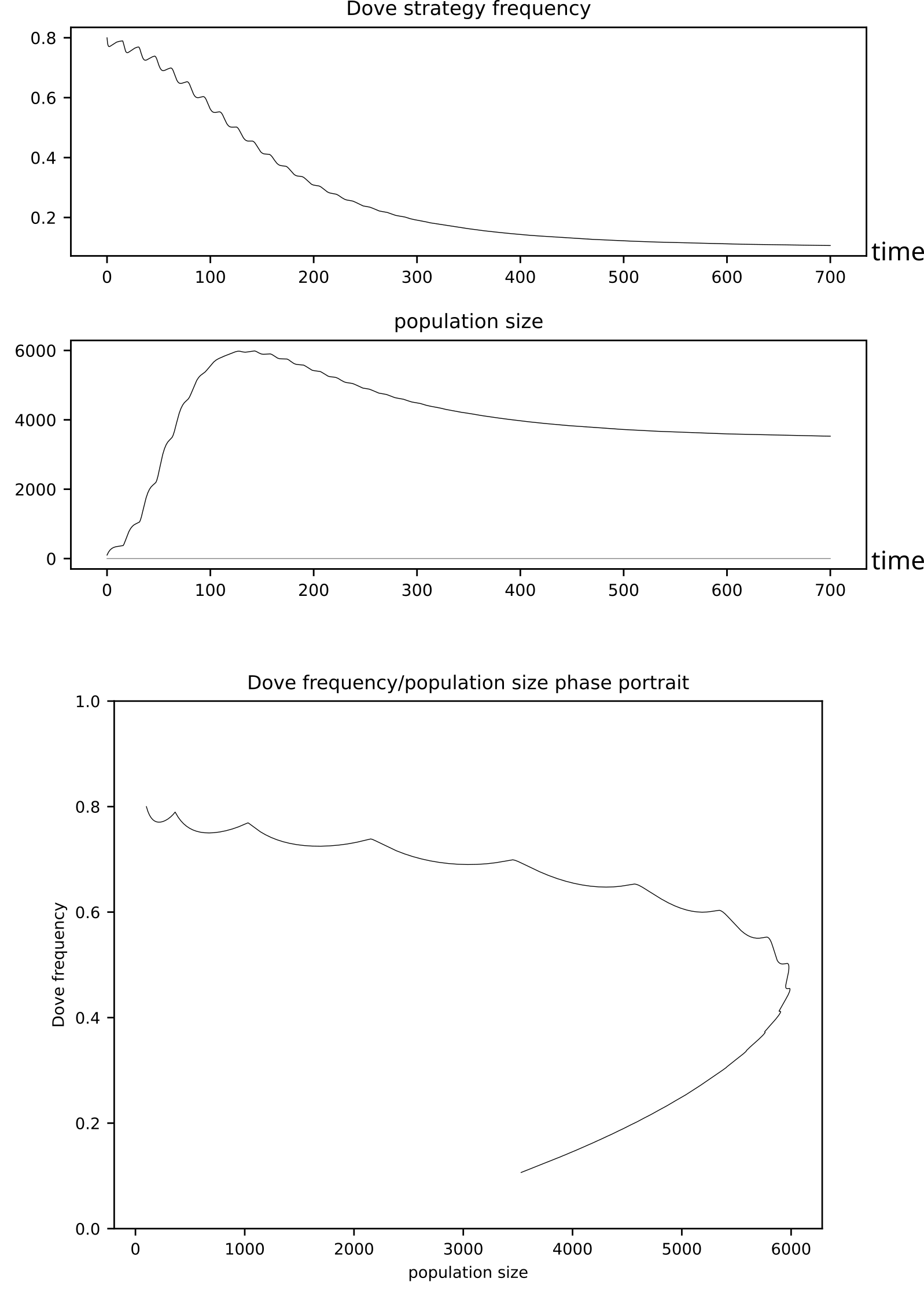}
	\caption{Case without juvenile mortality factors.Trajectory for parameters $F=0.9$, $d=1$, $\Phi =0.5$, $\Psi =0.2$
			and $\Omega =0.0001$ and the constant initial history $q_{d}=0.8$ and $n=100$.	}
			\label{fig:Fig1}
\end{figure}

Let us consider the version of the model with $\mathbf{D}(n(t))=1$. In this
case we need to suppress the unlimited exponential growth by linear density
dependent adult mortality of type $\mathbf{C}$ which is $\Omega n(t)$.
Figure \ref{fig:Fig1} shows the trajectories of the model with the delay $\gamma =15$
and\ for the parameters $F=0.9$, $d=1$, $\Phi =0.5$, $\Psi =0.2$ and $\Omega
=0.0001$. The initial history is constant and equals $q_{d}=0.8$ and $n=100$
.\ This case is equivalent to the classical textbook replicator dynamics,
the only difference is the introduction of the delay. However, as we can see
in Figure \ref{fig:Fig1} the dynamics is more sophisticated showing small scale
oscillations and the overshoot of the equilibrium.

\bigskip

\bigskip

%\DeclareGraphicsExtensions{.eps} 

\begin{figure}[tbp] 
	\includegraphics[height=19cm]{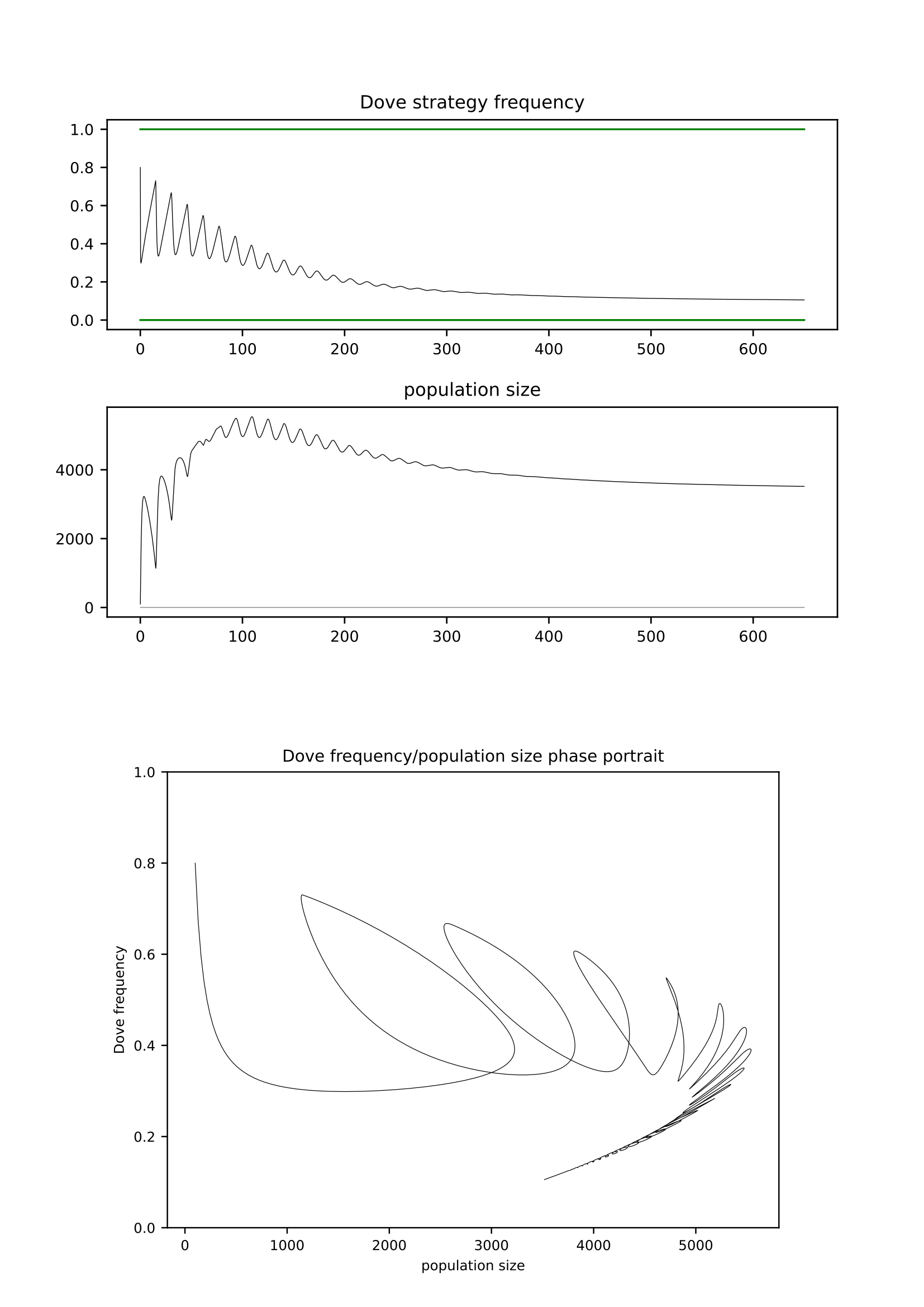}
	\caption{Case with non-constant initial history.
		Initial frequency linearly grows from $0.35$ to $0.8$ and the population
		size declines linearly from $3100$ to $100$.	}
	    	\label{fig:Fig2}
\end{figure}

The trajectories are very sensitive for initial histories. This is shown in Figure \ref{fig:Fig2} where the initial frequency linearly grows from $0.35$ to $0.8$
and the population size declines linearly from $3100$ to $100$. This
situation can be interpreted as the rapid ecological catastrophe where Hawks
suffered greater mortality than Doves. As we can see trajectories are
dramatically different than for the constant history.\bigskip 

\begin{figure}[tbp] 
	\includegraphics[height=19cm]{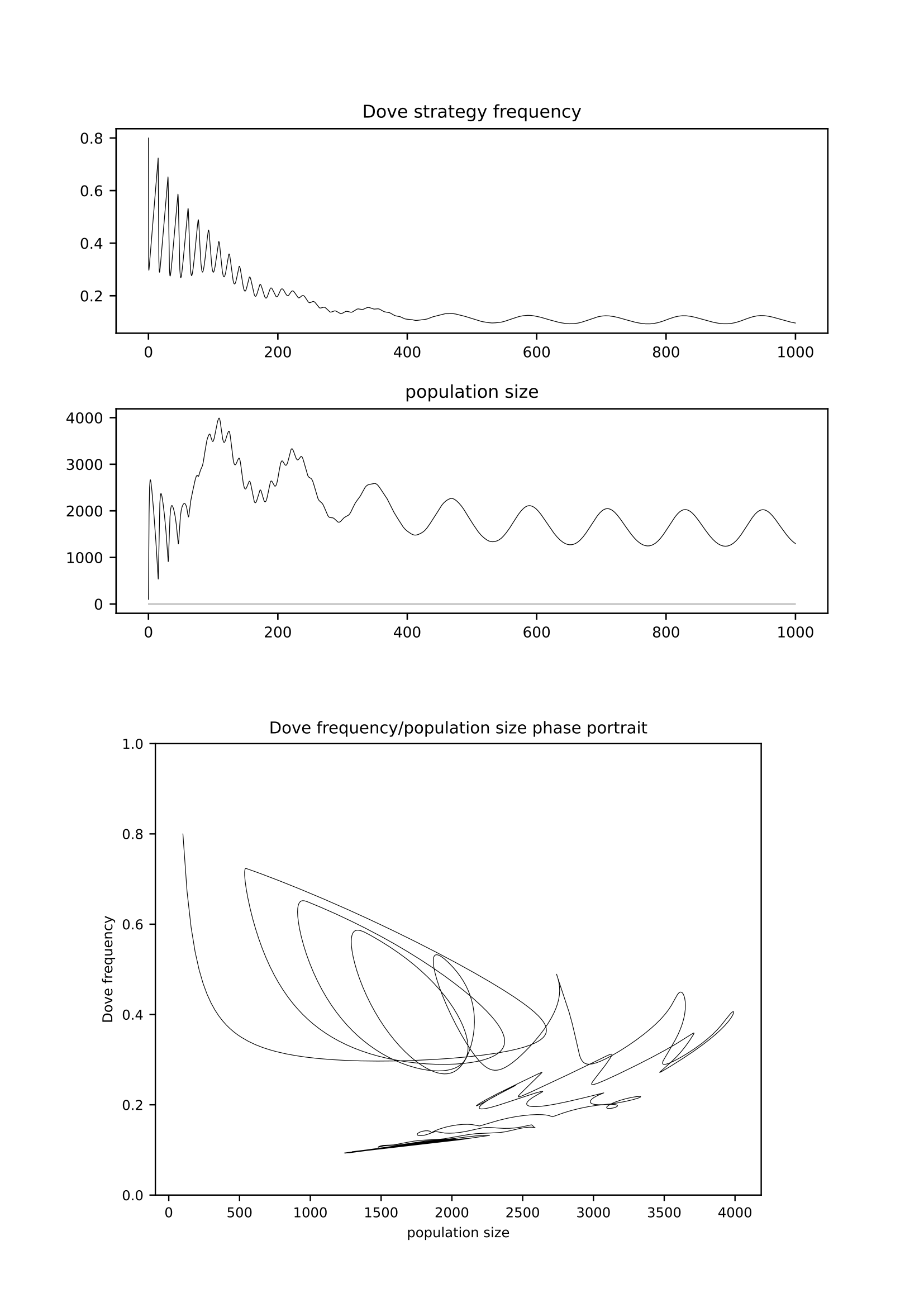}
	\caption{Trajectories with added periodic
		background mortality with amplitude $\protect\alpha =0.2$ and period $
		\protect\theta =120$.	}
		\label{fig:Fig3}
\end{figure}

 Without delay the standard textbook
replicator dynamics is closed and nothing can affect the dynamics of the
strategy frequencies. We can see that in
the case of the model with the delay in fertility payoffs, this is not the
case anymore even without additional density dependent juvenile recruitment mortality. When we add some periodic mortality (additional adult
mortality of type $\mathbf{A}$ with amplitude $\alpha =0.2$ and period $
\theta =120$) we can observe induced oscillations of the strategy
frequencies in addition to the oscillations of the population size (Fig. \ref{fig:Fig3}).

\begin{figure}[tbp] 
	\includegraphics[height=10cm]{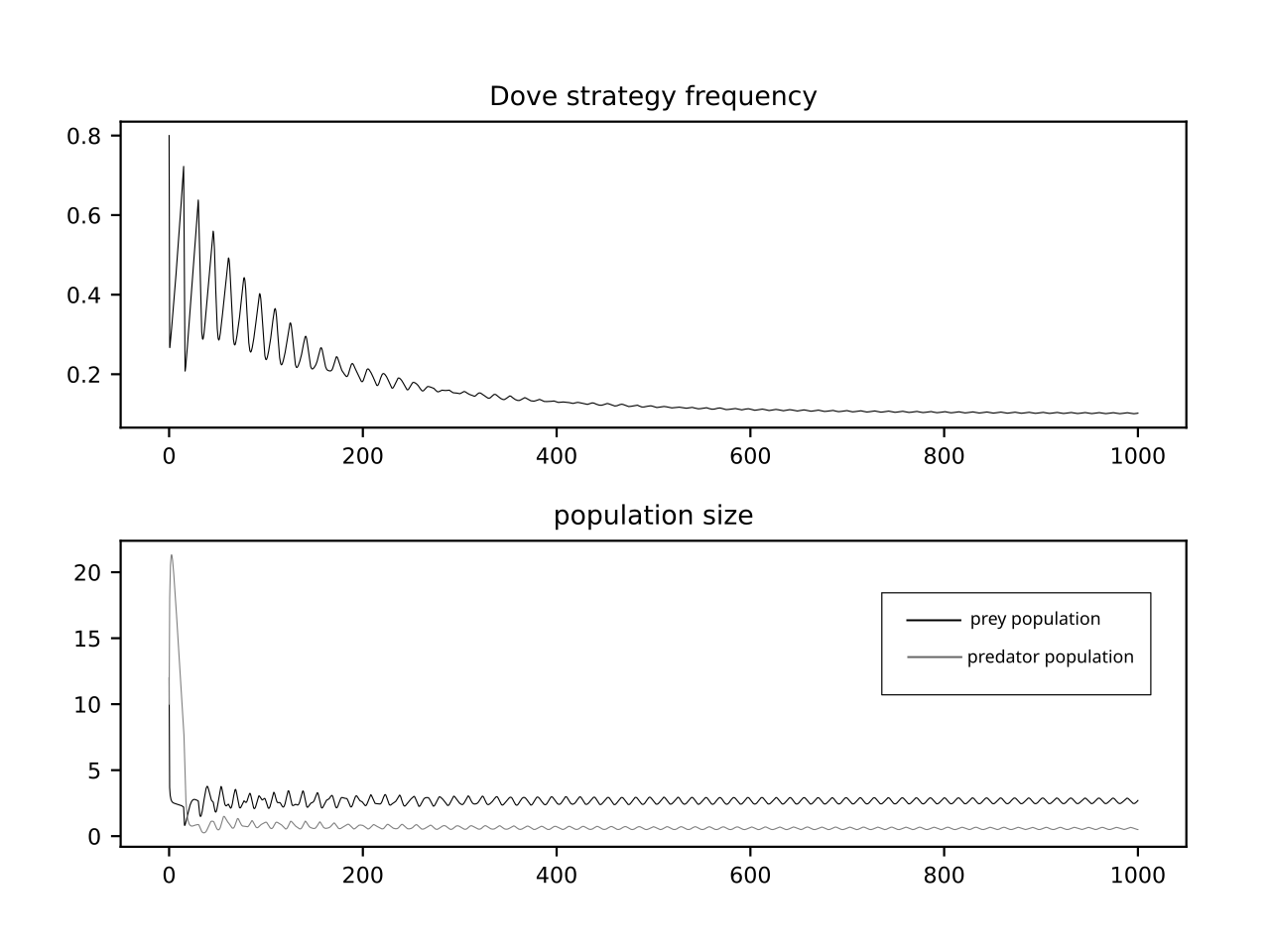}
	\caption{Case with added predator pressure.
		Parameters of the predator-prey subsystem are $b_{p}=0.3$ ,$\ d_{p}=0.8$ and 
		$p=0.6$.	}
		\label{fig:Fig4}
\end{figure}

\begin{figure}[tbp] 
	\includegraphics[height=10cm]{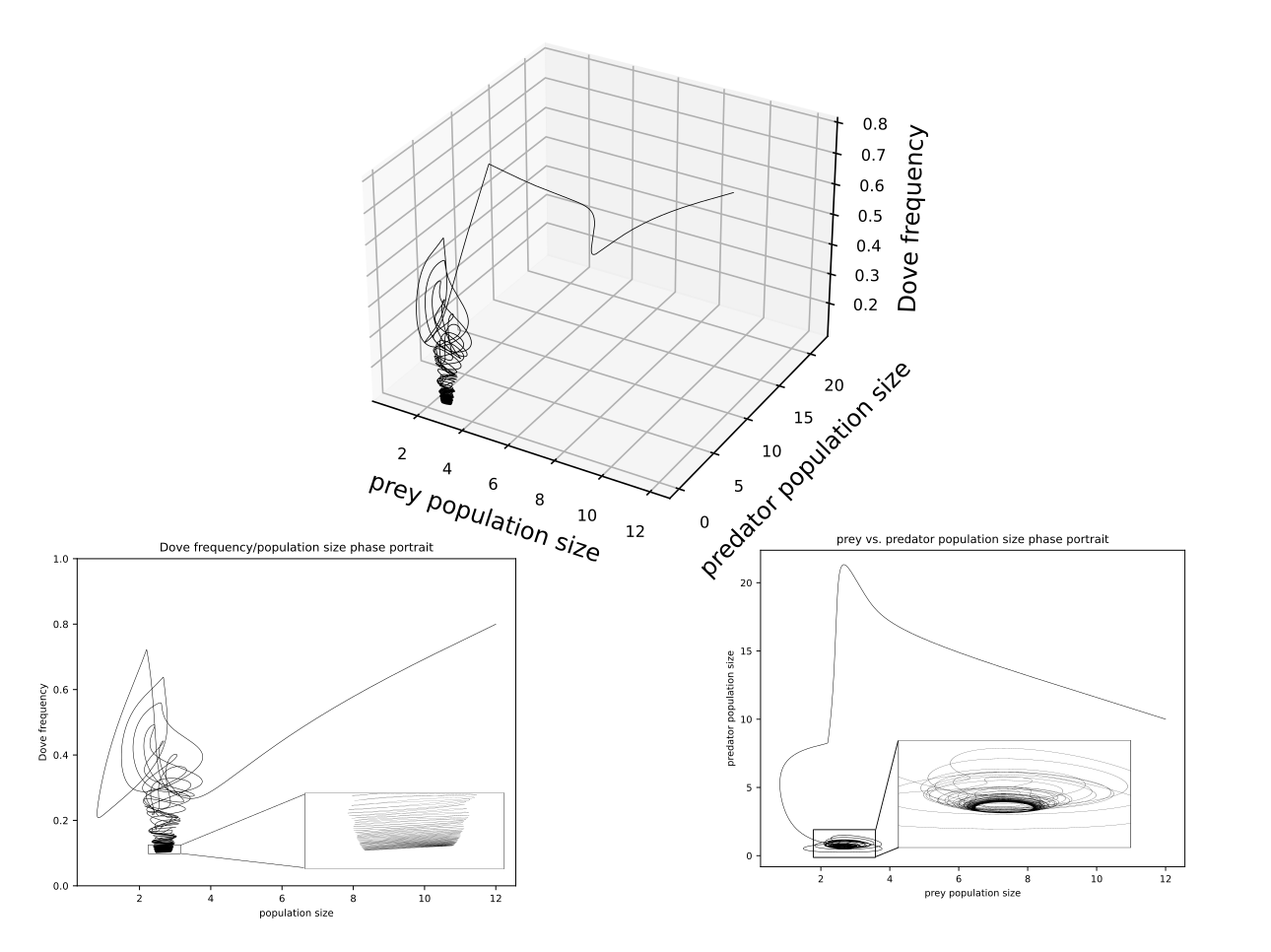}
	\caption{Phase portraits of the system
		from Fig.4}
		\label{fig:Fig5}
\end{figure}

\begin{figure}[tbp] 
	\includegraphics[height=10cm]{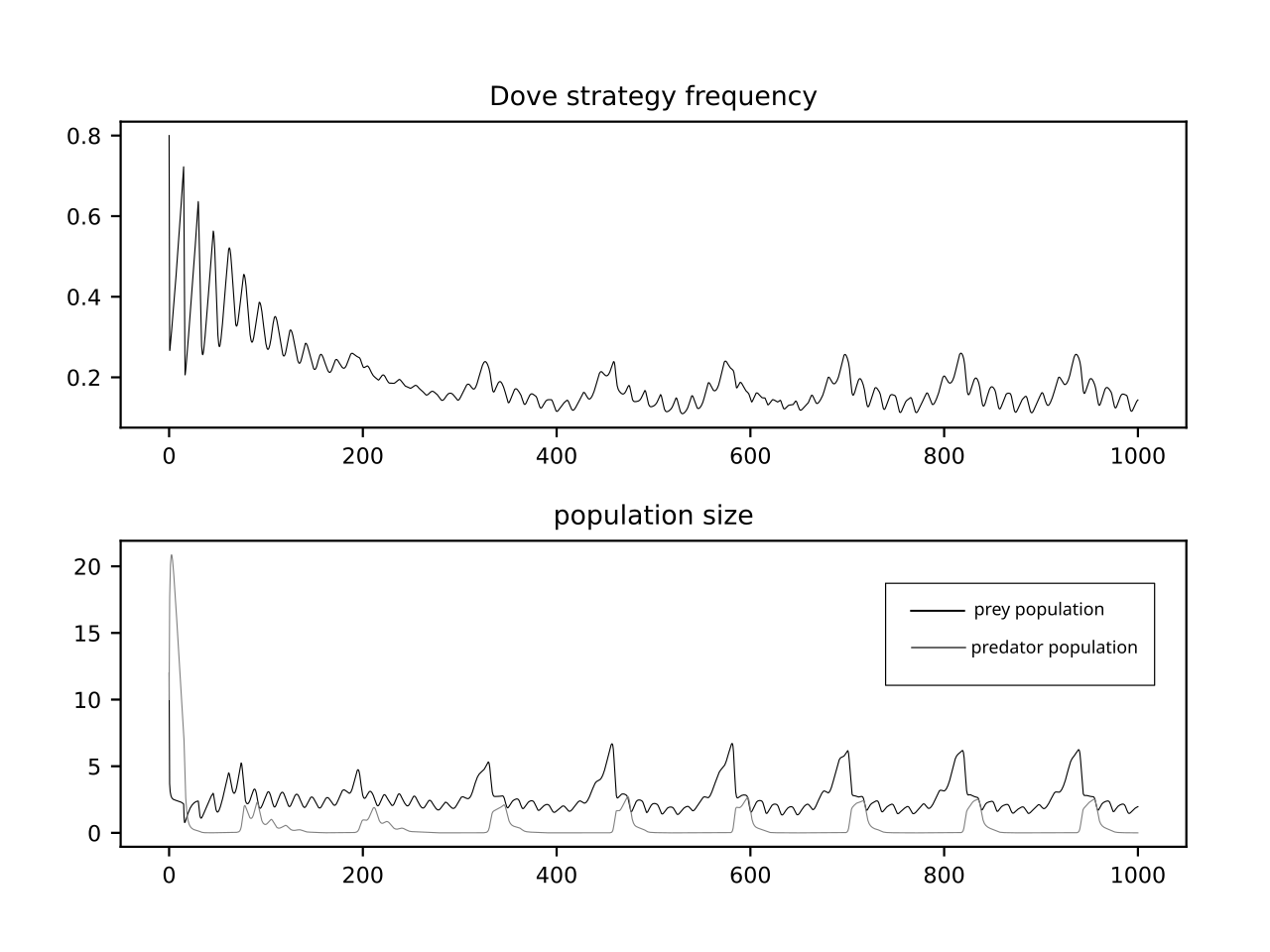}
	\caption{ System with predator pressure
		from Figure 5 with additional seasonal mortality (with amplitude $\protect
		\alpha =0.25$ and period $\protect\theta =120$)	}
		\label{fig:Fig6}
\end{figure}

\begin{figure}[tbp] 
	\includegraphics[height=10cm]{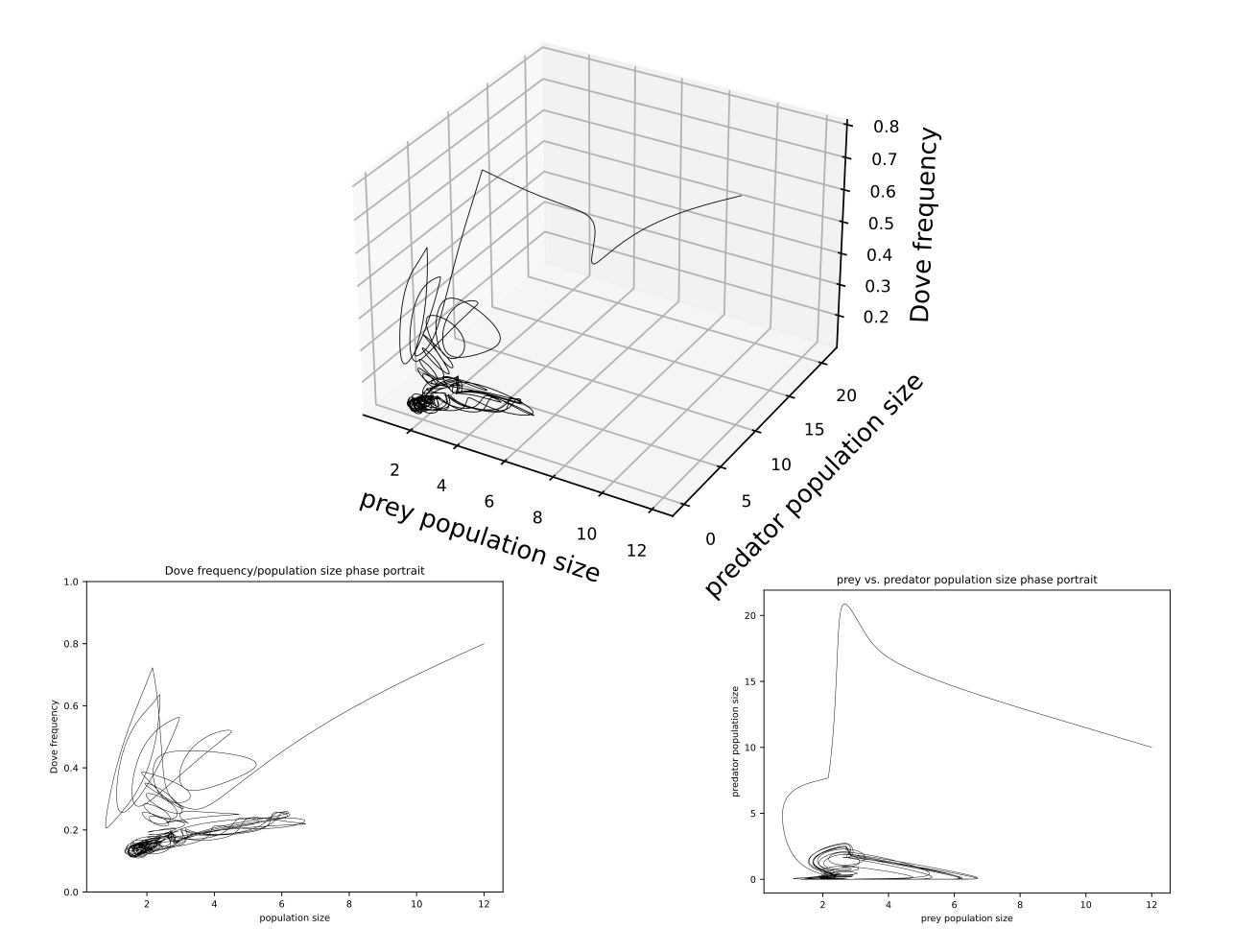}
	\caption{ Phase portraits of the system
		from Fig.6.}
		\label{fig:Fig7}
\end{figure}

Similar behaviour we can observe when we add to our model additional
equation and mortality factor describing the predator pressure. In effect we
obtain combination of the replicator dynamics and simple Lotka-Volterra
system. The parameters of the prey-predator subsystem are $b_{p}=0.3$ ,$\
d_{p}=0.8$ and $p=0.6$. In this case instead of external oscillator, we have
additional feedback. Trajectories are depicted in Fig. \ref{fig:Fig4} and phase portraits
in Fig. \ref{fig:Fig5}. When we combine the predator pressure with seasonal mortality
(with amplitude $\alpha =0.25$ and period $\theta =120$), the resulting
cycling behaviour is even more complex (see Fig. \ref{fig:Fig6} and \ref{fig:Fig7}). 

\subsection{Case with the logistic juvenile recruitment mortality, type b)
juvenile survival}

\begin{figure}[tbp] 
	\includegraphics[height=19cm]{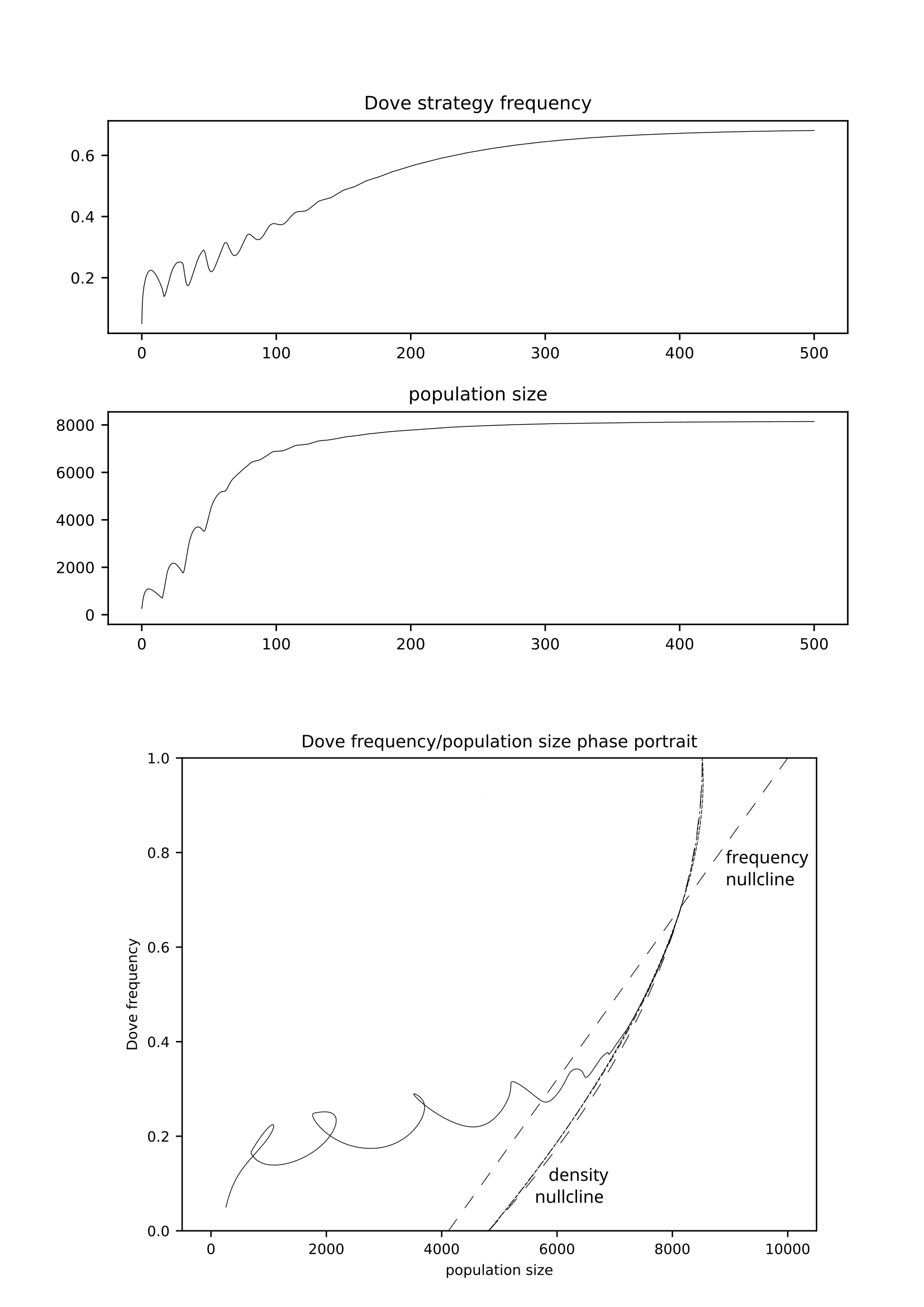}
	\caption{Model with logistic juvenile
		recruitment survival. Parameters are $F=1.7$, $d=1$, $\Phi =0.5$, $\Psi =0.2$
		\ and the delay $\protect\gamma =15$. We assume the non constant history,
		frequency declining from $0.225$ to $0.05$ and the population size declining
		from $560$ to $260$. Nullclines are shown in dashed lines.}
		\label{fig:Fig8}
\end{figure}

This is the model similar to the previous works on demographic games \cite{argbr1,argbr2,argbr3}, but with added
delay. Model parameters are $F=1.7$, $d=1$, $\Phi =0.5$, $\Psi =0.2$ \ and
the delay $\gamma =15$. We assume the non constant history, frequency
declining from $0.225$ to $0.05$ and the population size declining from $560$
to $260$. Numerical simulations show that when trajectory reaches the
neighbourhood of the area limited by nullclines, then the fluctuations
resulting from the delay disappear and the system behaves similarly to the
model without delay (Fig \ref{fig:Fig8}). This may be result of the slowdown of the pace
of convergence in the area limited by nullclines. In effect, impact of
the delay is weaker due to smaller differences in the values of the
population state at time $t$ and $t-\gamma $.

\begin{figure}[tbp] 
	\includegraphics[height=19cm]{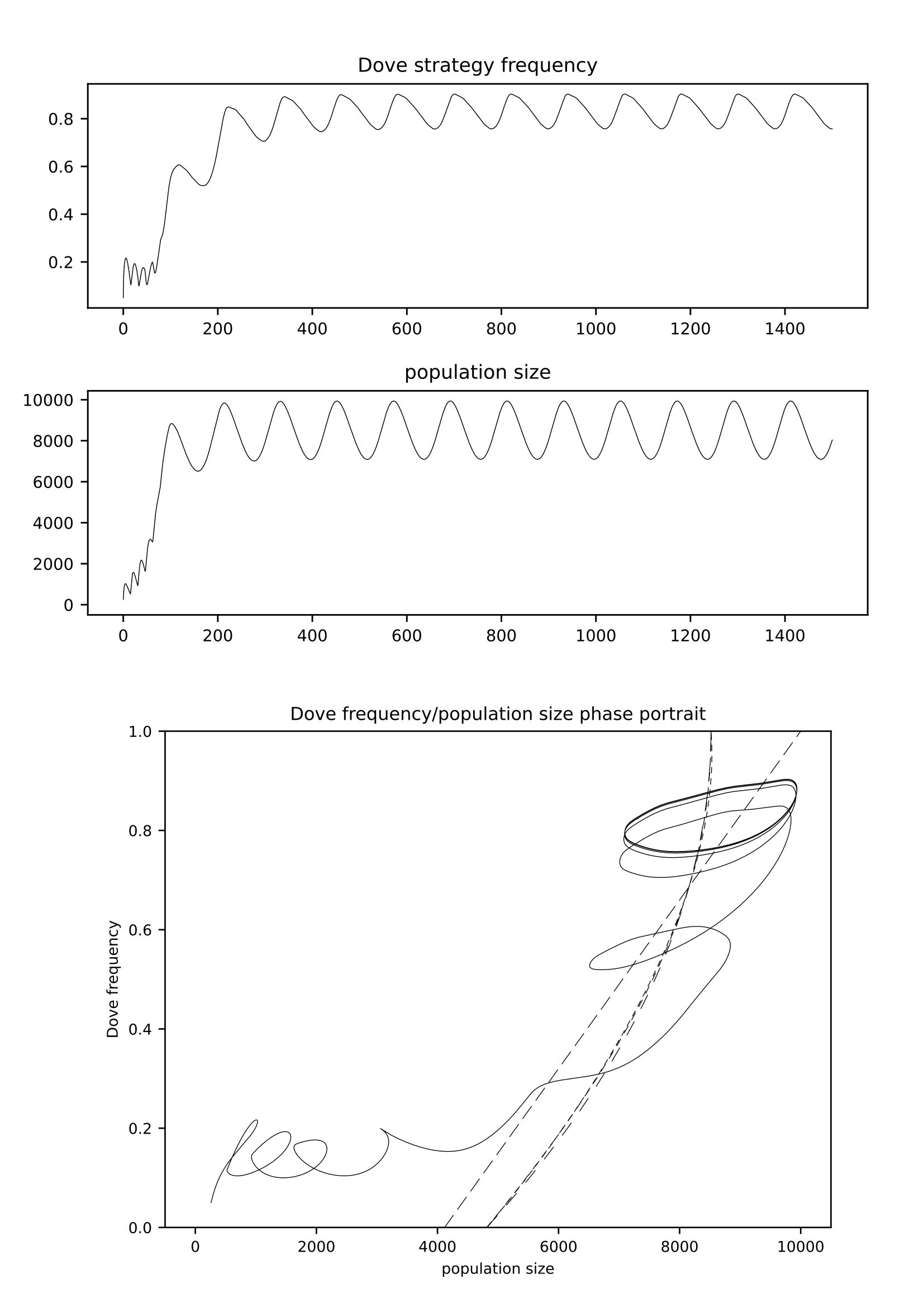}
	\caption{Model from Fig.8 with constant
		background mortality replaced (by setting $\Psi =0$) by periodic mortality
		with amplitude $\protect\alpha =0.4$ and period $\protect\theta =12$. We can
		observe the overshoot of the intersection of the nullclines (dashed lines).}
		\label{fig:Fig9}
\end{figure}

When we replace the constant background mortality by seasonal mortality
fluctuating around the same value by setting $\Psi =0$ and adding $\alpha
=0.25$ and period $\theta =120$, we observe the overshoot of the
intersection and the limit cycle is significantly above it (Fig. \ref{fig:Fig9} ).

\begin{figure}[tbp] 
	\includegraphics[height=7cm]{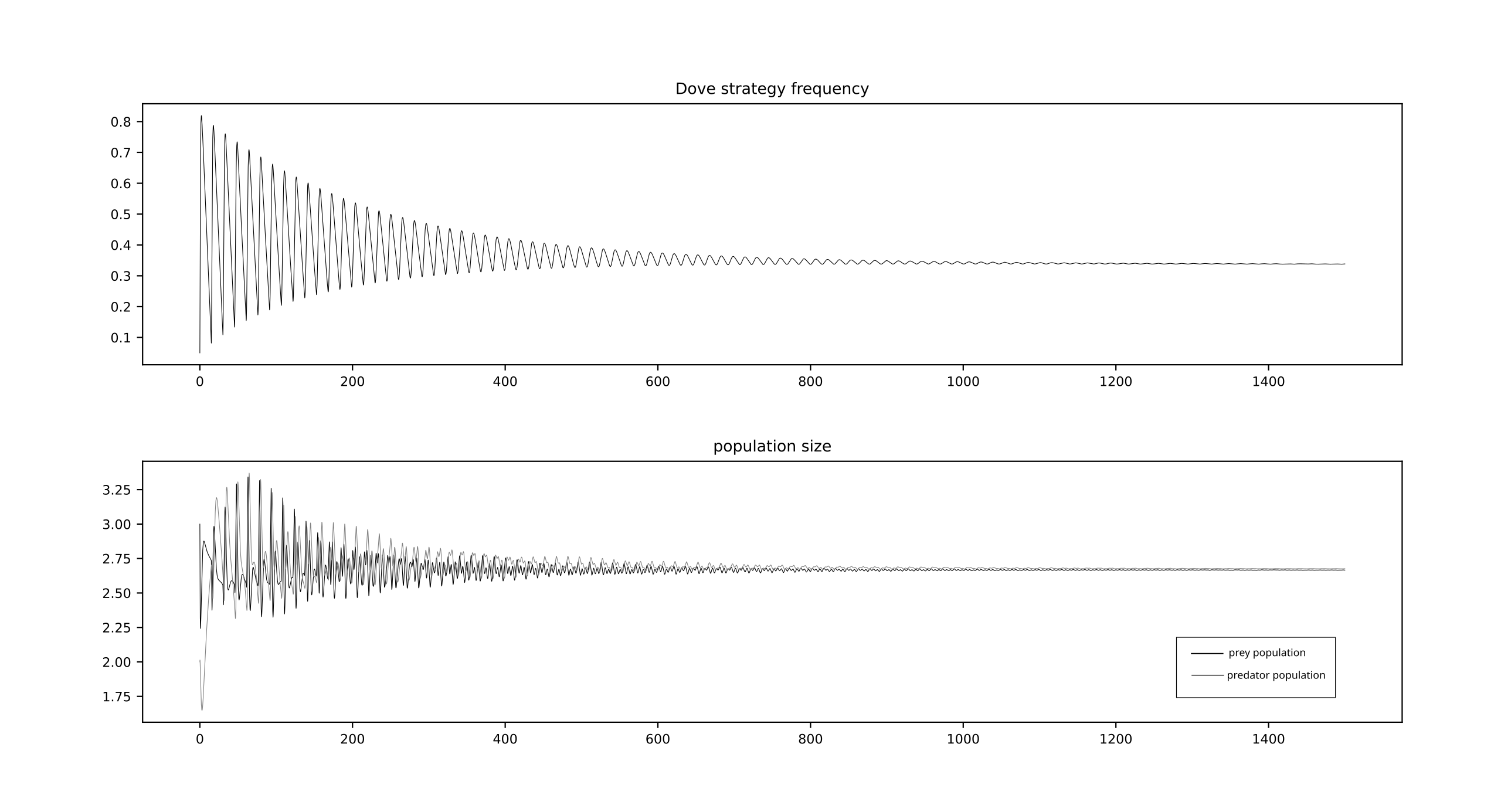}
	\caption{Model
		with added predator pressure. Parameters are $F=0.7$, $d=1$, $\Phi =2$, $
		\Psi =0.2$, carrying capacity $K=50$ and with delay $\protect\gamma =15$.
		Parameters for the predator subsystem are $b_{p}=0.3$ ,$\ d_{p}=0.8$ and $
		p=0.6$. Initial history where frequency declines linearly from $0.95$ to $
		0.05$ and population size linearly increases from $1.5$ to $3$. }
		\label{fig:Fig10}
\end{figure}

\begin{figure}[tbp] 
	\includegraphics[height=10cm]{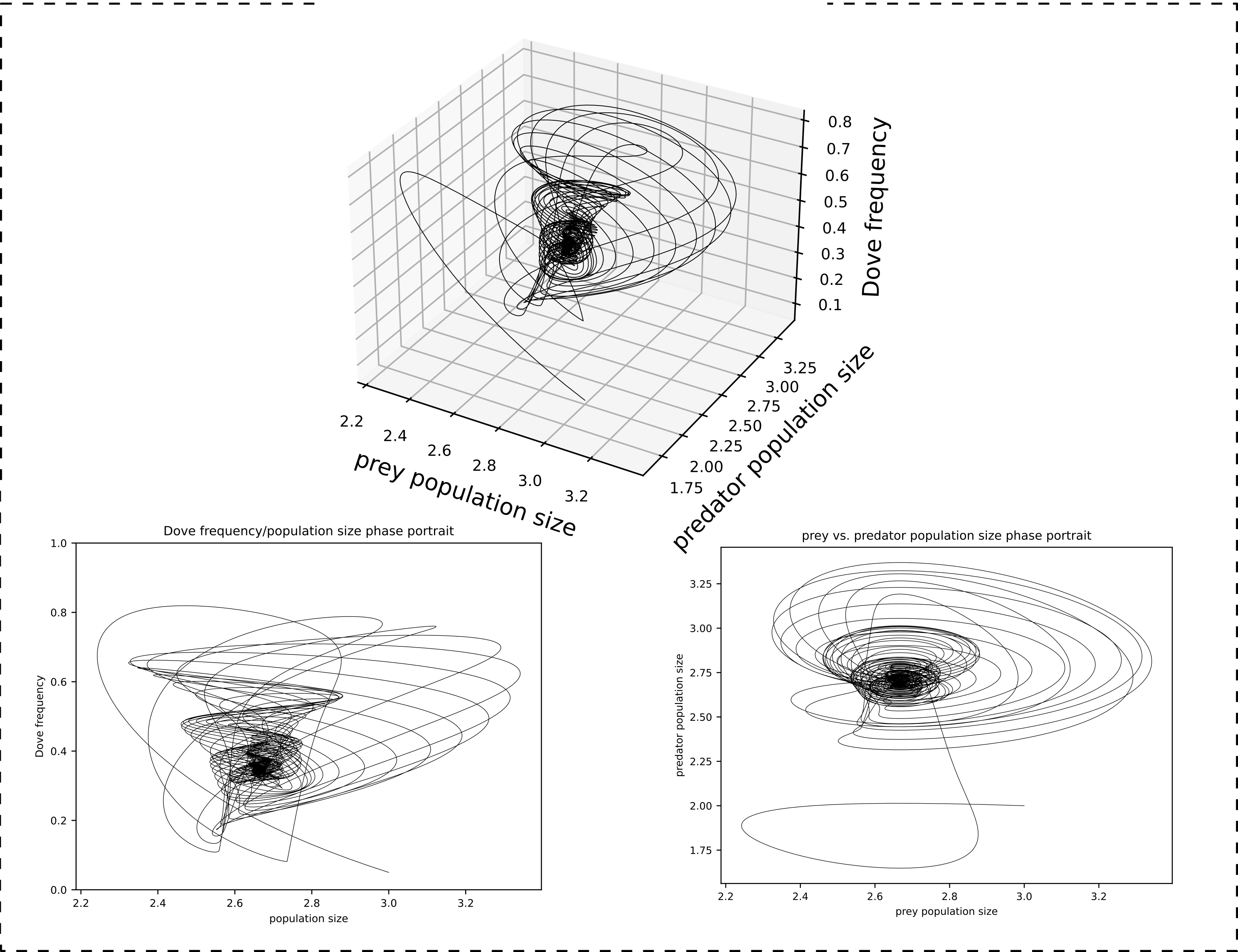}
	\caption{Phase portraits of the system from Fig. 11. }
	\label{fig:Fig11}
\end{figure}

\begin{figure}[tbp] 
	\includegraphics[height=7cm]{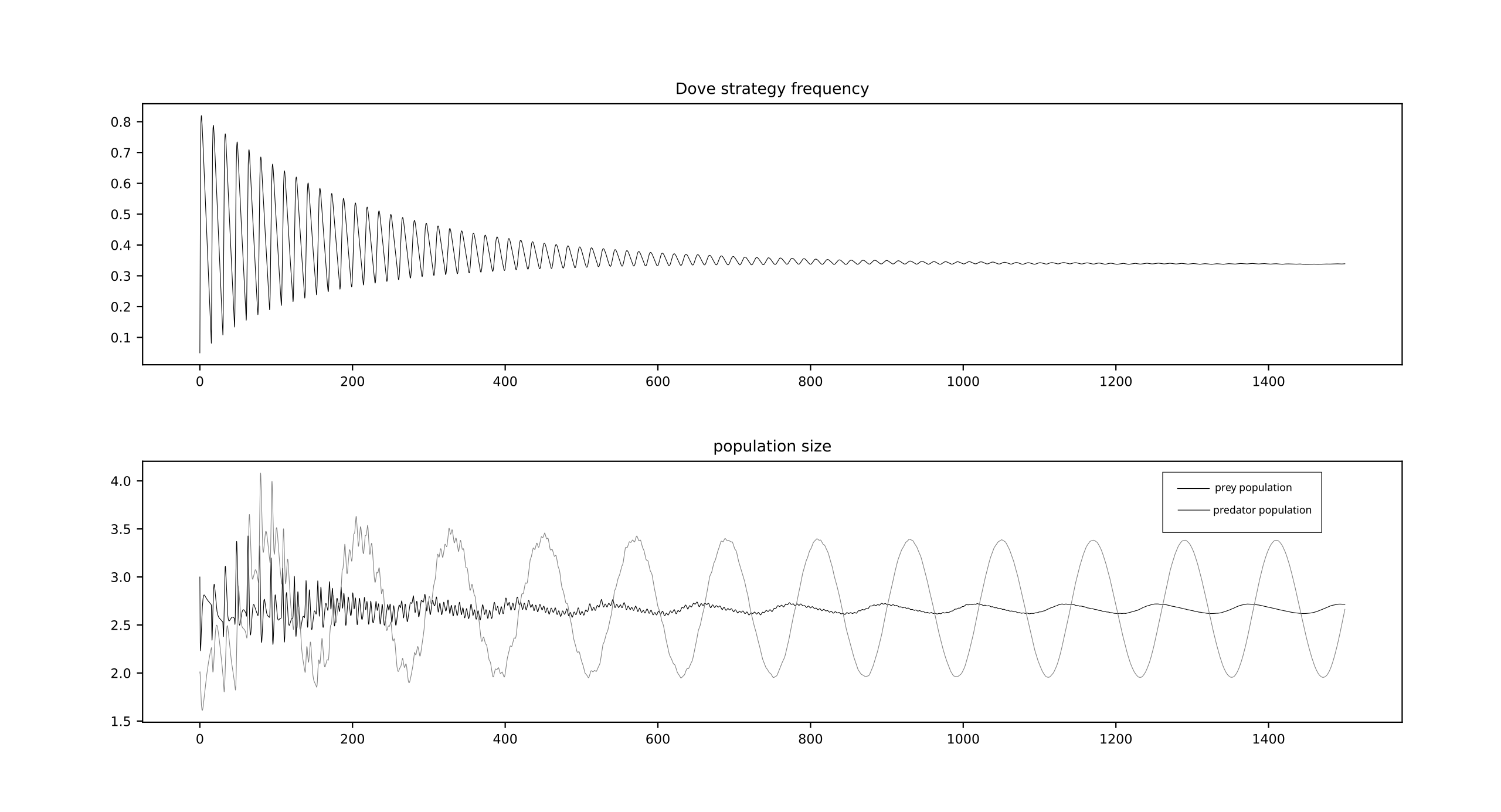}
	\caption{System from Figures 10 and 11 with replaced
		constant mortality with with periodic mortality by setting $\Psi =0$ and
		amplitude $\protect\alpha =0.4$ and period $\protect\theta =12$. }
		\label{fig:Fig12}
\end{figure}

\begin{figure}[tbp] 
	\includegraphics[height=12cm]{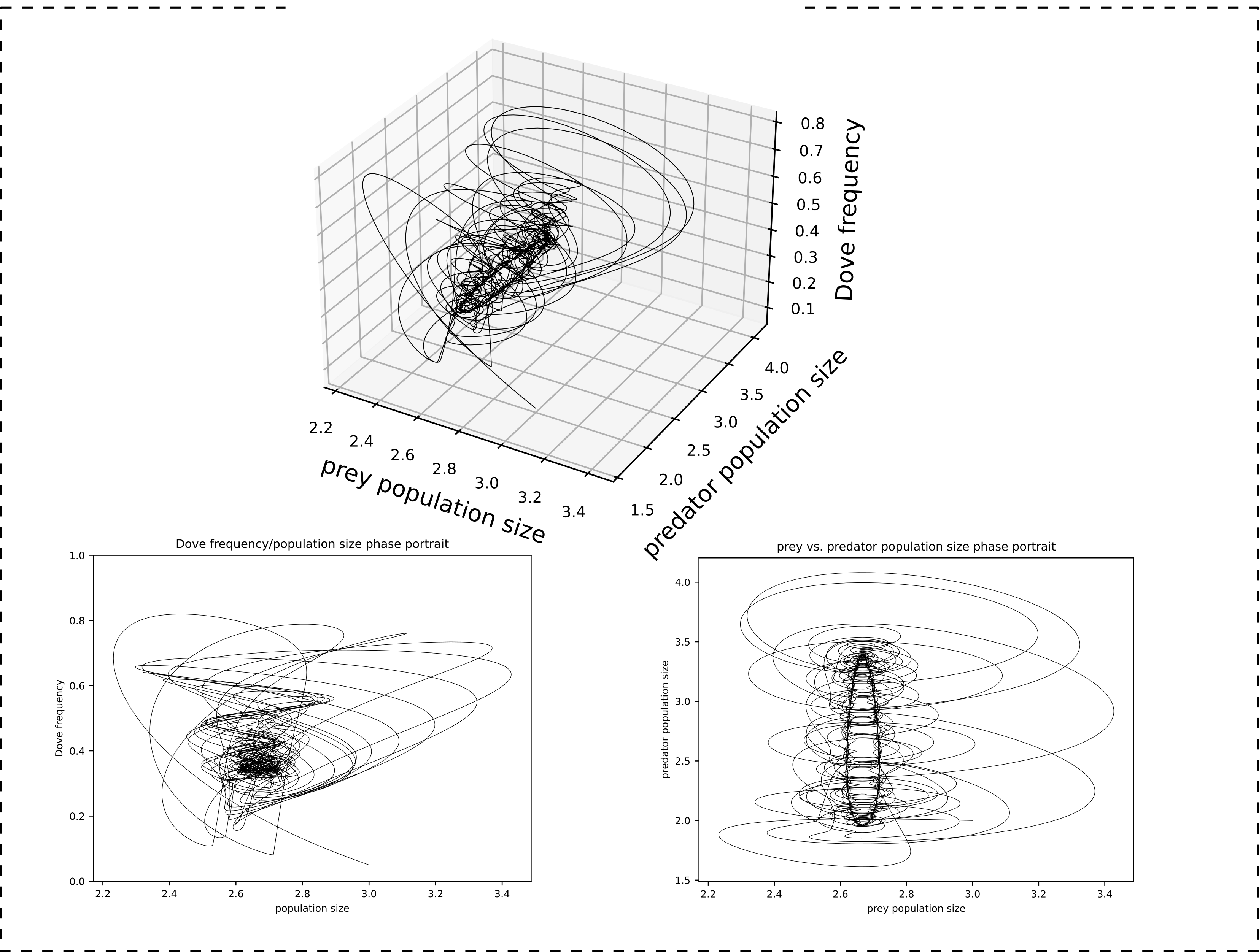}
	\caption{Phase portraits of the system from Fig. 11. }
	\label{fig:Fig13}
\end{figure}

We also can combine our model with predator impact for parameters $F=0.7$, $
d=1$, $\Phi =2$, $\Psi =0.2$, carrying capacity $K=50$ and with delay $
\gamma =15$. We change initial history where frequency declines linearly
from $0.95$ to $0.05$ and population size linearly increases from $1.5$ to $
3 $. Initial number of predators equals $2$. Parameters for the predator
subsystem are again $b_{p}=0.3$ ,$\ d_{p}=0.8$ and $p=0.6$. The obtained
cyclic trajectories are very sophisticated (Fig. \ref{fig:Fig10} and Fig. \ref{fig:Fig11}). We obtain an interesting
pattern, when we replace the constant mortality with periodic
mortality by setting $\Psi =0$ and amplitude $\alpha =0.4$ and period $
\theta =12$. Seasonality do not alters the frequency dynamics significantly,
but surprisingly induces greater oscillations on the predator population
(Fig. \ref{fig:Fig12} and \ref{fig:Fig13}).

\subsection{Case with the delayed juvenile recruitment mortality, type c)
juvenile survival \protect\bigskip}

\begin{figure}[tbp] 
	\includegraphics[height=12cm]{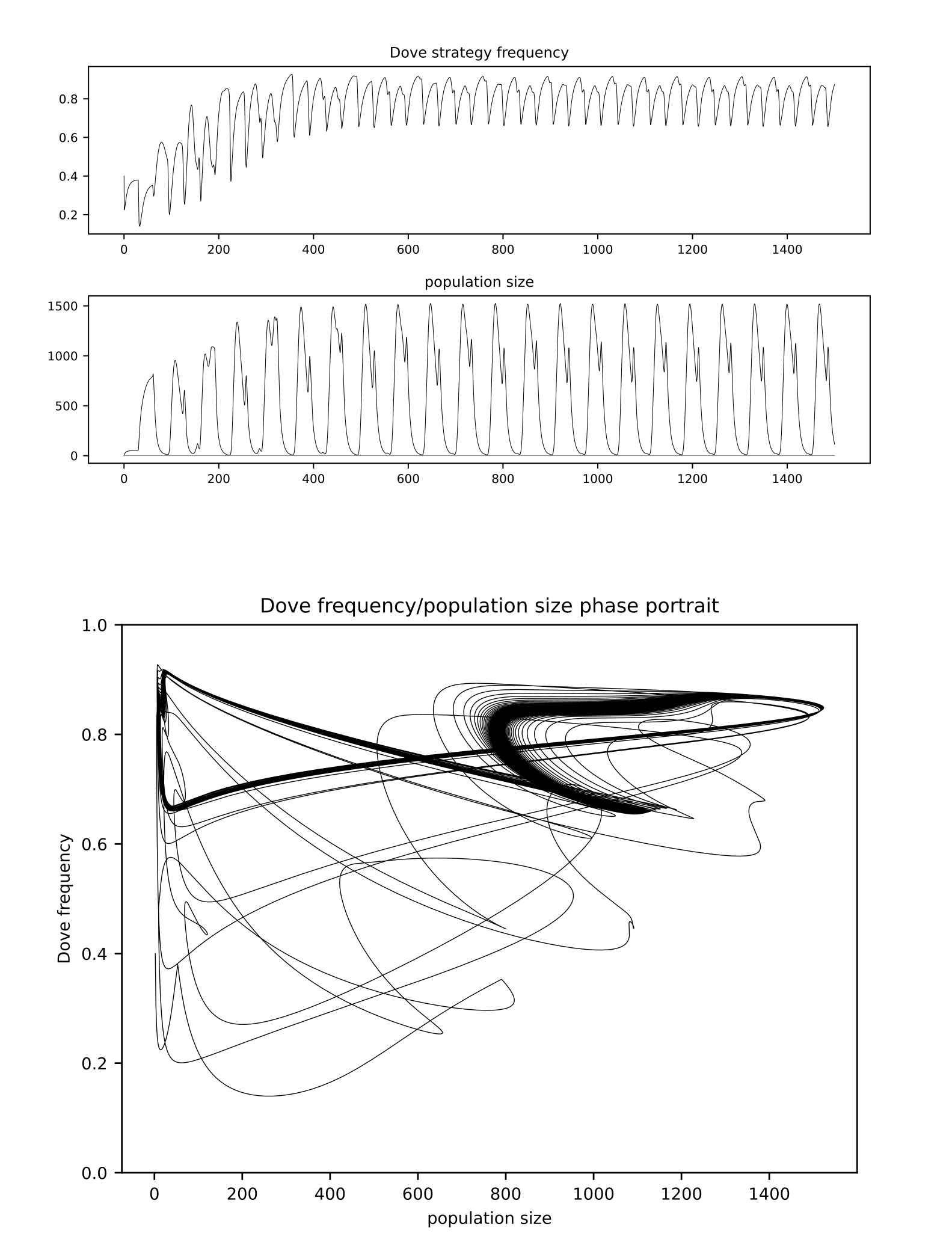}
	\caption{Model with delayed juvenile
		recruitment $\mathbf{D}(n(t))=u^{n(t-\protect\tau )/z}$ with parameters $
		u=0.6$ and $z=50$. Payoff parameters are $F=18$, $d=1$, $\Phi =0.5$, $\Psi
		=0.158$ and delay $\protect\gamma =30$. Constant history with frequency $0.4$
		and populations size $2$. We can observe bifurcation leading to the complex
		cycle, forming the Elvis-Presley-shaped pattern. }
		\label{fig:Fig14}
\end{figure}

Figure \ref{fig:Fig14} shows the resulting complex cyclic trajectories. This version of
the model is very sensitive to small changes of the parameters.

In the case of delayed juvenile recruitment $\mathbf{D}(n(t))=u^{n(t-\gamma
)/z}$ we can observe the cascade of bifurcations. Let us set the parameters $
F=18$, $d=1$, $\Phi =0.5$, $\Psi =0.158$, parameters of juvenile survival
function are $u=0.6$ and $z=50$ and delay $\gamma =30$. Initial history
is constant with frequency $0.4$ and populations size $2$.

\begin{figure}[tbp] 
	\includegraphics[height=19cm]{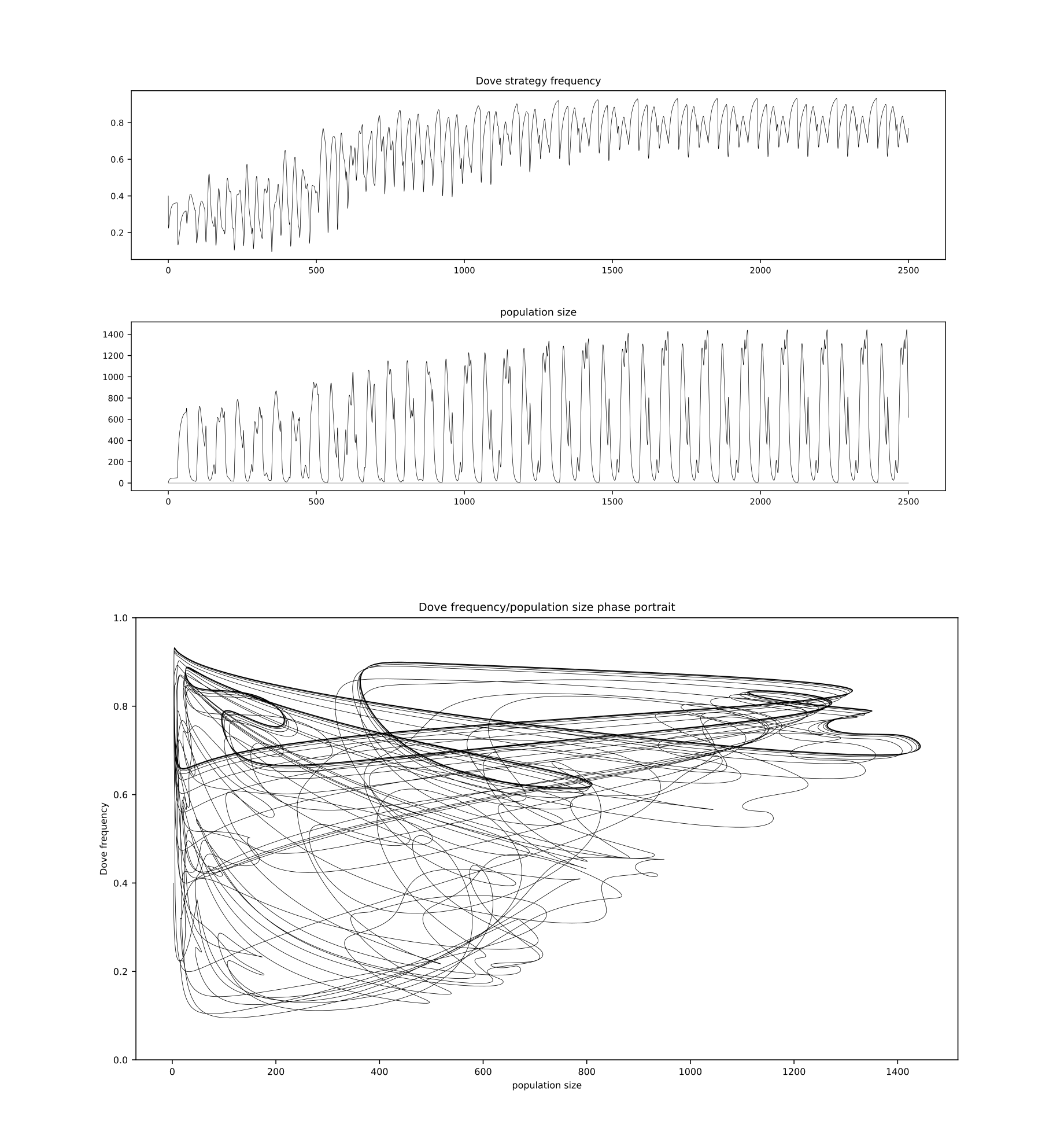}
	\caption{Model from Fig.14 with only altered background mortality $\Psi =0.184$. We can observe dramatical increase in complexity of trajectories. }
	\label{fig:Fig15}
\end{figure}

\begin{figure}[tbp] 
	\includegraphics[height=19cm]{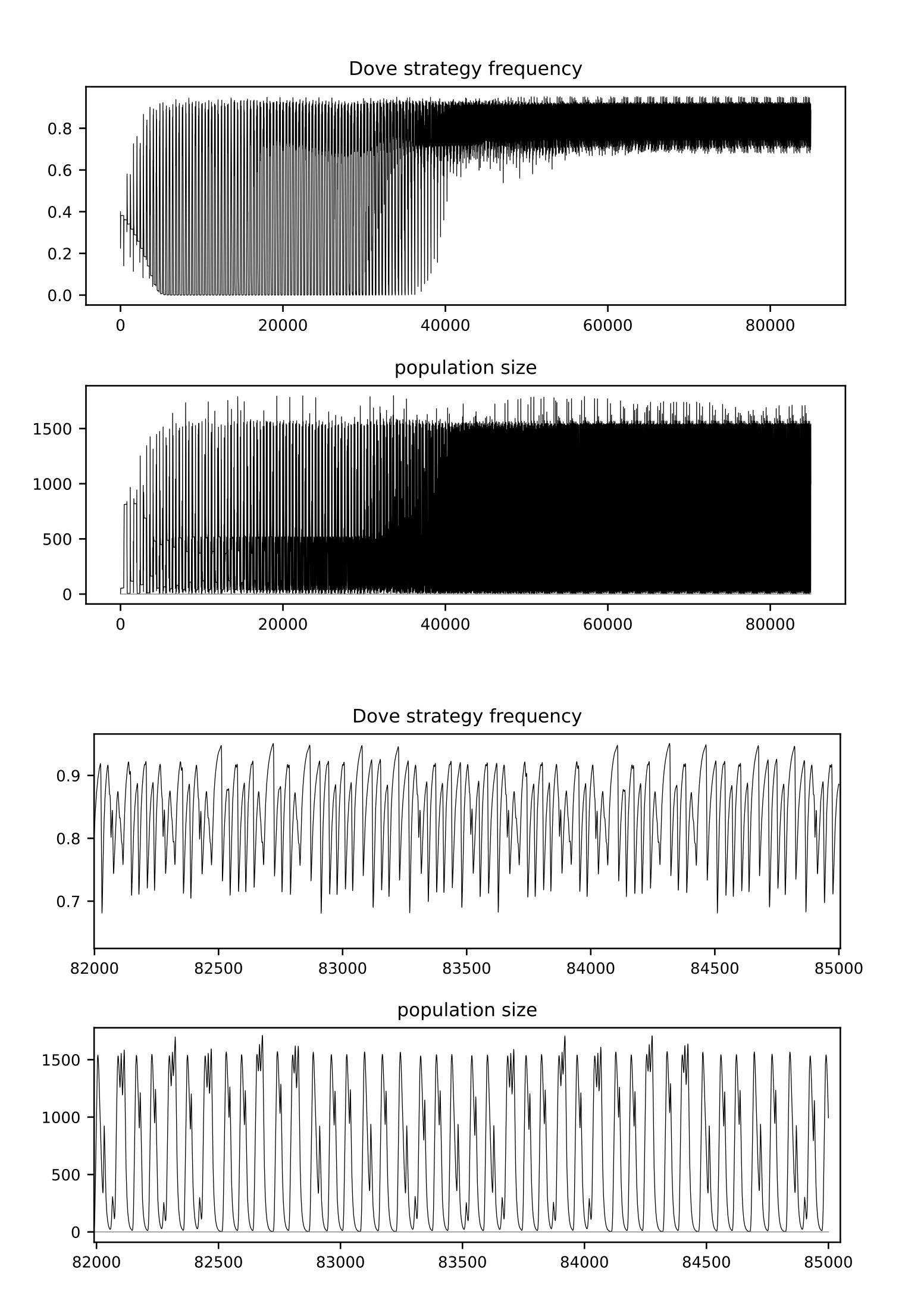}
	\caption{System from Fig. 15 with long delay $\protect\gamma =394.7$. Long transient leading to the rapid regime shift can be observed. The underlying mechanism remains unknown, and suggests possible route for a future research. }
	\label{fig:Fig16}
\end{figure}

\begin{figure}[tbp] 
	\includegraphics[height=14cm]{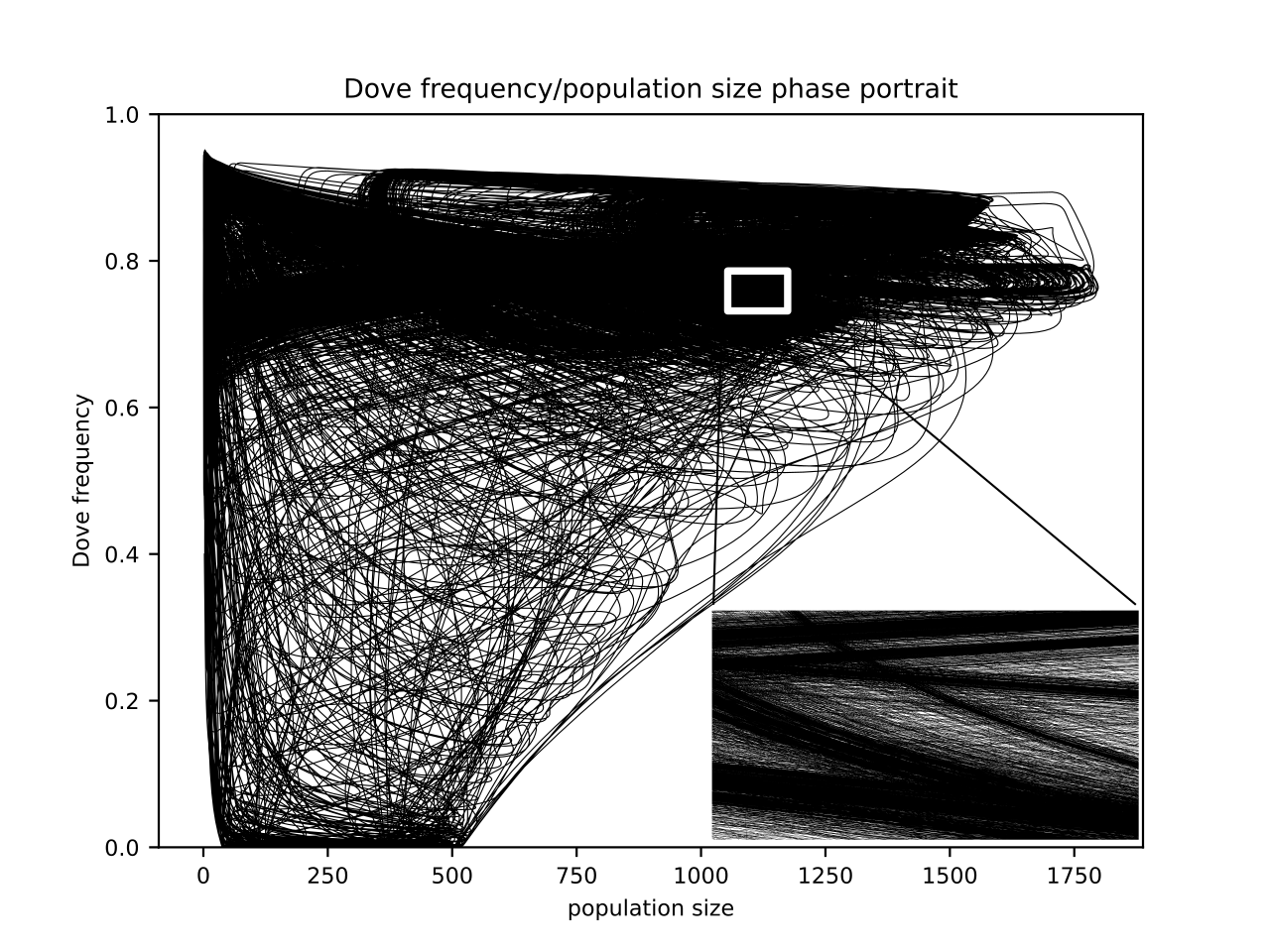}
	\caption{Phase portrait for the system from Fig. \ref{fig:Fig16}. }	
	\label{fig:Fig17}
\end{figure}

For example when we set the background mortality for $\Psi =0.184$ we can
obtain even more complex patterns (Fig. \ref{fig:Fig15}). We can set the significantly
greater delay $\gamma =394.7$. Since single time unit is the average per
capita time between interactions, this long delay is biologically plausible
for long living organisms. In this example we want to show maximal
complexity of the obtained trajectories, which are depicted in Fig. \ref{fig:Fig16} and Fig.
\ref{fig:Fig17}.

\begin{figure}[tbp] 
	\includegraphics[height=19cm]{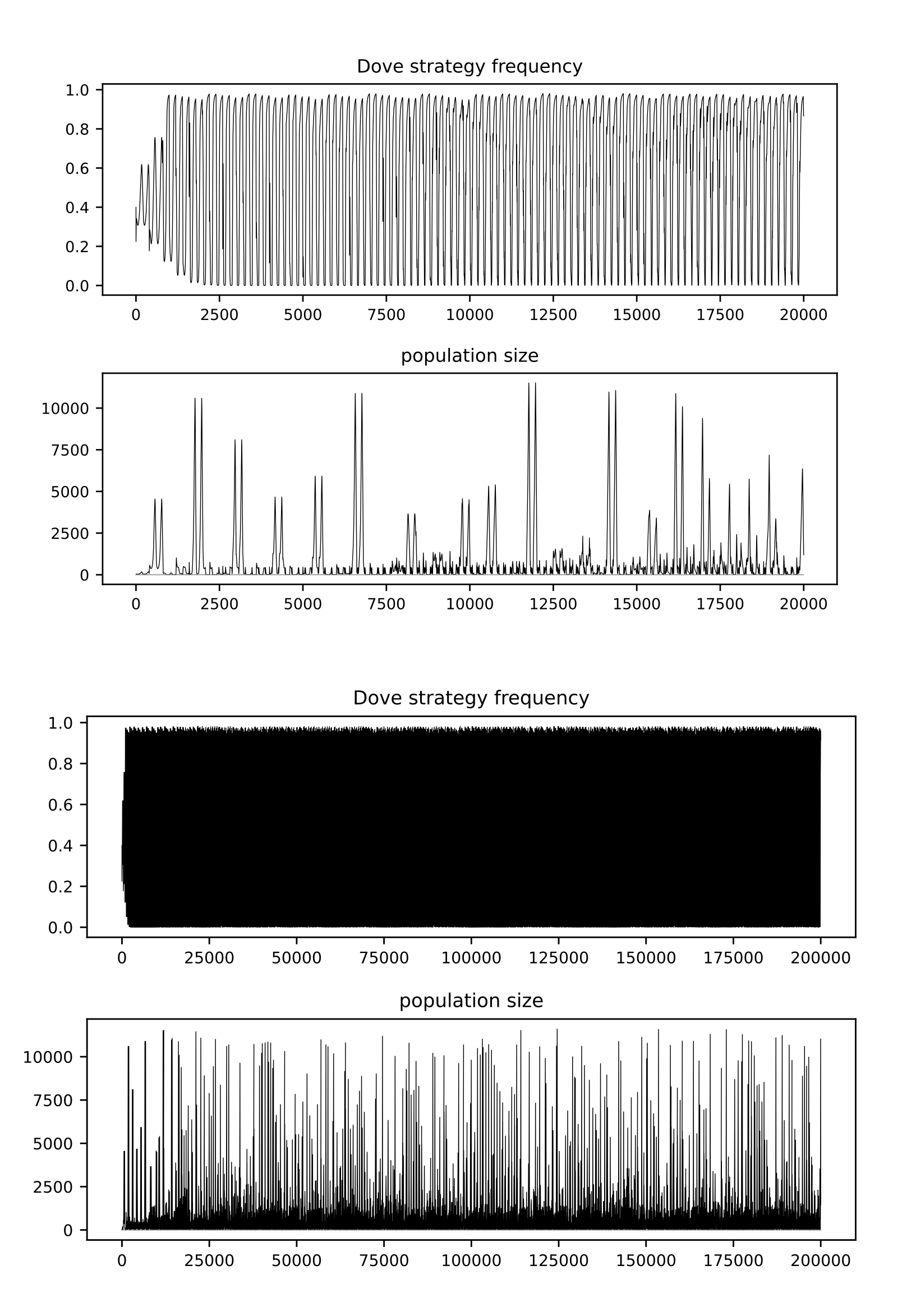}
	\caption{System from Fig. \ref{fig:Fig16}  and \ref{fig:Fig17} with
		added periodic mortality by setting mortality $\Psi =0.008$ and for
		fluctuating part the amplitude $\protect\alpha =0.4$ and period $\protect
		\theta =12$. We can observe that mechanism of the strategic selection is completely unefficient. In addition, the ecological dynamics shows the "boom and bust" pattern. }
		\label{fig:Fig18}
\end{figure}

\begin{figure}[tbp]
	\includegraphics[height=10cm]{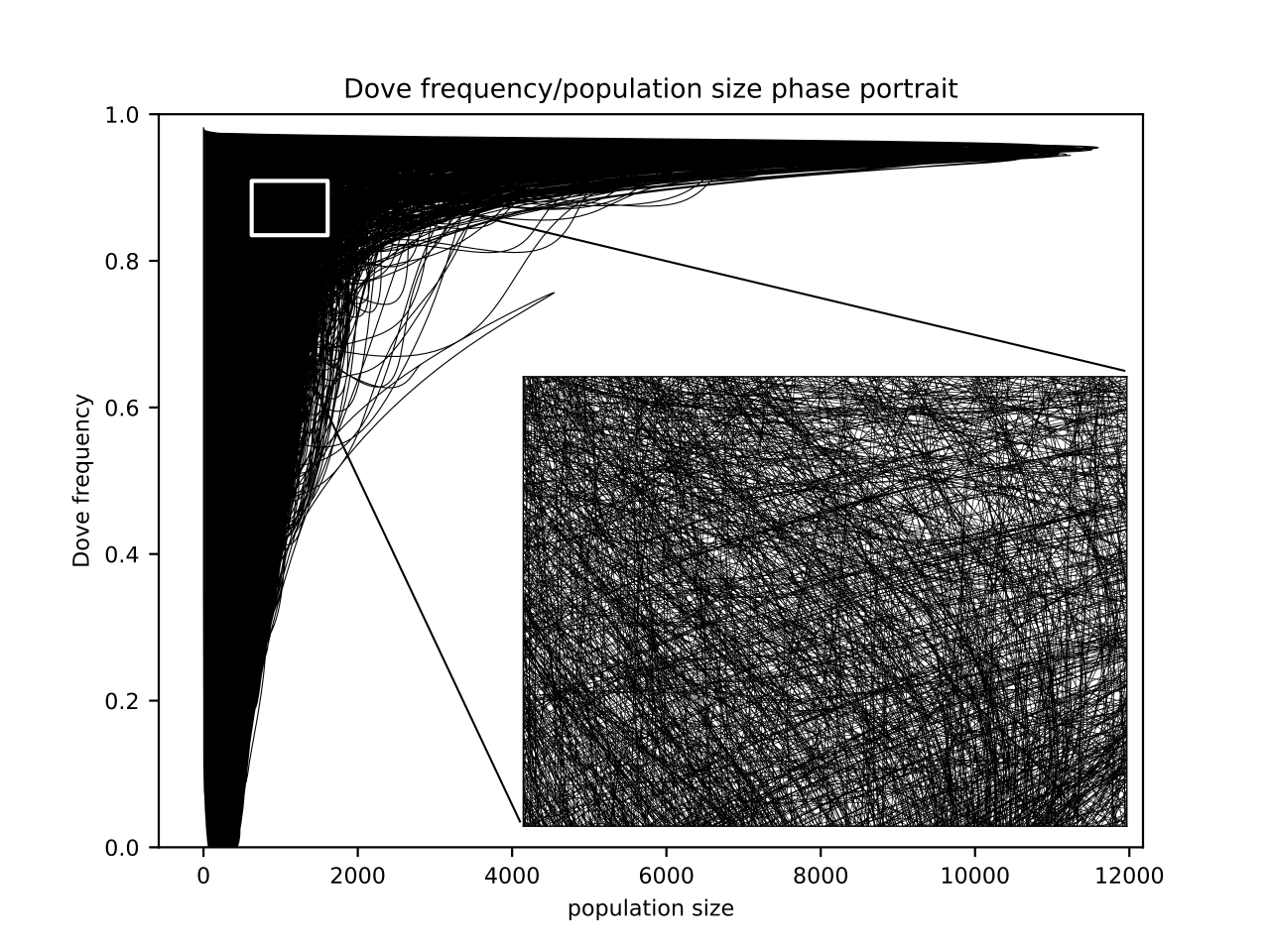}
	\caption{Phase portrait for the system from Fig 18. }
	\label{fig:Fig19} 
\end{figure}

It is interesting, what happens when we divide the background mortality into the
constant and the fluctuating part by setting $\Psi =0.008$ and for
fluctuating part the amplitude $\alpha =0.4$ and period $\theta =12$. The
obtanied trajectories show extreme impact on the dynamics of strategy
frequencies (Fig. \ref{fig:Fig18} and Fig. \ref{fig:Fig19}).

Adding the predator pressure modelled by simple Lotka-Volterra system leads
to the similar behaviour than in the case of logistic juvenile survival
without delay. Therefore presenting it in detail is pointless.\bigskip

\bigskip

\FloatBarrier

\subsection{Bifurcation analysis}
\label{sec:bif-discussion}

The parameters of systems with density dependence driven by adult mortality
or logistic juvenile recruitment survival in a long run generate very simple
dynamics (steady states and periodic orbits under seasonality), therefore we do not
present bifurcation diagrams for those. We start from the application of the Theorem 1 and present a plot of the
surfaces of the first bifurcation and the loss of stability of the stable
point as the functions of the model parameters ($F$, $d$ and $\Phi $, $\Psi $
, see Figure \ref{fig:BifSurf}). The simplification of the system and derivation of the
respective coefficients is in the Appendix 4. The plots are obtained
numerically by simple algorithm: firstly, checking that the rest point exists and
chosing the smallest positive root of the characteristic polynomial (if it
exists). The Scilab source code is available in Supplementary Materials and in repository \cite{GitHubTraj}.

\begin{figure}[tbp]
	\includegraphics[height=10cm]{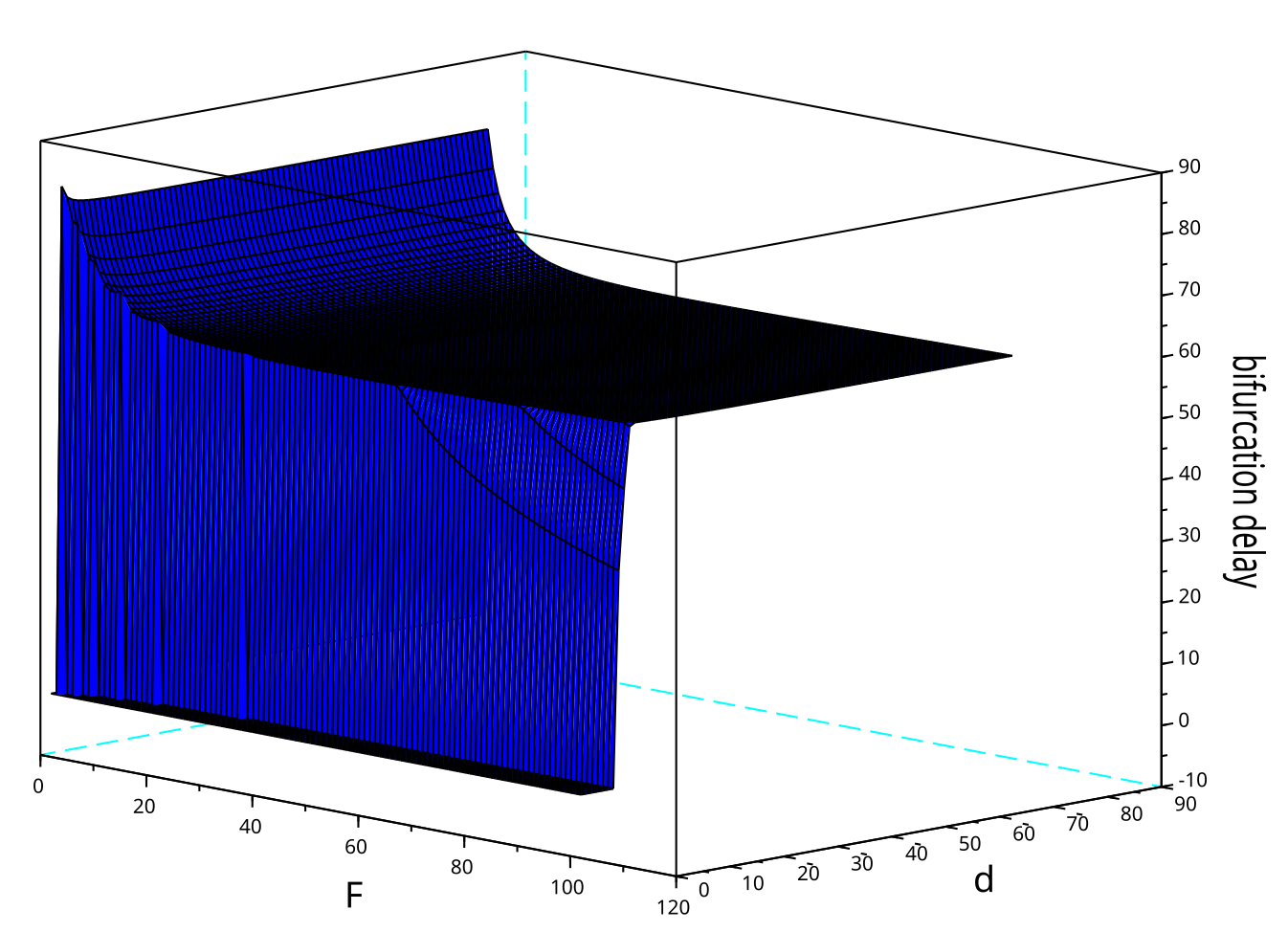}
	\includegraphics[height=10cm]{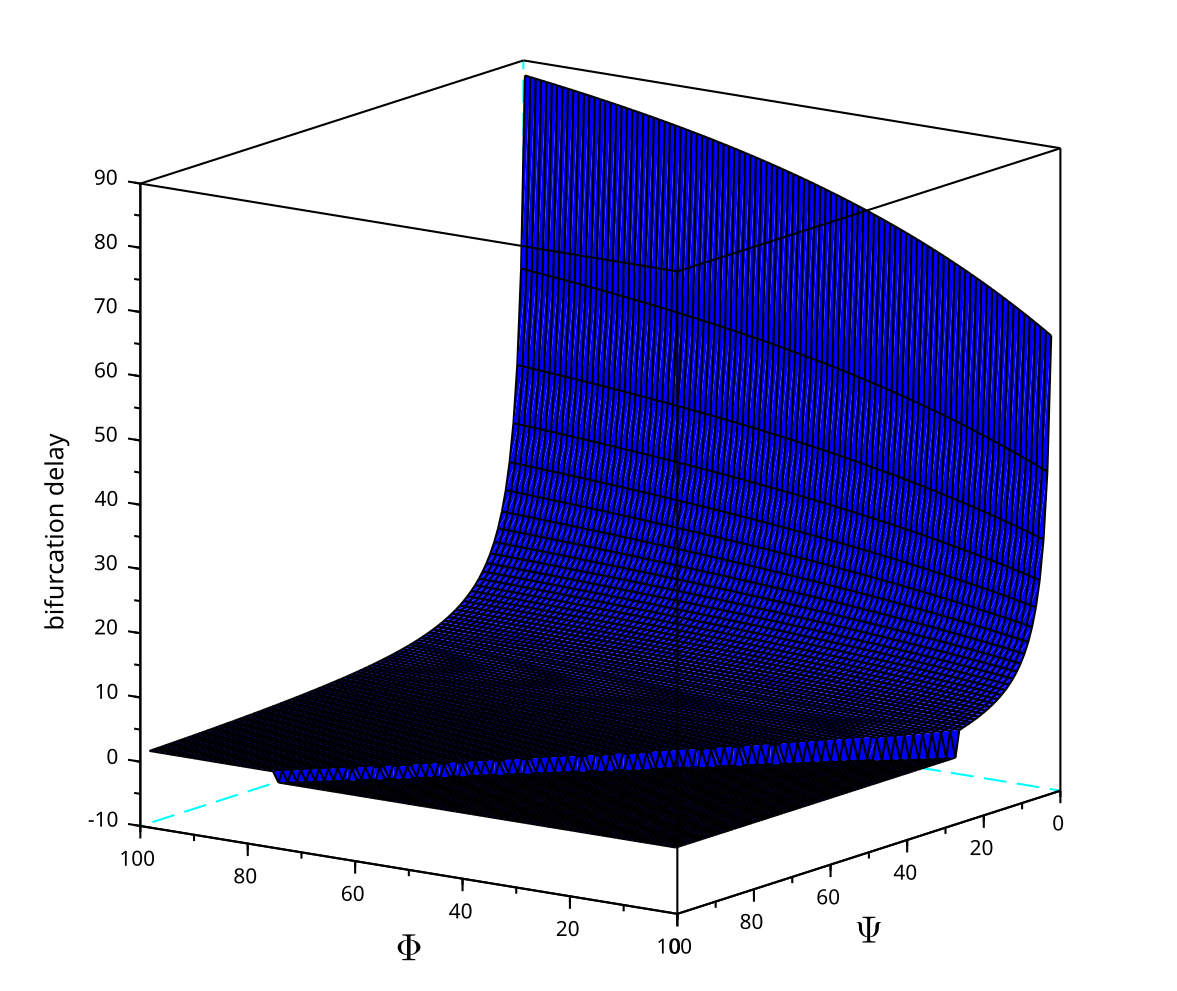}
			\caption{Plot of the bifurcation surfaces resulting from the application of the Theorem 1. Panel a) shows the plot over values of the focal game parameters $F$ and $d$ for fixed values $\Psi=0.0115$ and $\Phi=0.1$. Panel b) shows the plot over values of the background parameters $\Phi$ and $\Psi$ for fixed values $F=0.6$ and $d=1$ }	
			\label{fig:BifSurf} 
\end{figure}

\bigskip

\begin{figure}[tbp]
\center{
	\includegraphics[width=12cm]{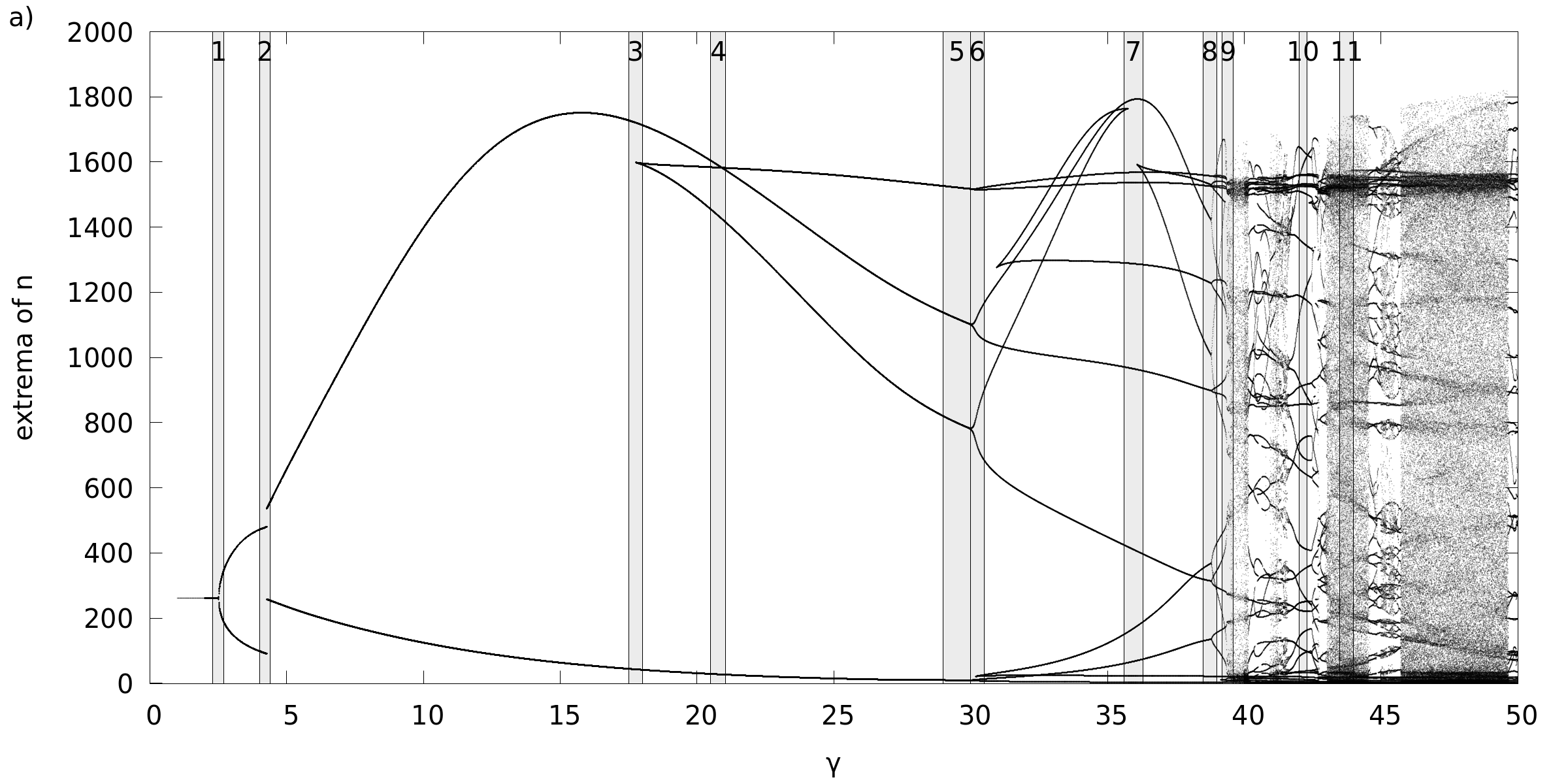}
	
	\includegraphics[width=12cm]{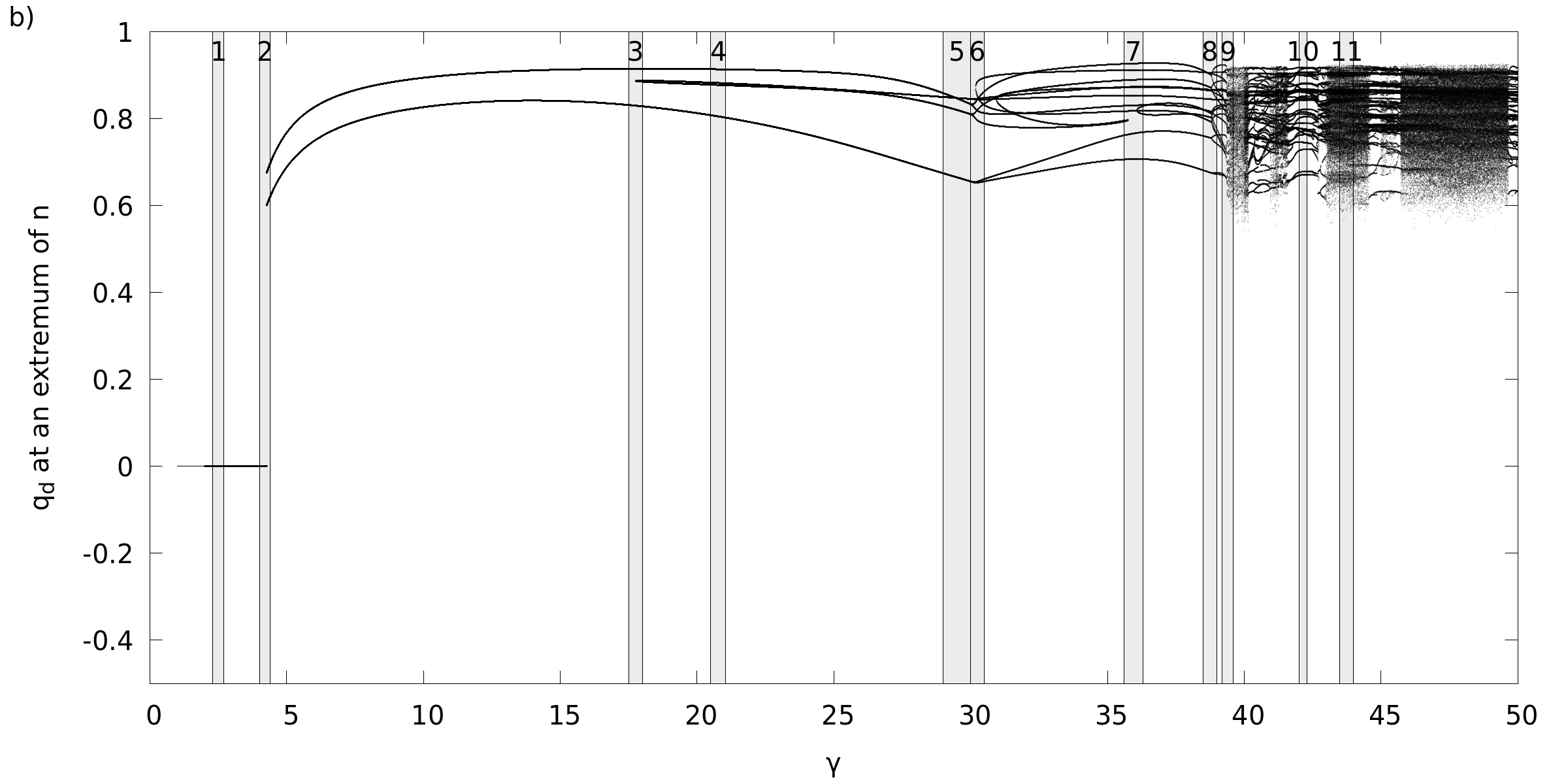}
}
\caption{ An extrema bifurcation diagram for Dove system with third kind of $
D(n(t))$ function. For each value of $\protect\gamma $ we record all extrema
of the function $n(t)$ (i.e. values $n(t_{i})$ where $n^{\prime }(t_{i})=0$)
after the numerical solution approaches the apparent attractor. The values
of $q_{d}$ are recorded at the same points $t_{i}$. The grey bars numbered
1--11 corresponds to the plots discussed in Section~\protect\ref
{sec:bif-discussion} }
\label{fig:bifdiag} 
\end{figure}

\begin{figure}[tbp]
\center{
\includegraphics[height=4cm]{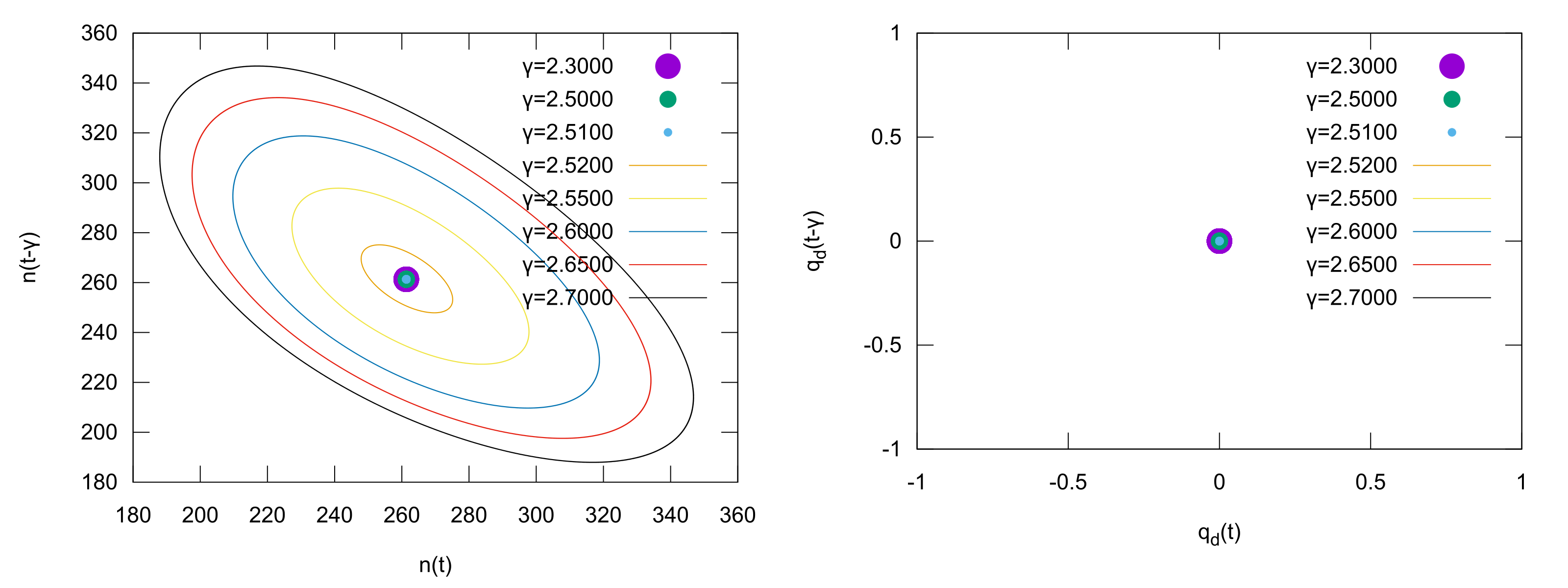}}
\caption{ A Hopf bifurcation in $n$ happens for some $\protect\gamma^* \in
[2.51,2.52]$ (region~1 in Fig~.\protect\ref{fig:bifdiag}). The fixed point
for three values of $\protect\gamma$ are shown as overlapping dots. Then a
periodic orbit appears for $\protect\gamma > 2.51$ that grows in amplitude.
The corresponding trajectories of $n$ and $q$ are drawn in the same colours.}
\label{fig:periodic-biff-1}
\end{figure}

\begin{figure}[tbp]
\center{
\includegraphics[height=4cm]{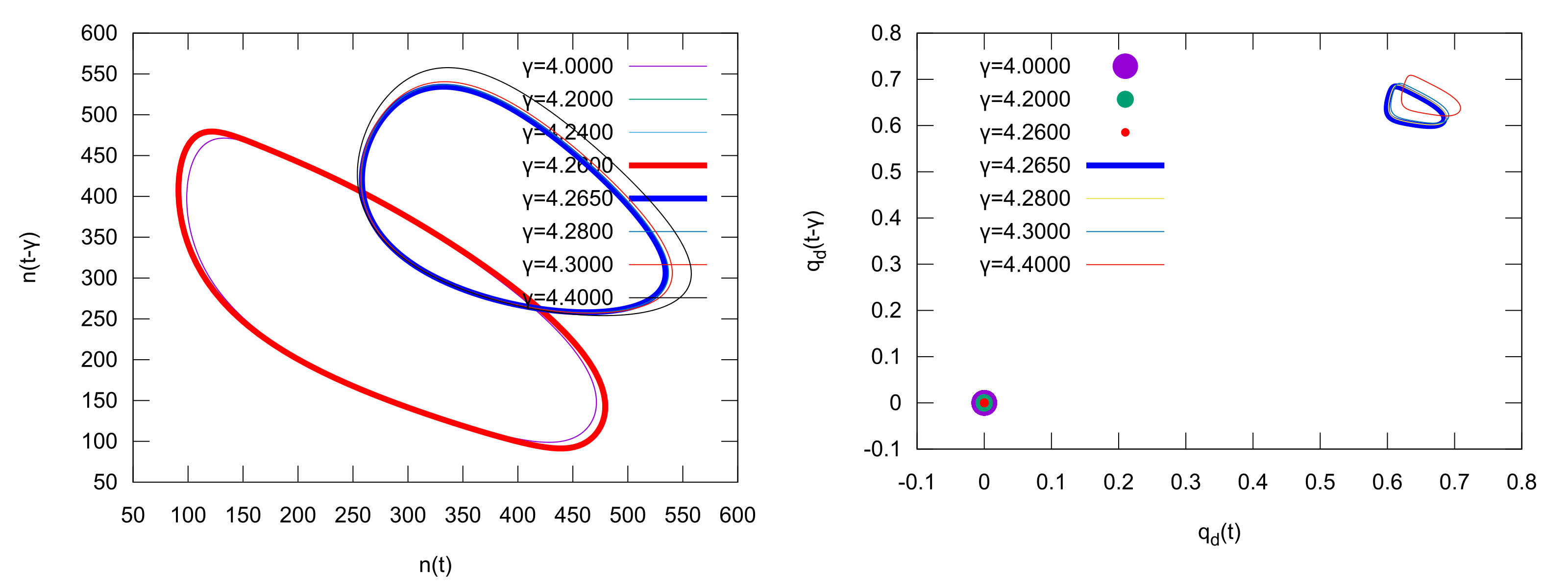}}
\caption{ A periodic orbit in $q_d$ appears out of thin air for some $%
\protect\gamma \in [4.260, 4.265]$ (region~2 in Fig~.\protect\ref%
{fig:bifdiag}). In the right panel we see the red dot that corresponds to a
constant value trajectory of $q_d$ for $\protect\gamma = 4.260$ that
corresponds to a periodic orbit in $n$ shown as a thick red line on the
left. The thick, blue periodic orbit $g_d$ for $\protect\gamma = 4.265$ in
the right panel corresponds to the thick, blue orbit for $n$ on the left.
The corresponding trajectories of $n$ and $q$ are drawn in the same colours.}
\label{fig:periodic-biff-2}
\end{figure}

\begin{figure}[tbp]
\center{
\includegraphics[height=4cm]{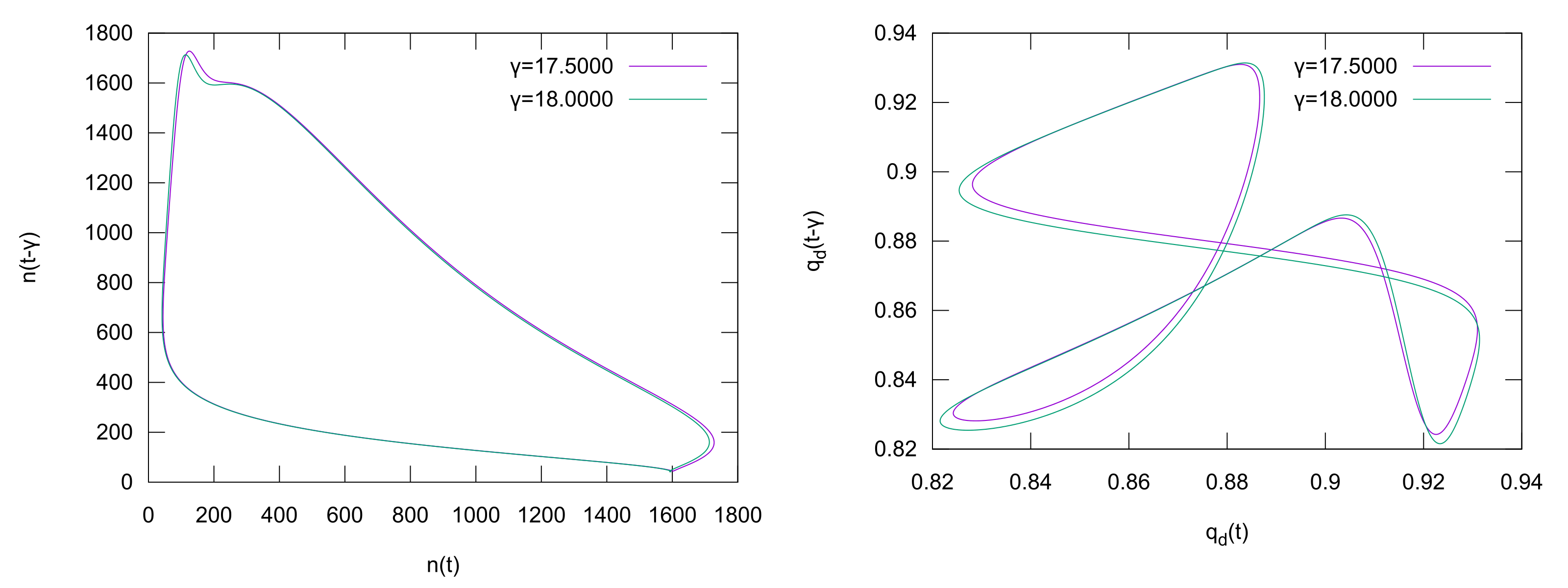}
\includegraphics[height=4cm]{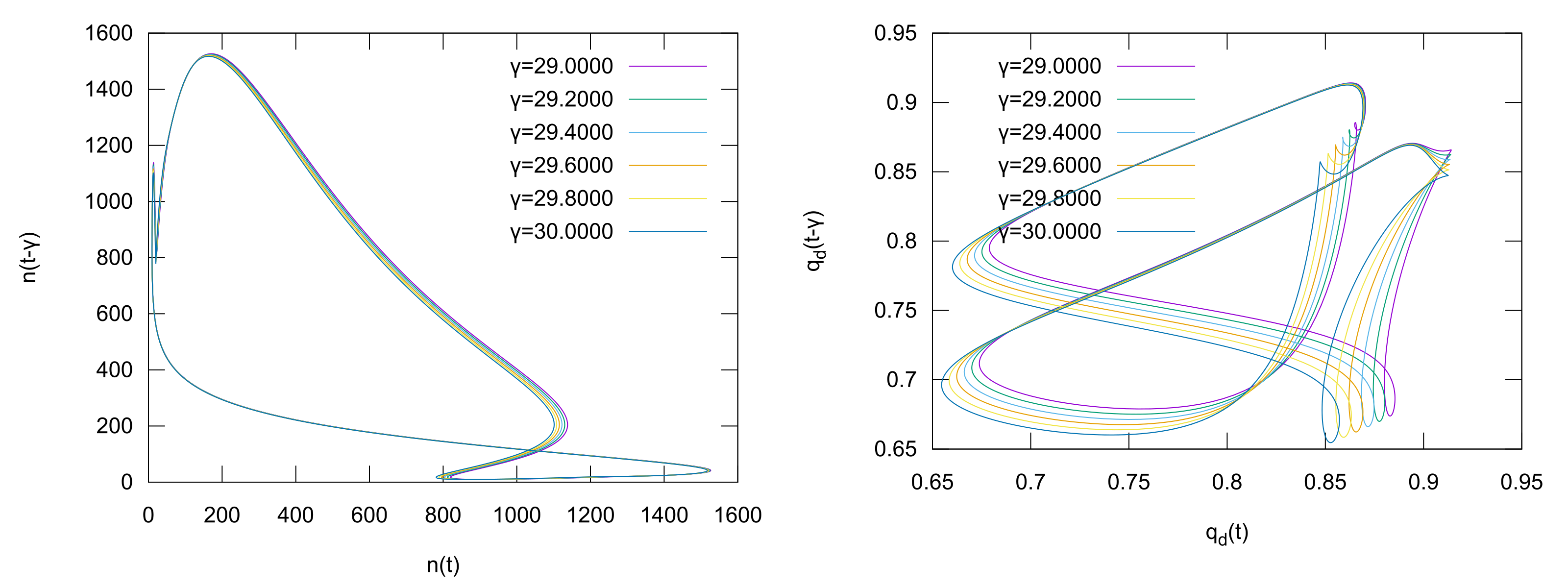}
}
\caption{ First \emph{kink} in $n$ appears (top panels, region~3 in Fig~.%
\protect\ref{fig:bifdiag}) and then another one in $q_{d}$ (bottom panels,
region~5 in Fig~.\protect\ref{fig:bifdiag}). The small kink in $n$ can be
seen in the right-bottom corner of the first top panel, while the kink in $%
q_{d}$ is larger, and can be seen in the top-right corner of the solution in
bottom-right panel. The corresponding trajectories of $n$ and $q$ are drawn
in the same colours.}
\label{fig:periodic-biff-3-and-5} 
\end{figure}

\begin{figure}[tbp]
\center{
\includegraphics[height=4cm]{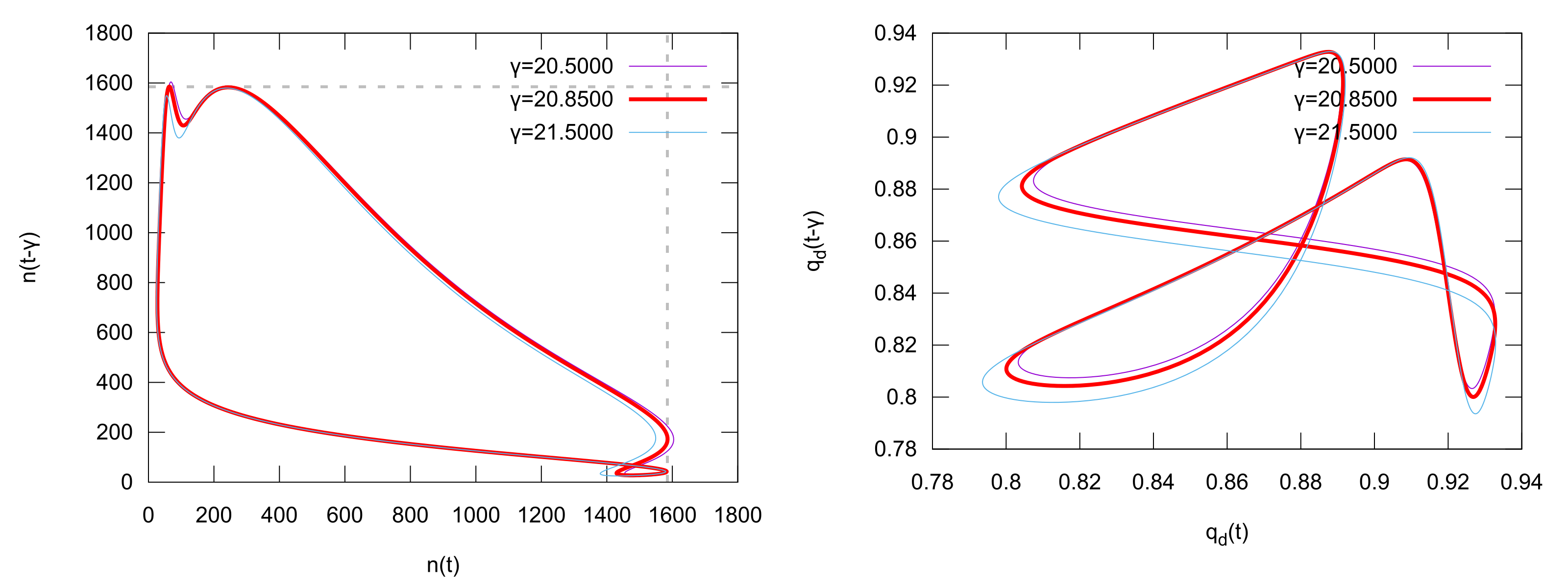}
\includegraphics[height=4cm]{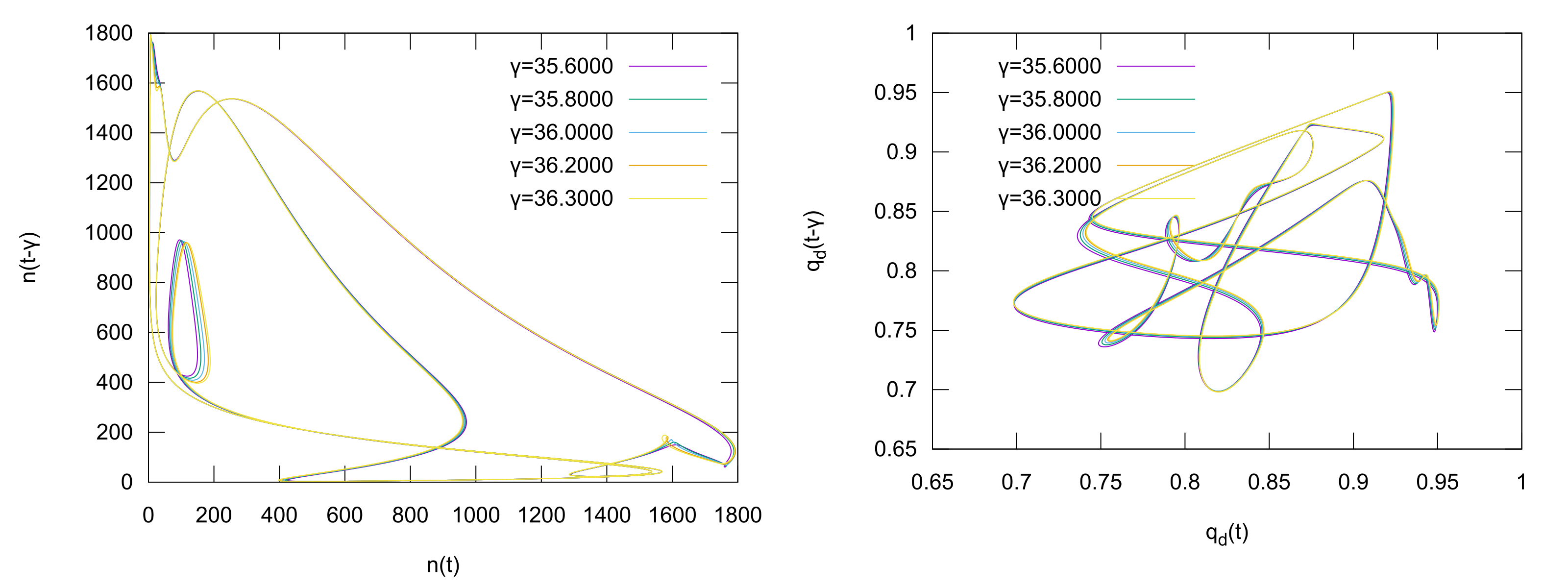}
}
\caption{ The crossing and folds in periodic orbits (regions~4 and 7 in in
Fig~.\protect\ref{fig:bifdiag}). There is no apparent change in the dynamics
and those are kind of artefacts that are caused by using the extrema
bifurcation diagrams. The corresponding trajectories of $n$ and $q$ are
drawn in the same colours.}
\label{fig:periodic-biff-4-and-7} 
\end{figure}

\begin{figure}[tbp] 
\center{
\includegraphics[height=4cm]{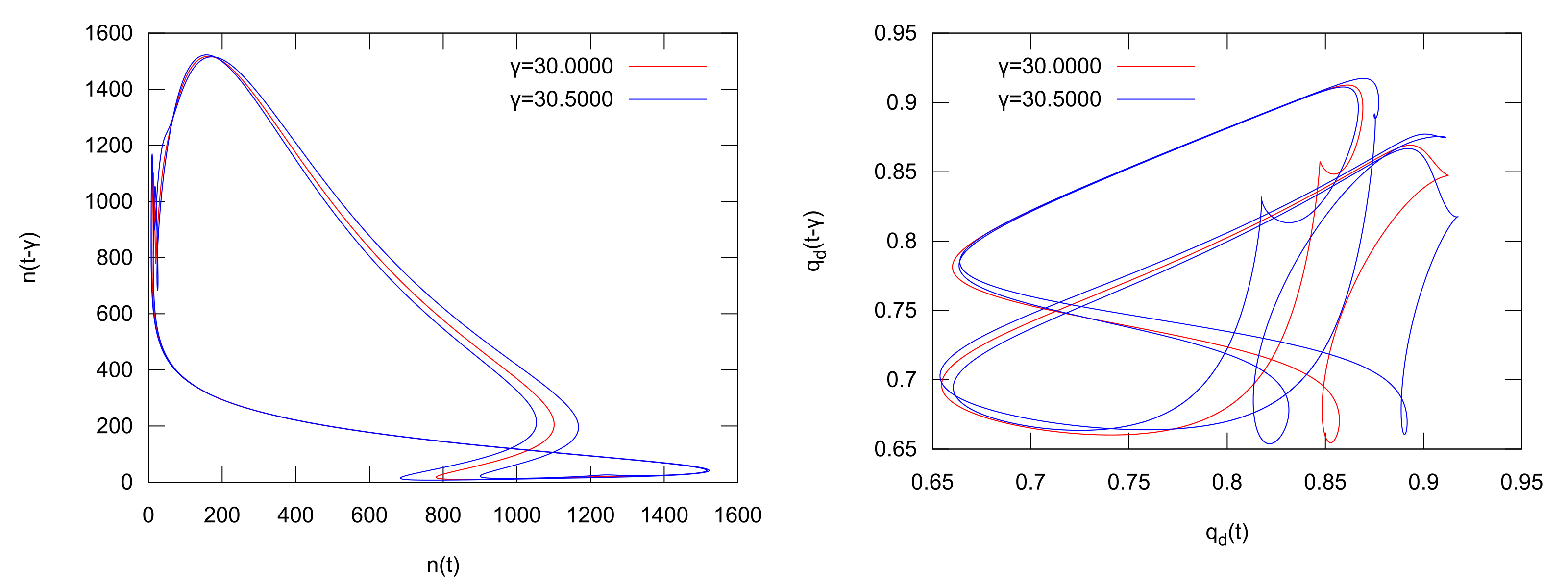}
\includegraphics[height=4cm]{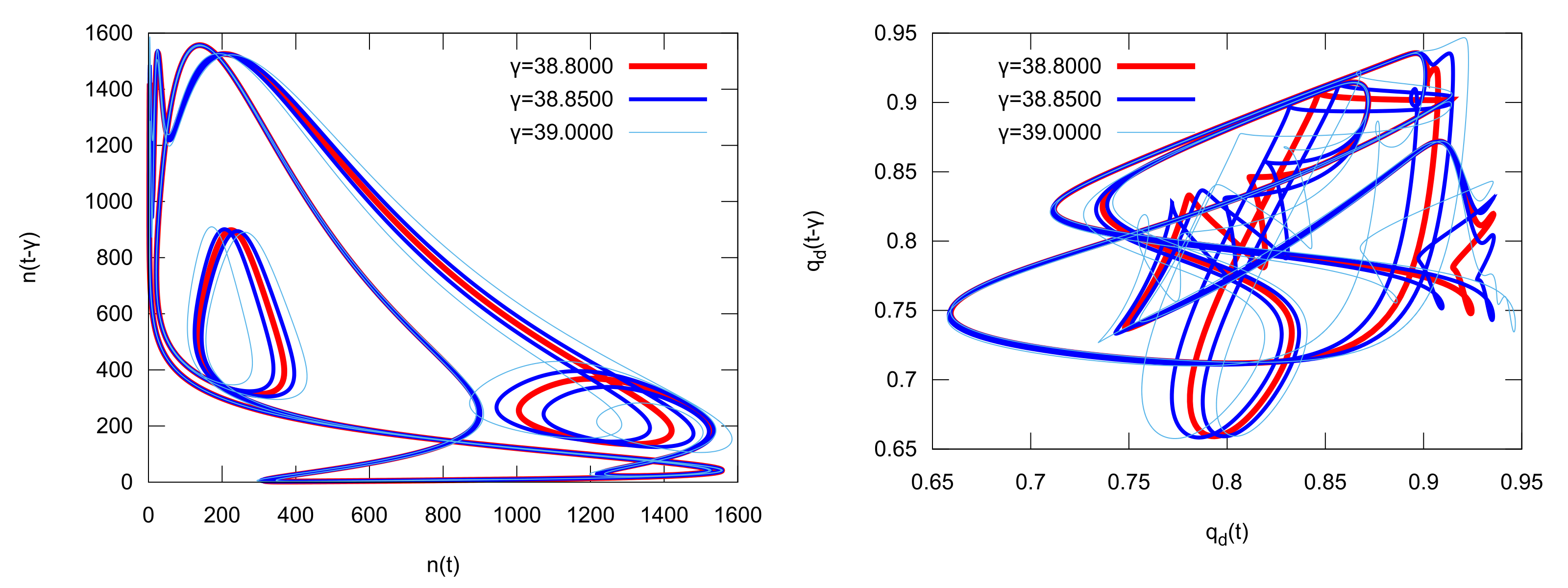}
}
\caption{ Some examples of a period doubling bifurcation (region~6 and 8 in
Fig~.\protect\ref{fig:bifdiag}). The trajectory before the period doubling
is drawn in red, while the trajectory after the bifurcation is drawn in
blue. The corresponding trajectories of $n$ and $q$ are drawn in the same
colours.}
\label{fig:periodic-biff-6-and-8}
\end{figure}

\begin{figure}[tbp]
\center{
\includegraphics[height=4cm]{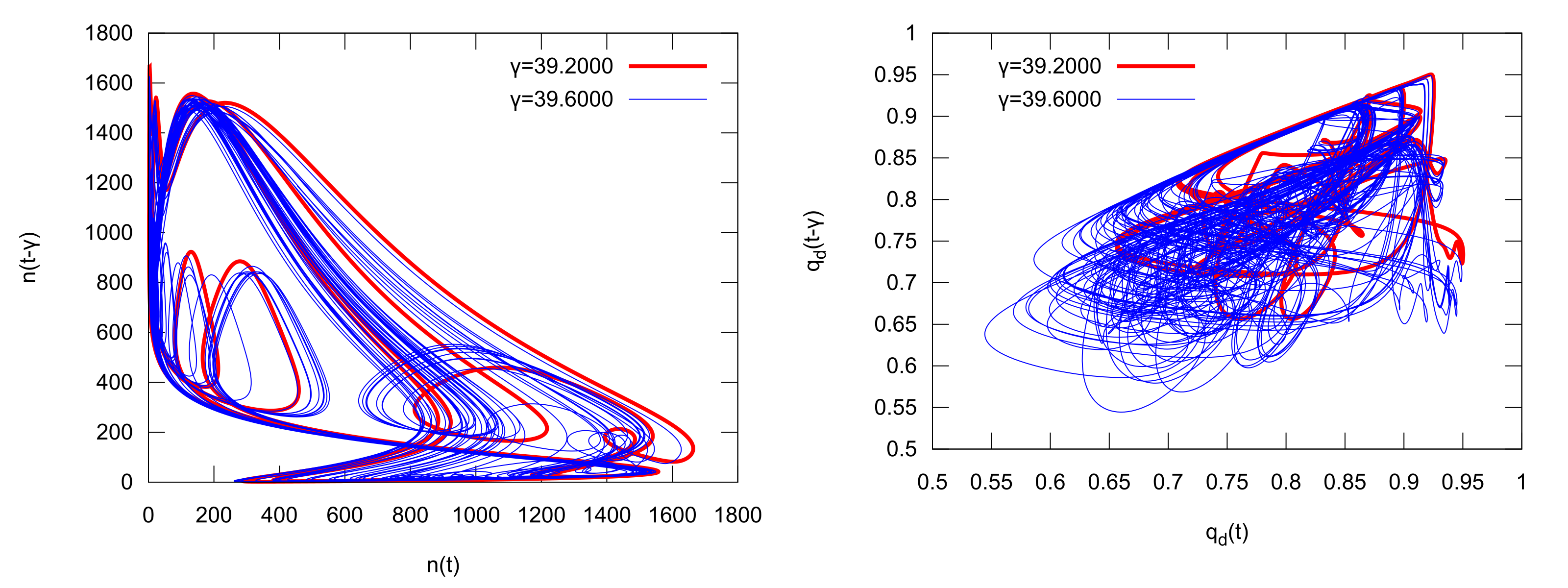}}
\caption{ The onset of chaos (region~9 in Fig~.\protect\ref{fig:bifdiag}).
The trajectory prior to entering the chaotic regime is shown in red. It
appears complex, but still resembles a closed curve. The chaotic attractor,
shown in blue, is similar to known DDE attractors from the literature. The
corresponding trajectories of $n$ and $q$ are drawn in the same colours. }
\label{fig:periodic-biff-9}
\end{figure}

\begin{figure}[tbp]
\center{
\includegraphics[height=4cm]{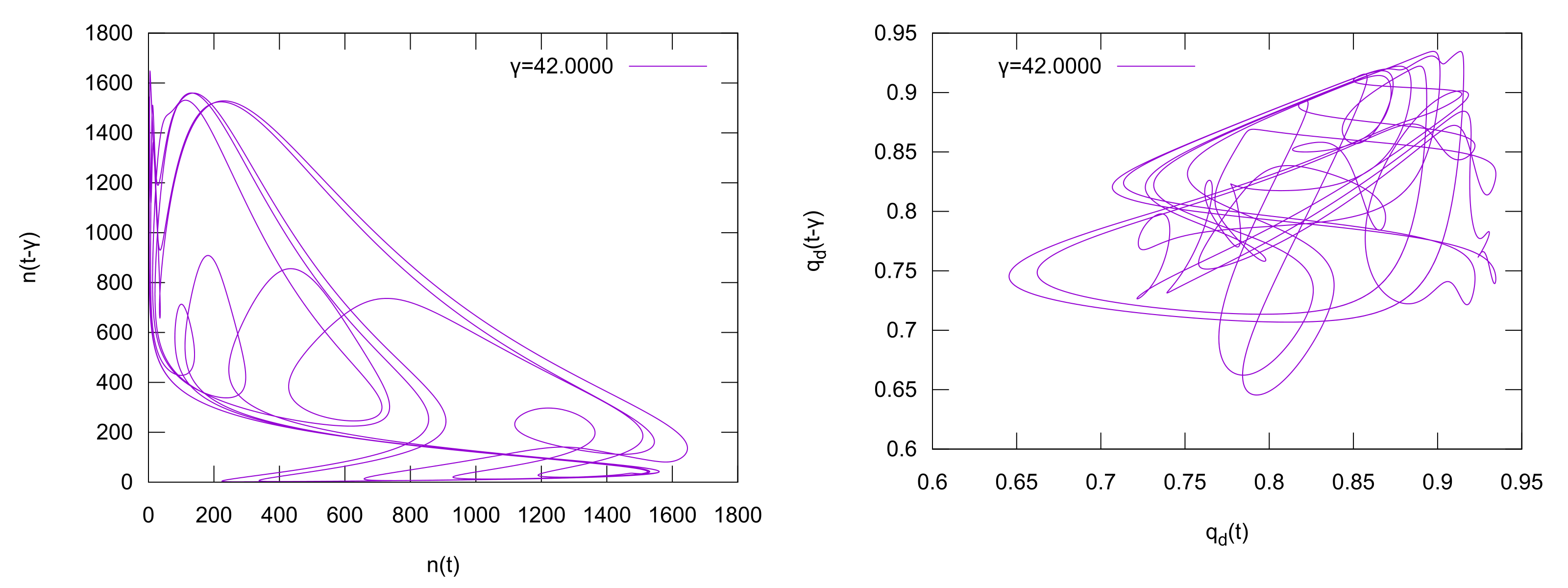}}
\caption{ A periodic window (region~10 in Fig~.\protect\ref{fig:bifdiag}).
The orbit, while complicated, looks like a closed curve. The corresponding
trajectories of $n$ and $q$ are drawn in the same colours. }
\label{fig:periodic-biff-10}
\end{figure}

\begin{figure}[tbp]
\center{
\includegraphics[height=4cm]{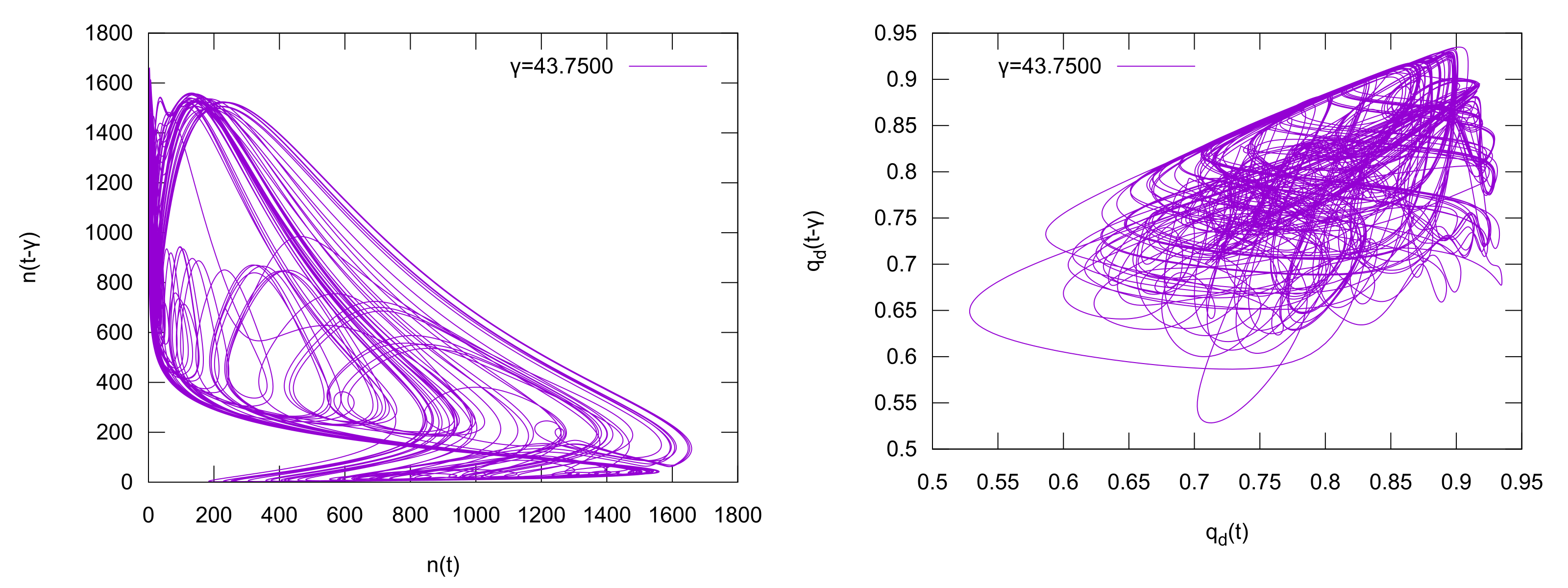}
\includegraphics[height=4cm]{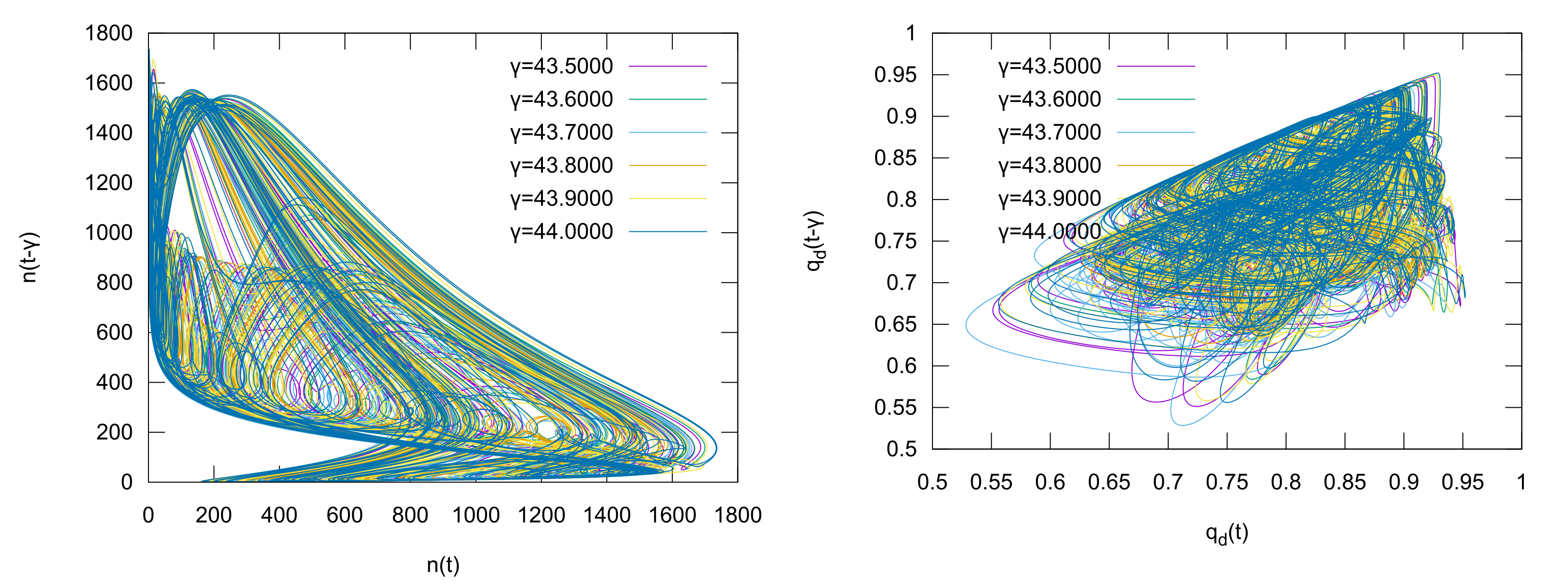}
}
\caption{ A chaotic attractor (region~11 in Fig~.\protect\ref{fig:bifdiag}).
In the top panels we have an apparent chaotic attractor for one value of
parameter $\protect\gamma $, while in the bottom panels we see the
superposition of attractors for a range of parameters. It seems that the
attractor is changing its shape significantly on this short interval of the
parameter. The corresponding trajectories of $n$ and $q$ are drawn in the
same colours.}
\label{fig:periodic-biff-11} 
\end{figure}

As we can see, for certain parameter values the equilibrium point may be not affected by delays, while for another it may loose stability in effect of bifurcation. Thus the question arises, what happens for even longer delays than the critical bifurcation value? We can analyze the bifurcation structure for the interesting case when the initial bifurcation avoids the extinction of the Dove strategy. The exreme values $n_{i}^{\gamma }=n^{\gamma
}(t_{i})$, indicating bifurcations, are plotted in Figure \ref{fig:bifdiag}  against the corresponding $
\gamma $. We also present the corresponding values $q_{d,i}^{\gamma
}=q_{d}^{\gamma }(t_{i})$ in the lower panel of Fig.~\ref{fig:bifdiag}. The
system parameters, time values $S$, $T$, and the number of points $N$ are
provided in the accompanying source codes \cite{SzczelinaCode}.
Interested readers may use those codes to recompute the bifurcation diagrams
presented in this paper. In what follows, we present the orbits in the
standard $(u(t),u(t-\gamma ))$ projection in Figures~\ref
{fig:periodic-biff-1}--\ref{fig:periodic-biff-11}, which is suitable for
visualizing periodic motion, as periodic solutions appear in these
coordinates as closed loops. Several interesting events occur on the route
to chaos in this system. The first Hopf bifurcation takes place in region~1,
for $\gamma \in \lbrack 2.3,2.7]$ (exact value computed according to Theorem 1 is $\gamma=2.5147879$). The steady state loses stability and a
periodic orbit in $n(t)$ emerges, see Fig.~\ref{fig:periodic-biff-1}. Then,
for some $\gamma \in \lbrack 4.260,4.265]$, there is an abrupt change in
dynamics, resulting in the emergence of a closed cycle in $q_{d}$ and a
shift in the trajectory of $n$. A more detailed analysis of this bifurcation
is necessary but lies outside the scope of the present paper. For larger $
\gamma $, the picture becomes more complex; many of the appearing,
disappearing, and intersecting curves are caused by \emph{kinks} and/or are
artefacts of the extrema-based bifurcation diagram, as discussed earlier;
see Figures~\ref{fig:periodic-biff-3-and-5} and \ref%
{fig:periodic-biff-4-and-7}. Period-doubling bifurcations are also observed,
as shown in Fig.~\ref{fig:periodic-biff-6-and-8}. These eventually lead to
the onset of chaos for values of $\gamma \in \lbrack 39.2,39.6]$. The
stability regions, such as region~10, feature complex attracting periodic
orbits, typical in systems with unimodal feedback (see, e.g., \cite
{Benedek}). This suggests the presence of many additional
structures, such as an infinite number of unstable periodic orbits (see,
e.g., \cite{Gierzkiewicz}). Finally, in Fig.~\ref
{fig:periodic-biff-11}, we show data from the chaotic region~11, where the
dynamics closely resembles those observed in classical equations such as the
Mackey-Glass or Lasota-Ważewska systems.

\bigskip

\subsection{Long transients are even longer under delay}

\begin{figure}[tbp]

	\includegraphics[height=10cm]{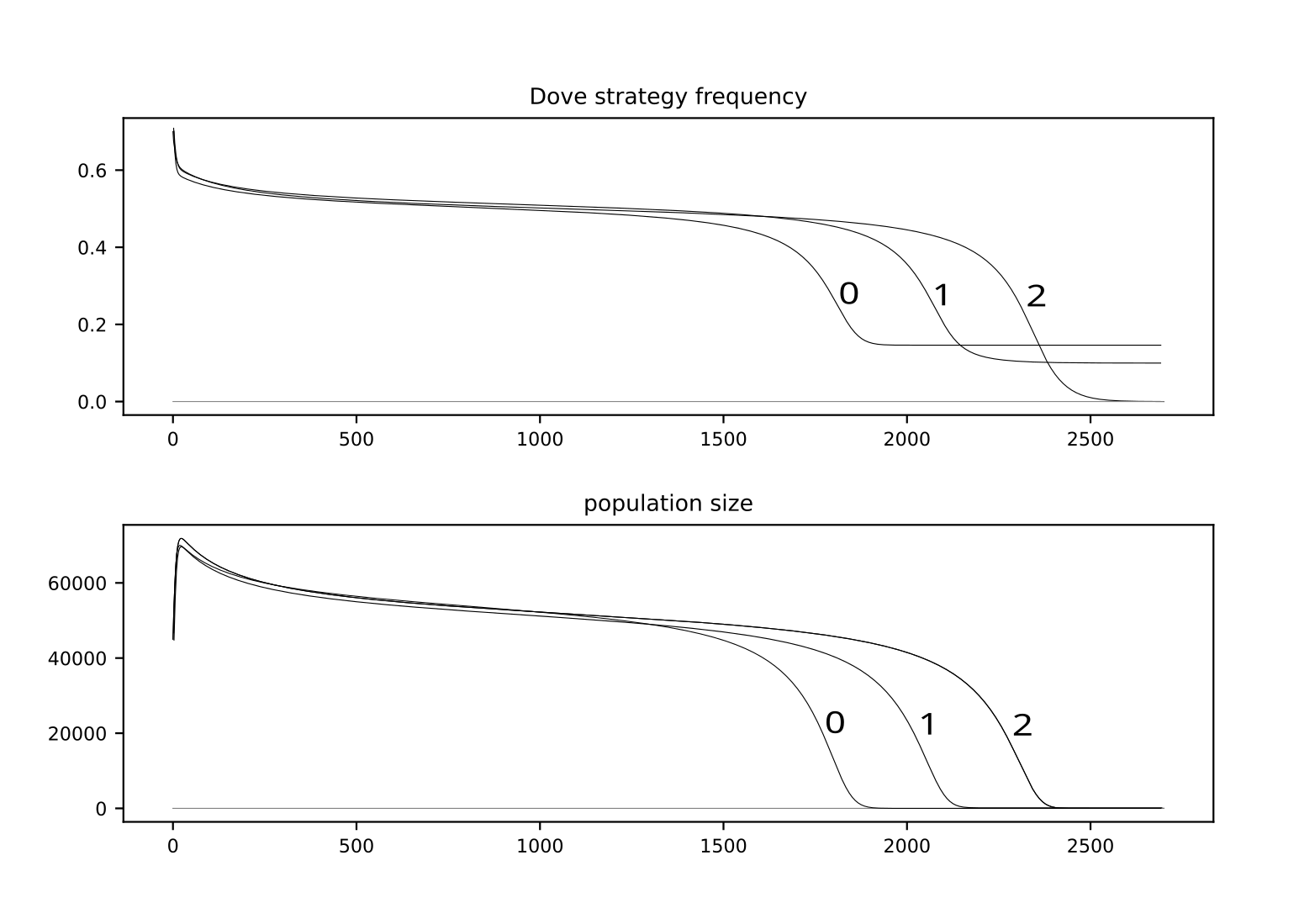}
	\caption{Trajectories od the model generating the ghost attractor and a long transient behavior (parameters: $F=0.85$, $d=1$, $\Phi=0$ and $\Psi=0.1253$). Trajectories are generated for the undelayed system ($\gamma=0$), and two cases with short delays ($\gamma=1$ and $\gamma=2$). We observe that the classical rest point $q_{d}=0.15$, $n=0$ resulting from the intersection of the frequency nullcline with the trivial density nullcline $n=0$ looses stability and shifts towards $q_{d}=0$ with the increase of the delay. In addition we observe the significant increase of the duration of the long transient phase. }
		\label{fig:LTshort} 
\end{figure}

\begin{figure}[tbp]

	\includegraphics[height=10cm]{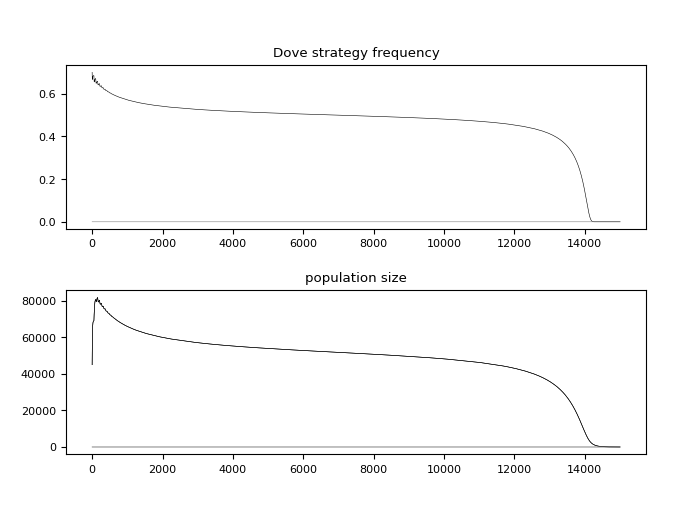}
	\caption{The exmple with the same parameters as in the Figure \ref{fig:LTshort}, but with very long delay $\gamma=50$. We observe that the duration of the long transient dramatically increased. Thus delays stabilize the ghost attractors. }
		\label{fig:LTlong} 
\end{figure}

In the previous section we observed that the classical rest points under juvenile recruitment with delay  can be highly sensitive to the impact of the delay. In this section we examine the performance of "ghost attractors" responsible for the long transient behavior. "Ghost attractors" emerge after a saddle-node bifurcation, when the nullcline intersections collide and disappear. Then nullclines disconnect and form a narrow channel, still attracting the trajectories and catching them for some time. In our Hawk-Dove model this happens when $\Delta$ is negative but still very close to zero. We observe the paradoxical effect that the classical rest points are very fragile with respect to the impact of delays while ghost attractors (objects that do not exist from the point of view of the classical theory) are getting stronger and their duration increases with the delay. This is depicted in Figure \ref{fig:LTshort} for the short delays and in Figure \ref{fig:LTlong} for the very long delays.

\FloatBarrier

\section{Discussion}

In this paper we introduced the delays resulting from maturation time of the
juveniles into the eco-evolutionary demographic replicator dynamics \cite{argbr1,argbr2,argbr3}.
 It was shown that the dynamics with delayed fertility rewards
produces complex trajectories and is very sensitive to the external
factors, even in the simplest case without juvenile recruitment mortality.
In general we tested three types of juvenile recruitment: neutral
recruitment (juvenile survival equals 1 and the growth is regulated by mortality of adult individuals) and two types of density dependent
juvenile survival functions: classical logistic term depending on carrying
capacity (interpreted as maximal population load) and suppression term with
delay, which should be rather interpreted as the fraction of produced eggs
(while fertility payoff describes the maximal potential reproductive
output). In addition the model was tested for the response of seasonal
mortality, explicit predator presure (modelled by simple Lotka-Volterra
system) and combination of both factors. The resulting trajectories, instead of
gradual convergence to the equilibria, as in the classical models, show
complicated oscillations before reaching the restpoint. Then the system,
similarly to the classical ODE models gradually converges to the stable
equilibrium, in area limited by the nullclines. The most extreme phenomena are
observed with juvenile survival with the delay. Then we can observe extreme
fluctuations, mostly in the population size which can literally oscillate
between $3$ and $11000$. We can observe oscillations in the strategy
frequencies too.\ When we add seasonal mortality those frequency
oscillations can take the most extreme form, running from nearly $0$ to
nearly $1$, which is quite unusual for the Hawk-Dove game. In addition, combination of the juvenile suppression with delay and the long maturation time can overturn the strictly game theoretic predictions and prevent the extinction of Doves. Then we observe the extremely complex cyclic or even chaotic dynamics.  

Maturation time is very important life-history parameter and the results
from this paper show that aspects related to life history theory \cite{Stearns,Roff},
 which is based on demography \cite{Metcalf1,Metcalf2}, can seriously affect the trajectories of population dynamics.
Therefore, the investigations of the relationships between those fields
should be crucial. In the future research we can relax the assumption of
uniform delays. Different behavioural types may need different phenotypes,
which may need different maturation sizes due to for example different body
sizes. In the next step, more detailed models with explicit age or stage
structure \cite{argbr4age,argRudState} can be
used for confirmation of the predictions obtained in this paper. This can be
important from the point of view of the newly emerging eco-evolutionary
synthesis \cite{Post,Pelletier,Hanski,Hendry1,Hendry2}.

\bigskip 

\subsection{Appendix 1}

Recall the fertility payoffs are 
\begin{eqnarray*}
V_{H} &=&(1-q_{d})0.5F+q_{d}F=\left( 1+q_{d}\right) 0.5F \\
V_{D} &=&q_{d}0.5F+(1-q_{d})0=q_{d}0.5F \\
\bar{V} &=&(1-q_{d})V_{H}+q_{d}V_{D}=(1-q_{d})\left( 1+q_{d}\right)
0.5F+q_{d}^{2}0.5F \\
&=&\left[ (1-q_{d})\left( 1+q_{d}\right) +q_{d}^{2}\right] 0.5F=0.5F
\end{eqnarray*}%
\ Mortality payoffs are 
\begin{eqnarray*}
D_{H} &=&\left( 1-q_{d}\right) 0.5d\text{ \ and }\ D_{D}=0 \\
\bar{D} &=&\left( 1-q_{d}\right) D_{H}+q_{d}D_{D}=\left( 1-q_{d}\right)
^{2}0.5d
\end{eqnarray*}

Bracketed terms are $\left( V_{D}-\bar{V}\right) =-\left( 1-q_{d}\right)
0.5F $ and $\left( D_{D}-\bar{D}\right) =-\left( 1-q_{d}\right) ^{2}0.5d$
leading to the system%
\begin{eqnarray}
\frac{dq_{d}}{dt} &=&0.5q_{d}(t)\left( 1-q_{d}(t)\right) \left( \left(
1-q_{d}(t)\right) d-\left( 1-\frac{n(t)}{K}\right) F\right) \\
\frac{dn}{dt} &=&n(t)\left( \left( 0.5F+\Phi \right) \left( 1-\frac{n(t)}{K}
\right) -\left[ \left( 1-q_{d}(t)\right) ^{2}0.5d+\Psi \right] \right) .
\end{eqnarray}

\textbf{\ Calculation of the nullclines}: Nullclines are solutions of zeros
of the r.h.s. of the equations:

\begin{equation}
\left( 1-q_{d}\right) d-\left( 1-\frac{n}{K}\right) F=0\text{ \ \ }%
\Longrightarrow \text{ \ \ }q_{d}=1-\left( 1-\frac{n}{K}\right) \frac{F}{d}.
\end{equation}

Nullcline for the population size will be 
\begin{gather}
\left( 0.5F+\Phi \right) \left( 1-\frac{n}{K}\right) -\left[ \left(
1-q_{d}\right) ^{2}0.5d+\Psi \right] =0 \\
1-\frac{n}{K}=\frac{\left( 1-q_{d}\right) ^{2}0.5d+\Psi }{0.5F+\Phi }
\label{logisticterm} \\
n(t)=\left[ 1-\frac{\left( 1-q_{d}\right) ^{2}0.5d+\Psi }{0.5F+\Phi }\right]
K.
\end{gather}
\bigskip

\textbf{\ Calculation of the rest points}: The rest points will be
intersections of both nullclines.

\begin{eqnarray*}
q_{d} &=&1-\left( \frac{\left( 1-q_{d}\right) ^{2}0.5d+\Psi }{0.5F+\Phi }
\right) \frac{F}{d} \\
q_{d} &=&1-\frac{\left( 1-q_{d}\right) ^{2}0.5+\frac{\Psi }{d}}{\left( 0.5+
\frac{\Phi }{F}\right) } \\
\left( 1-q_{d}\right) \left( 0.5+\frac{\Phi }{F}\right) &=&\left(
1-q_{d}\right) ^{2}0.5+\frac{\Psi }{d} \\
\left( 0.5+\frac{\Phi }{F}\right) -q_{d}\left( 0.5+\frac{\Phi }{F}\right)
&=&\left( 1-q_{d}\right) ^{2}0.5+\frac{\Psi }{d} \\
\left( 0.5+\frac{\Phi }{F}\right) -q_{d}\left( 0.5+\frac{\Phi }{F}\right)
&=&\left( 1-2q_{d}+q_{d}^{2}\right) 0.5+\frac{\Psi }{d} \\
\left( 0.5+\frac{\Phi }{F}\right) -q_{d}\left( 0.5+\frac{\Phi }{F}\right)
&=&0.5-q_{d}+0.5q_{d}^{2}+\frac{\Psi }{d} \\
0.5q_{d}^{2}+\left( \frac{\Phi }{F}-0.5\right) q_{d}&-&\frac{\Phi }{F}+
\frac{\Psi }{d}=0
\end{eqnarray*}
\bigskip

Let us solve this equation. The $\Delta $ will provide the condition of
existence of the nullcline intersections

\begin{eqnarray*}
\Delta &=&\left( \frac{\Phi }{F}-0.5\right) ^{2}-2\left( \frac{\Psi }{d}-
\frac{\Phi }{F}\right) =\frac{\Phi }{F}^{2}-\frac{\Phi }{F}+0.25-2\frac{\Psi 
}{d}+2\frac{\Phi }{F}=\frac{\Phi }{F}^{2}+\frac{\Phi }{F}-2\frac{\Psi }{d}+0.25 \\
&=&\left( \frac{\Phi }{F}+0.5\right) ^{2}-2\frac{\Psi }{d},
\end{eqnarray*}

Then for $\Delta >0$ we have two rest point frequencies: 
\begin{eqnarray*}
\check{q}_{d} &=&0.5-\frac{\Phi }{F}-\sqrt{\left( \dfrac{\Phi }{F}\right)
^{2}+\dfrac{\Phi }{F}-2\dfrac{\Psi }{d}+0.25} \\
\hat{q}_{d} &=&0.5-\frac{\Phi }{F}+\sqrt{\left( \dfrac{\Phi }{F}\right) ^{2}+
\dfrac{\Phi }{F}-2\dfrac{\Psi }{d}+0.25}
\end{eqnarray*}
Let's now substitute those values into the density nullcline and calculate
the respective population sizes
\begin{eqnarray*}
\left( 1-q_{d}\right) ^{2} &=&\left( 0.5+\frac{\Phi }{f}\pm \sqrt{\Delta }
\right) ^{2} \\
&=&\left( 0.5+\frac{\Phi }{f}\right) ^{2}\pm \left( 1+2\frac{\Phi }{f}
\right) \sqrt{\Delta }+\Delta  \\
&=&\left( 0.5+\frac{\Phi }{f}\right) ^{2}\pm \left( 1+2\frac{\Phi }{f}
\right) \sqrt{\Delta }+\left( \dfrac{\Phi }{f}+0.5\right) ^{2}-2\dfrac{\Psi 
}{d} \\
&=&2\left( 0.5+\frac{\Phi }{f}\right) ^{2}\pm \left( 1+2\frac{\Phi }{f}
\right) \sqrt{\Delta }-2\dfrac{\Psi }{d} \\
&=&2\left[ \left( 0.5+\frac{\Phi }{f}\right) ^{2}\pm \left( 0.5+\frac{\Phi }{f}
\right) \sqrt{\Delta }-\dfrac{\Psi }{d} \right]
\end{eqnarray*}
\bigskip 

then the juvenile survival term is
\begin{eqnarray*}
\mathbf{D}(n) &=&\dfrac{\left( 1-q_{d}\right) ^{2}0.5d+\Psi }{0.5f+\Phi } \\
&=&\frac{d}{f}\dfrac{\left( 1-q_{d}\right) ^{2}0.5+\frac{\Psi }{d}}{0.5+
\frac{\Phi }{f}} \\
&=&\frac{d}{f}\dfrac{\left( 0.5+\frac{\Phi }{f}\right) ^{2}\pm \left( 0.5+
\frac{\Phi }{f}\right) \sqrt{\Delta }}{0.5+\frac{\Phi }{f}} \\
&=&\frac{d}{f}\left( 0.5+\frac{\Phi }{f}\right) \pm \sqrt{\Delta }
\end{eqnarray*}
For the logistic suppression $\mathbf{D}(n)=\left(1-\frac{n}{K}\right)$ the density in the rest points will be%
\begin{equation*}
n=\left[ 1-\frac{d}{f}\left( 0.5+\frac{\Phi }{f}\right) \mp \sqrt{\Delta }
\right] K
\end{equation*}
\bigskip 

Now we can infer the stability of the rest points. According to the Lemma 2
in \cite{argbr3} when the derivative with respect to $n$ of the r.h.s. of
the frequency equation (\ref{repq}) is positive (in our case it equals to $
0.5q_{d}\left( 1-q_{d}\right) F/K>0$), then the slopes along the $q_{d}$
axis of the frequency ($S_{q}$) and density ($S_{n}$) nullclines should
satisfy condition $S_{n}<S_{q}$. Nullcline derivatives are $\dfrac{dq_{d}(n)
}{dn}=\dfrac{F}{dK}$ (then the slope along $q_{d}$ axis is $S_{q}=\dfrac{dK}{
F}$ ) and $S_{n}=\dfrac{d\tilde{n}(q_{d})}{dq_{d}}=\dfrac{\left(
1-q_{d}\right) }{0.5F+\Phi }dK$ \ Then the condition $S_{n}<S_{q}$ is

\begin{equation*}
\dfrac{\left( 1-q_{d}\right) }{0.5F+\Phi }dK<\dfrac{dK}{F}\Longrightarrow
\left( 1-q_{d}\right) <0.5+\dfrac{\Phi }{F}\Longrightarrow q_{d}>0.5-\dfrac{
\Phi }{F}.
\end{equation*}

Therefore, intersection $q_{d}<0.5-\dfrac{\Phi }{F}$ is a saddle point while
intersection $q_{d}>0.5-\dfrac{\Phi }{F}$ is stable attractor. The condition
on nullcline slopes is clearly visible on numerical solutions.\bigskip

\subsection{Appendix 2\protect\bigskip}

Per capita fertility rate is

\begin{eqnarray*}
\frac{n_{1}(t-\gamma )}{n_{1}(t)}e_{1}Vq^{T}(t-\gamma ) &=&\frac{n(t-\gamma )
\dfrac{n_{1}(t-\gamma )}{n(t-\gamma )}}{n(t)\dfrac{n_{1}(t)}{n(t)}}%
e_{1}Vq^{T}(t-\gamma ) \\
&=&\frac{n(t-\gamma )q_{1}(t-\gamma )}{n(t)q_{1}(t)}e_{1}Vq^{T}(t-\gamma ).
\end{eqnarray*}

the average fertility rate is 
\begin{equation*}
\sum_{i}q_{i}(t)\frac{n(t-\gamma )q_{i}(t-\gamma )}{n(t)q_{i}(t)}%
e_{i}Vq^{T}(t-\gamma )=\frac{n(t-\gamma )}{n(t)}\sum_{i}q_{i}(t-\gamma
)e_{i}Vq^{T}(t-\gamma )
\end{equation*}
In effect the fertility bracketed term will be%
\begin{eqnarray*}
&&q_{1}(t)\left[ \frac{n(t-\gamma )q_{1}(t-\gamma )}{n(t)q_{1}(t)}%
e_{1}Vq^{T}(t-\gamma )-\frac{n(t-\gamma )}{n(t)}\sum_{i}q_{i}(t-\gamma
)e_{i}Vq^{T}(t-\gamma )\right] \\
&=&\frac{n(t-\gamma )}{n(t)}\left[ q_{1}(t-\gamma )e_{1}Vq^{T}(t-\gamma
)-q_{1}(t)\sum_{i}q_{i}(t-\gamma )e_{i}Vq^{T}(t-\gamma )\right]
\end{eqnarray*}
Note that the above dynamics is well defined and cannot escape the unit
interval. For $q_{1}(t)=0$ fertility bracket equals $q_{1}(t-\gamma
)V_{1}(t-\gamma )\geq 0$ while for $q_{1}(t)=1$ it is $\left[ q_{1}(t-\gamma
)V_{1}(t-\gamma )-\bar{V}(t-\gamma )\right] $\bigskip 
\begin{eqnarray*}
&&\left[ q_{1}(t-\gamma )V_{1}(t-\gamma )-\bar{V}(t-\gamma )\right] \\
&=&q_{1}(t-\gamma )V_{1}(t-\gamma )-q_{1}(t-\gamma )V_{1}(t-\gamma
)-\sum_{i\neq 1}q_{i}(t-\gamma )V_{i}(t-\gamma ) \\
&=&-\sum_{i\neq 1}q_{i}(t-\gamma )V_{i}(t-\gamma )\leq 0
\end{eqnarray*}

For two competing strategies the fertility bracketed term reduces to

\begin{eqnarray*}
&&q_{1}(t-\gamma )V_{1}(t-\gamma )-q_{1}(t)\bar{V}(t-\gamma ) \\
&=&q_{1}(t-\gamma )V_{1}(t-\gamma )-q_{1}(t)\left[ q_{1}(t-\gamma
)V_{1}(t-\gamma )+(1-q_{1}(t-\gamma ))V_{2}(t-\gamma ))\right] \\
&=&q_{1}(t-\gamma )V_{1}(t-\gamma )-q_{1}(t)q_{1}(t-\gamma )V_{1}(t-\gamma
)-q_{1}(t)(1-q_{1}(t-\gamma ))V_{2}(t-\gamma )) \\
&=&\left[ 1-q_{1}(t)\right] q_{1}(t-\gamma )V_{1}(t-\gamma
)-q_{1}(t)(1-q_{1}(t-\gamma ))V_{2}(t-\gamma )
\end{eqnarray*}

\subsection{Appendix 3\protect\bigskip}

\begin{proof}
Recall that the point $(n^{\ast },n_{1}^{\ast })$ is a stable point for the
system without delay ($\gamma =0$). Now we looking for $\gamma >0$ when the
system (\ref{eq-no})--(\ref{eq-n1o}) loses its stability. The linearized
system is of the form 
\begin{eqnarray}
&&\textbf{x}^{\prime }(t)=b_{11}\textbf{x}(t)+b_{12}\textbf{x}(t-\gamma
)+b_{13}\textbf{y}(t)+b_{14}\textbf{y}(t-\gamma ),  \label{eq-lin-n} \\
&&\textbf{y}^{\prime }(t)=b_{21}\textbf{x}(t)+b_{22}\textbf{x}(t-\gamma
)+b_{23}\textbf{y}(t)+b_{24}\textbf{y}(t-\gamma ).  \label{eq-lin-n1}
\end{eqnarray}
Next we substitute to this system $\textbf{x}(t)=e^{cti}$ and $\textbf{y}(t)=\beta e^{cti}$
and obtain a system of algebraic equations for $c$, $\beta $ and $\gamma $: 
\begin{eqnarray}
&&ci=b_{11}+b_{12}e^{-c\gamma i}+b_{13}\beta +b_{14}\beta e^{-c\gamma i},
\label{eq-lin-n-alg} \\
&&\beta ci=b_{21}+b_{22}e^{-c\gamma i}+b_{23}\beta +b_{24}\beta e^{-c\gamma
i}.  \label{eq-lin-n1-alg}
\end{eqnarray}
The bifurcation point, is the minimum value of $\gamma $ for which there is
a real number $c$ and a complex number $\beta $ satisfying system (\ref%
{eq-lin-n-alg})--(\ref{eq-lin-n1-alg}). Deriving $\beta $ from equations (
\ref{eq-lin-n-alg})--(\ref{eq-lin-n1-alg}), we get 
\begin{equation}
\frac{-ci+b_{11}+b_{12}e^{-c\gamma i}}{b_{13}+b_{14}e^{-c\gamma i}}=\frac{
b_{21}+b_{22}e^{-c\gamma i}}{-ci+b_{23}+b_{24}e^{-c\gamma i}}.
\label{eq-xgamma}
\end{equation}%
Let $z=e^{-c\gamma i}$. Then we have 
\begin{gather}
\frac{-ci+b_{11}+b_{12}z}{b_{13}+b_{14}z}=\frac{b_{21}+b_{22}z}{
-ci+b_{23}+b_{24}z}, \\
\left( b_{11}+b_{12}z-ci\right) \left( b_{23}+b_{24}z-ci\right) =\left(
b_{21}+b_{22}z\right) \left( b_{13}+b_{14}z\right) \\
b_{12}b_{24}z^{2}+\left( b_{11}b_{24}+b_{23}b_{12}\right) z-\left(
b_{12}+b_{24}\right) ciz-\left( b_{23}+b_{11}\right) ci+b_{11}b_{23}-c^{2}=
\\
b_{22}b_{14}z^{2}+\left( b_{14}b_{21}+b_{22}b_{13}\right) z+b_{21}b_{13}
\end{gather}
\ This leads to the quadratic equation on $z$
\begin{equation*}
\left( b_{12}b_{24}-b_{22}b_{14}\right) z^{2}+\left(
b_{11}b_{24}+b_{23}b_{12}-b_{14}b_{21}-b_{22}b_{13}-\left(
b_{12}+b_{24}\right) ci\right) z+b_{11}b_{23}-b_{21}b_{13}-\left(
b_{23}+b_{11}\right) ci-c^{2}=0,
\end{equation*}
which can be presented in the form: 
\begin{equation}
Az^{2}+\tilde{B}z+\tilde{C}=0,  \label{eq-xgamma2}
\end{equation}
where 
\begin{eqnarray*}
&&A=b_{12}b_{24}-b_{22}b_{14}, \\
&&\tilde{B}%
=b_{11}b_{24}+b_{12}b_{23}-b_{14}b_{21}-b_{22}b_{13}-(b_{12}+b_{24})ci, \\
&&\tilde{C}=b_{11}b_{23}-b_{21}b_{13}-(b_{11}+b_{23})ci-c^{2}
\end{eqnarray*}

Then $z$ is the solution of the above equation. Since $z\bar{z}=1$, we
obtain from Eq. (\ref{eq-xgamma2}) that 
\begin{equation}
Az+\tilde{B}+\tilde{C}\bar{z}=0.  \label{eq-xgamma3}
\end{equation}
Let $z=X+Yi$, where $X,Y$ are real numbers. Since $\func{Im}(\tilde{B}
)=-(b_{12}+b_{24})c$ and $\func{Im}(\tilde{C})=-(b_{11}+b_{23})c$ we have 

\begin{equation*}
Az=AX+AYi\text{ \ \ and }\tilde{C}\bar{z}=\Re (\tilde{C})X+\func{Im}(\tilde{C})Y
+\left( \func{Im}(\tilde{C})X-\Re (\tilde{C})Y\right)i
\end{equation*}

from Eq. (\ref{eq-xgamma3}) we obtain

\begin{eqnarray}
(A+\Re (\tilde{C}))X+\func{Im}(\tilde{C})Y &=&-\Re (\tilde{B}),
\label{system1a} \\
\func{Im}(\tilde{C})X+(A-\Re (\tilde{C}))Y &=&-\func{Im}(\tilde{B}),
\label{system1b}
\end{eqnarray}

The parts independent of $c$ of $A$, $\tilde{B}$ and $\tilde{C}$ are:

\begin{eqnarray*}
&&A=b_{12}b_{24}-b_{22}b_{14}, \\
&&B=b_{11}b_{24}+b_{12}b_{23}-b_{13}b_{22}-b_{14}b_{21}, \\
&&C=b_{11}b_{23}-b_{13}b_{21},
\end{eqnarray*}

thus $\Re (\tilde{B})=B$ and $\Re (\tilde{C})=C-c^{2}$. Then system (\ref{system1a},\ref{system1b}) is

\begin{eqnarray*}
(A+\left[ C-c^{2}\right] )X-(b_{11}+b_{23})cY &=&-B, \\
-(b_{11}+b_{23})cX+(A-\left[ C-c^{2}\right] )Y &=&(b_{12}+b_{24})c,
\end{eqnarray*}

then from the Cramer formulae we obtain:

\begin{eqnarray*}
W &=&\left\vert 
\begin{array}{cc}
A+\left[ C-c^{2}\right] & -(b_{11}+b_{23})c \\ 
-(b_{11}+b_{23})c & A-\left[ C-c^{2}\right]
\end{array}%
\right\vert =A^{2}-\left[ C-c^{2}\right] ^{2}-(b_{11}+b_{23})^{2}c^{2} \\
W_{X} &=&\left\vert 
\begin{array}{cc}
-B & -(b_{11}+b_{23})c \\ 
(b_{12}+b_{24})c & A-\left[ C-c^{2}\right]
\end{array}
\right\vert =-B(A-\left[ C-c^{2}\right] )+(b_{11}+b_{23})(b_{12}+b_{24})c^{2}
\\
W_{Y} &=&\left\vert 
\begin{array}{cc}
(A+\Re (C)) & -\Re (B) \\ 
-(b_{11}+b_{23})c & (b_{12}+b_{24})c
\end{array}
\right\vert =(A+\left[ C-c^{2}\right] )(b_{12}+b_{24})c-B(b_{11}+b_{23})c
\end{eqnarray*}

then

\begin{equation*}
X=\frac{W_{X}}{W}=\frac{-B(A-\left[ C-c^{2}\right]
)+(b_{11}+b_{23})(b_{12}+b_{24})c^{2}}{A^{2}-\left[ C-c^{2}\right]
^{2}-(b_{11}+b_{23})^{2}c^{2}},
\end{equation*}
\begin{equation*}
Y=\frac{W_{Y}}{W}=\frac{(A+\left[ C-c^{2}\right]
)(b_{12}+b_{24})c-B(b_{11}+b_{23})c}{A^{2}-\left[ C-c^{2}\right]
^{2}-(b_{11}+b_{23})^{2}c^{2}}.
\end{equation*}

Then from Euler's formula we have $z=e^{-c\gamma i}=\cos (-c\gamma )+i\sin
(-c\gamma )$, thus $X^{2}+Y^{2}=1$. This leads to the equation: 

\begin{eqnarray*}
&&\left[ -B(A-C+c^{2})+(b_{11}+b_{23})(b_{12}+b_{24})c^{2}\right] ^{2} \\
&&\hskip50pt+\left[ (A+C-c^{2})(b_{12}+b_{24})c-B(b_{11}+b_{23})c\right] ^{2}
\\
&&\hskip100pt=\left[ (b_{11}+b_{23})^{2}c^{2}+(\left[ C-c^{2}\right]
^{2}-A^{2}\right] ^{2}.
\end{eqnarray*}
then we can expand the respective parts of the above equation. First
bracketed term on the left hand side is:
\begin{eqnarray*}
&&\left[ (b_{11}+b_{23})(b_{12}+b_{24})c^{2}-B(A-C+c^{2})\right] ^{2} \\
&=&\left[ (b_{11}+b_{23})(b_{12}+b_{24})\right] ^{2}c^{4} \\
&&-2(b_{11}+b_{23})(b_{12}+b_{24})c^{2}B(A-C+c^{2}) \\
&&+B^{2}(A-C+c^{2})^{2} \\
&=&\left[ (b_{11}+b_{23})(b_{12}+b_{24})\right] ^{2}c^{4} \\
&&-2(b_{11}+b_{23})(b_{12}+b_{24})(A-C)Bc^{2} \\
&&-2(b_{11}+b_{23})(b_{12}+b_{24})Bc^{4} \\
&&+B^{2}\left( A-C\right) ^{2}+2B^{2}\left( A-C\right) c^{2}+B^{2}c^{4} \\
&=&\left( \left[ (b_{11}+b_{23})(b_{12}+b_{24})\right]
^{2}-2(b_{11}+b_{23})(b_{12}+b_{24})B+B^{2}\right) c^{4} \\
&&+\left( 2B^{2}\left( A-C\right)
-2(b_{11}+b_{23})(b_{12}+b_{24})(A-C)B\right) c^{2} \\
&&+B^{2}\left( A-C\right) ^{2} \\
&=&\left( (b_{11}+b_{23})(b_{12}+b_{24})-B\right) ^{2}c^{4} \\
&&+2B(A-C)\left( B-(b_{11}+b_{23})(b_{12}+b_{24})\right) c^{2} \\
&&+B^{2}\left( A-C\right) ^{2}
\end{eqnarray*}
second bracketed term on the left hand side is
\begin{eqnarray*}
&&\left[ (A+C-c^{2})(b_{12}+b_{24})c-B(b_{11}+b_{23})c\right] ^{2} \\
&=&c^{2}\left[
(A+C-c^{2})^{2}(b_{12}+b_{24})^{2}-2(A+C-c^{2})(b_{12}+b_{24})(b_{11}+b_{23})B+(b_{11}+b_{23})^{2}B^{2}
\right] \\
&=&c^{2}\left[ (\left( A+C\right) ^{2}-2\left( A+C\right)
c^{2}+c^{4})(b_{12}+b_{24})^{2}\right. \\
&&-2(A+C)(b_{12}+b_{24})(b_{11}+b_{23})B+2(b_{12}+b_{24})(b_{11}+b_{23})Bc^{2}
\\
&&\left. +(b_{11}+b_{23})^{2}B^{2}\right] \\
&=&c^{2}\left[ \left( A+C\right) ^{2}(b_{12}+b_{24})^{2}-2\left( A+C\right)
(b_{12}+b_{24})^{2}c^{2}+(b_{12}+b_{24})^{2}c^{4}\right. \\
&&+2(b_{12}+b_{24})(b_{11}+b_{23})Bc^{2} \\
&&\left. +(b_{11}+b_{23})^{2}B^{2}-2(A+C)(b_{12}+b_{24})(b_{11}+b_{23})B 
\right] \\
&=&c^{2}\left[ (b_{12}+b_{24})^{2}c^{4}\right. \\
&&+\left[ 2(b_{12}+b_{24})(b_{11}+b_{23})B-2\left( A+C\right)
(b_{12}+b_{24})^{2}\right] c^{2} \\
&&\left. +\left( A+C\right)
^{2}(b_{12}+b_{24})^{2}-2(A+C)(b_{12}+b_{24})(b_{11}+b_{23})B+(b_{11}+b_{23})^{2}B^{2} 
\right] \\
&=&c^{2}\left[ (b_{12}+b_{24})^{2}c^{4}\right. \\
&&+\left[ 2(b_{12}+b_{24})(b_{11}+b_{23})B-2\left( A+C\right)
(b_{12}+b_{24})^{2}\right] c^{2} \\
&&\left. +\left( \left( A+C\right) (b_{12}+b_{24})-(b_{11}+b_{23})B\right)
^{2}\right] \\
&=&(b_{12}+b_{24})^{2}c^{6} \\
&&+\left[ 2(b_{12}+b_{24})(b_{11}+b_{23})B-2\left( A+C\right)
(b_{12}+b_{24})^{2}\right] c^{4} \\
&&+\left( \left( A+C\right) (b_{12}+b_{24})-(b_{11}+b_{23})B\right) ^{2}c^{2}
\end{eqnarray*}
third bracketed term located on the right hand side is
\begin{eqnarray*}
&&\left[ (b_{11}+b_{23})^{2}c^{2}+(C-c^{2})^{2}-A^{2}\right] ^{2} \\
&=&\left[ (b_{11}+b_{23})^{2}c^{2}+C^{2}-2Cc^{2}+c^{4}-A^{2}\right] ^{2} \\
&=&\left[ c^{4}+\left( (b_{11}+b_{23})^{2}-2C\right) c^{2}+C^{2}-A^{2}\right]
^{2} \\
&=&\left( c^{2}+\left( (b_{11}+b_{23})^{2}-2C\right) \right) ^{2}c^{4} \\
&&+2\left( c^{4}+\left( (b_{11}+b_{23})^{2}-2C\right) c^{2}\right) \left(
C^{2}-A^{2}\right) \\
&&+\left( C^{2}-A^{2}\right) ^{2} \\
&=&\left( c^{4}+2\left( (b_{11}+b_{23})^{2}-2C\right) c^{2}+\left(
(b_{11}+b_{23})^{2}-2C\right) ^{2}\right) c^{4} \\
&&+2\left( C^{2}-A^{2}\right) c^{4} \\
&&+2\left( (b_{11}+b_{23})^{2}-2C\right) \left( C^{2}-A^{2}\right) c^{2} \\
&&+\left( C^{2}-A^{2}\right) ^{2} \\
&=&c^{8} \\
&&+2\left( (b_{11}+b_{23})^{2}-2C\right) c^{6} \\
&&+\left[ \left( (b_{11}+b_{23})^{2}-2C\right) ^{2}+2\left(
C^{2}-A^{2}\right) \right] c^{4} \\
&&+2\left( (b_{11}+b_{23})^{2}-2C\right) \left( C^{2}-A^{2}\right) c^{2} \\
&&+\left( C^{2}-A^{2}\right) ^{2}
\end{eqnarray*}

Now we can calculate the coefficients of the polynomial, for $c^{8}$ we have 
$1$. For $c^{6}$ we have:
\begin{equation*}
P=2\left( (b_{11}+b_{23})^{2}-2C\right) -(b_{12}+b_{24})^{2}
\end{equation*}

For $c^{4}$ we have
\begin{eqnarray*}
Q &=&\left( (b_{11}+b_{23})^{2}-2C\right) ^{2}+2\left( C^{2}-A^{2}\right) \\
&&-2(b_{12}+b_{24})\left[ (b_{11}+b_{23})B-\left( A+C\right) (b_{12}+b_{24})
\right] \\
&&-\left( (b_{11}+b_{23})(b_{12}+b_{24})-B\right) ^{2}
\end{eqnarray*}

For $c^{2}$ we have
\begin{eqnarray*}
R &=&2\left( C^{2}-A^{2}\right) \left( (b_{11}+b_{23})^{2}-2C\right) \\
&&-\left( \left( A+C\right) (b_{12}+b_{24})-(b_{11}+b_{23})B\right) ^{2} \\
&&-2B(A-\tilde{C})\left( B-(b_{11}+b_{23})(b_{12}+b_{24})\right)
\end{eqnarray*}

and free term is
\begin{equation*}
S=\left( C^{2}-A^{2}\right) ^{2}-B^{2}\left( A-C\right) ^{2}
\end{equation*}%
Assume that $u=c^{2}$, then we obtain the equation:
\begin{equation}
u^{4}+Pu^{3}+Qu^{2}+Ru+S=0
\end{equation}

where
\begin{eqnarray*}
&&P=2\left( (b_{11}+b_{23})^{2}-2C\right) -(b_{12}+b_{24})^{2}c^{6}, \\
&&Q=\left( (b_{11}+b_{23})^{2}-2C\right) ^{2}-2(b_{12}+b_{24})\left[
(b_{11}+b_{23})B-\left( A+C\right) (b_{12}+b_{24})\right] \\
&&\hskip15pt{}+2\left( C^{2}-A^{2}\right) -\left(
(b_{11}+b_{23})(b_{12}+b_{24})-B\right) ^{2}, \\
&&R=2\left( C^{2}-A^{2}\right) \left( (b_{11}+b_{23})^{2}-2C\right) -2B(A-C) 
\left[ B-(b_{11}+b_{23})(b_{12}+b_{24})\right] \\
&&\hskip15pt{}-\left[ (A+C)(b_{12}+b_{24})-(b_{11}+b_{23})B\right] ^{2}, \\
&&S=(C^{2}-A^{2})^{2}-B^{2}(A-C)^{2}.
\end{eqnarray*}

Thus the constant $c$ satisfies the equation (\ref{eq-xgamma4}). We solve
equation (\ref{eq-xgamma4}) and find $c^{2}$, which satisfies (\ref
{eq-xgamma4}). Then we calculate $X$. Since $X=\cos (c\gamma )$, we finally
obtain (\ref{eq-xgamma5}).

QED
\end{proof}

\subsection{Appendix 4}

No we consider some applications of the general result of Theorem~\ref
{th:stab+bif} to some special cases of our model. Consider the following
system 
\begin{eqnarray}
&&\hskip-10mmn^{\prime }(t)=(0.5F+\Phi )n(t-\gamma )\mathbf{D}(n(t-\gamma
))-0.5d\frac{(n(t)-n_{1}(t))^{2}}{n(t)}-\Psi n(t),  \label{eq-n} \\
&&\hskip-10mmn_{1}^{\prime }(t)=\left[ 0.5F\frac{n_{1}(t-\gamma )}{%
n(t-\gamma )}+\Phi \right] \mathbf{D}(n(t-\gamma ))n_{1}(t-\gamma )-\Psi
n_{1}(t).  \label{eq-n1}
\end{eqnarray}
Then for $\gamma =0$ we have \ 
\begin{eqnarray*}
&&\hskip-10mmf_{1}(n,n_{\gamma },n_{1},n_{1\gamma })=(0.5F+\Phi )n_{\gamma }%
\mathbf{D}(n_{\gamma })-0.5d\frac{(n-n_{1})^{2}}{n}-\Psi n, \\
&&\hskip-10mmf_{2}(n,n_{\gamma },n_{1},n_{1\gamma })=\left[ 0.5F\frac{%
n_{1,\gamma }}{n_{\gamma }}+\Phi \right] \mathbf{D}(n_{\gamma })n_{1\gamma
}-\Psi n_{1}.
\end{eqnarray*}

Consider the case $\mathbf{D}(n^{\ast })=e^{-\nu n^{\ast }}$ (where $\nu=-\ln (u)/2$, see (\ref{DelSupp})). Then 
\begin{equation*}
n^{\ast }=-\frac{1}{\nu }\log \mathbf{D}(n^{\ast })=-\frac{1}{\nu }
\log \left( \frac{(0.5F+\Phi )d\pm \sqrt{\Delta }}{F^{2}}\right) .
\end{equation*}
and 
\begin{equation*}
b_{11}=\frac{\partial f_{1}}{\partial n}=-0.5d\frac{n^{\ast 2}-n_{1}^{\ast 2}
}{n^{\ast 2}}-\Psi ,\quad b_{12}=\frac{\partial f_{1}}{\partial n_{\gamma }}
=(0.5F+\Phi )(1-\nu n^{\ast })e^{-\nu n^{\ast }},
\end{equation*}
\begin{equation*}
b_{13}=\frac{\partial f_{1}}{\partial n_{1}}=d\left( 1-\frac{n_{1}^{\ast }}{
n^{\ast }}\right) ,\quad b_{14}=\frac{\partial f_{1}}{\partial n_{1\gamma }}
=0,
\end{equation*}
\begin{equation*}
b_{21}=\frac{\partial f_{2}}{\partial n}=0,\quad b_{22}=\frac{\partial f_{1}
}{\partial n_{\gamma }}=-\left[ 0.5F\frac{n_{1}^{\ast 2}}{n^{\ast 2}}
+0.5F\nu \frac{n_{1}^{\ast 2}}{n^{\ast }}+\nu \Phi n_{1}^{\ast }\right]
e^{-\nu n^{\ast }},
\end{equation*}
\begin{equation*}
b_{23}=\frac{\partial f_{2}}{\partial n_{1}}=-\Psi ,\quad b_{24}=\frac{
\partial f_{2}}{\partial n_{1\gamma }}=\left[ F\frac{n_{1}^{\ast }}{n^{\ast }
}+\Phi \right] e^{-\nu n^{\ast }}.
\end{equation*}

\end{document}